\documentclass[screen=true, 10pt, acmsmall]{acmart}
\usepackage[utf8]{inputenc}
\usepackage[T1]{fontenc}

\newcommand{\packageGraphicx}{\usepackage{graphicx}}
\newcommand{\packageHyperref}{\usepackage{hyperref}}
\newcommand{\renewrmdefault}{\renewcommand{\rmdefault}{ptm}}
\newcommand{\packageRelsize}{\usepackage{relsize}}
\newcommand{\packageMathabx}{\usepackage{mathabx}}
\newcommand{\packageWasysym}{
  \let\leftmoon\relax \let\rightmoon\relax \let\fullmoon\relax \let\newmoon\relax \let\diameter\relax
  \usepackage{wasysym}}

\newcommand{\packageTextcomp}{\usepackage{textcomp}}
\newcommand{\packageFramed}{\usepackage{framed}}
\newcommand{\packageHyphenat}{\usepackage[htt]{hyphenat}}
\newcommand{\packageColor}{\usepackage[usenames,dvipsnames]{color}}
\newcommand{\doHypersetup}{\hypersetup{bookmarks=true,bookmarksopen=true,bookmarksnumbered=true}}
\newcommand{\packageTocstyle}{\IfFileExists{tocstyle.sty}{\usepackage{tocstyle}\usetocstyle{standard}}{}}
\newcommand{\packageCJK}{\IfFileExists{CJK.sty}{\usepackage{CJK}}{}}
\renewcommand\packageColor\relax
\renewcommand\packageTocstyle\relax
\renewcommand\packageMathabx{\ifx\bigtimes\undefined \usepackage{mathabx} \else \relax \fi}

\renewcommand{\renewrmdefault}{}

\packageGraphicx
\packageHyperref
\renewrmdefault
\packageRelsize
\packageMathabx
\packageWasysym
\packageTextcomp
\packageFramed
\packageHyphenat
\packageColor
\doHypersetup
\packageTocstyle
\packageCJK

\newcommand{\sectionNewpage}{}

\newcommand{\preDoc}{}
\newcommand{\postDoc}{}

\newcommand{\BookRef}[2]{\emph{#2}}
\newcommand{\ChapRef}[2]{\SecRef{#1}{#2}}
\newcommand{\SecRef}[2]{section~#1}
\newcommand{\PartRef}[2]{part~#1}
\newcommand{\BookRefUC}[2]{\BookRef{#1}{#2}}
\newcommand{\ChapRefUC}[2]{\SecRefUC{#1}{#2}}
\newcommand{\SecRefUC}[2]{Section~#1}
\newcommand{\PartRefUC}[2]{Part~#1}

\newcommand{\BookRefLocal}[3]{\hyperref[#1]{\BookRef{#2}{#3}}}
\newcommand{\ChapRefLocal}[3]{\hyperref[#1]{\ChapRef{#2}{#3}}}
\newcommand{\SecRefLocal}[3]{\hyperref[#1]{\SecRef{#2}{#3}}}
\newcommand{\PartRefLocal}[3]{\hyperref[#1]{\PartRef{#2}{#3}}}
\newcommand{\BookRefLocalUC}[3]{\hyperref[#1]{\BookRefUC{#2}{#3}}}
\newcommand{\ChapRefLocalUC}[3]{\hyperref[#1]{\ChapRefUC{#2}{#3}}}
\newcommand{\SecRefLocalUC}[3]{\hyperref[#1]{\SecRefUC{#2}{#3}}}
\newcommand{\PartRefLocalUC}[3]{\hyperref[#1]{\PartRefUC{#2}{#3}}}

\newcommand{\BookRefUN}[1]{\BookRef{}{#1}}

\newcommand{\SecRefUN}[1]{``#1''}

\newcommand{\BookRefLocalUN}[2]{\hyperref[#1]{\BookRefUN{#2}}}

\newcommand{\SecRefLocalUN}[2]{\hyperref[#1]{\SecRefUN{#2}}}

\newcommand{\SectionNumberLink}[2]{\hyperref[#1]{#2}}

\newcommand{\Scribtexttt}[1]{{\texttt{#1}}}
\newcommand{\textsub}[1]{$_{\hbox{\textsmaller{#1}}}$}

\newcommand{\Smaller}[1]{\textsmaller{#1}}

\newcommand{\planetName}[1]{PLane\hspace{-0.1ex}T}

\newcommand{\Stttextmore}{{\fontencoding{T1}\selectfont>}}

\newcommand{\Stttextbar}{{\fontencoding{T1}\selectfont|}}

\makeatletter
\def\empty@finalstrut#1{%
  \unskip\ifhmode\nobreak\fi\vrule\@width\z@\@height\z@\@depth\z@}
\def\no@strut{\global\setbox\@arstrutbox\hbox{%
    \vrule \@height\z@
           \@depth\z@
           \@width\z@}%
    \gdef\@endpbox{\empty@finalstrut\@arstrutbox\par\egroup\hfil}%
}%
\def\yes@strut{\global\setbox\@arstrutbox\hbox{%
    \vrule \@height\arraystretch \ht\strutbox
           \@depth\arraystretch \dp\strutbox
           \@width\z@}%
    \gdef\@endpbox{\@finalstrut\@arstrutbox\par\egroup\hfil}%
}%
\def\@mkpream#1{\@firstamptrue\@lastchclass6
  \let\@preamble\@empty\def\empty@preamble{\add@ins}%
  \let\protect\@unexpandable@protect
  \let\@sharp\relax\let\add@ins\relax
  \let\@startpbox\relax\let\@endpbox\relax
  \@expast{#1}%
  \expandafter\@tfor \expandafter
    \@nextchar \expandafter:\expandafter=\reserved@a\do
       {\@testpach\@nextchar
    \ifcase \@chclass \@classz \or \@classi \or \@classii \or \@classiii
      \or \@classiv \or\@classv \fi\@lastchclass\@chclass}%
  \ifcase \@lastchclass \@acol
      \or \or \@preamerr \@ne\or \@preamerr \tw@\or \or \@acol \fi}
\def\@addamp{%
  \if@firstamp
    \@firstampfalse
    \edef\empty@preamble{\add@ins}%
  \else
    \edef\@preamble{\@preamble &}%
    \edef\empty@preamble{\expandafter\noexpand\empty@preamble &\add@ins}%
  \fi}
\newif\iftw@hlines \tw@hlinesfalse
\def\@xhline{\ifx\reserved@a\hline
               \tw@hlinestrue
             \else\ifx\reserved@a\Hline
               \tw@hlinestrue
             \else
               \tw@hlinesfalse
             \fi\fi
      \iftw@hlines
        \aftergroup\do@after
      \fi
      \ifnum0=`{\fi}%
}
\def\do@after{\emptyrow[\the\doublerulesep]}
\def\emptyrow{\noalign\bgroup\@ifnextchar[\@emptyrow{\@emptyrow[\z@]}}
\def\@emptyrow[#1]{\no@strut\gdef\add@ins{\vrule \@height\z@ \@depth#1 \@width\z@}\egroup%
\empty@preamble\\
\noalign{\yes@strut\gdef\add@ins{\vrule \@height\z@ \@depth\z@ \@width\z@}}%
}
\def\tabrow#1{\noalign\bgroup\@ifnextchar[{\@tabrow{#1}}{\@tabrow{#1}[]}}
\def\@tabrow#1[#2]{\no@strut\egroup#1\ifx.#2.\\\else\\[#2]\fi\noalign{\yes@strut}}
\def\endpltstabular{\crcr\egroup\egroup \egroup}
\expandafter \let \csname endpltstabular*\endcsname = \endpltstabular
\def\pltstabular{\let\@halignto\@empty\@pltstabular}
\@namedef{pltstabular*}#1{\def\@halignto{to#1}\@pltstabular}
\def\@pltstabular{\leavevmode \bgroup \let\@acol\@tabacol
   \let\@classz\@tabclassz
   \let\@classiv\@tabclassiv \let\\\@tabularcr\@stabarray}
\def\@stabarray{\m@th\@ifnextchar[\@sarray{\@sarray[c]}}
\def\@sarray[#1]#2{%
  \bgroup
  \setbox\@arstrutbox\hbox{%
    \vrule \@height\arraystretch\ht\strutbox
           \@depth\arraystretch \dp\strutbox
           \@width\z@}%
  \@mkpream{#2}%
  \edef\@preamble{%
    \ialign \noexpand\@halignto
      \bgroup \@arstrut \@preamble \tabskip\z@skip \cr}%
  \let\@startpbox\@@startpbox \let\@endpbox\@@endpbox
  \let\tabularnewline\\%
    \let\@sharp##%
    \set@typeset@protect
    \lineskip\z@skip\baselineskip\z@skip
    \@preamble}

\makeatother

\newenvironment{bigtabular}{\begin{pltstabular}}{\end{pltstabular}}
\newcommand{\SBoxedLeft}{\textcolor[rgb]{0.6,0.6,1.0}{\vrule width 3pt\hspace{3pt}}}

\newlength{\stabLeft}
\newcommand{\bigtableleftpad}{\hspace{\stabLeft}}
\newcommand{\atItemizeStart}[0]{\addtolength{\stabLeft}{\labelsep}
                                \addtolength{\stabLeft}{\labelwidth}}

\newenvironment{SingleColumn}{\begin{list}{}{\topsep=0pt\partopsep=0pt%
\listparindent=0pt\itemindent=0pt\labelwidth=0pt\leftmargin=0pt\rightmargin=0pt%
\itemsep=0pt\parsep=0pt}\item}{\end{list}}

\newcommand{\Sendabbrev}[1]{#1\@}

\newenvironment{Subflow}{\begin{list}{}{\topsep=0pt\partopsep=0pt%
\listparindent=0pt\itemindent=0pt\labelwidth=0pt\leftmargin=0pt\rightmargin=0pt%
\itemsep=0pt}\item}{\end{list}}

\newenvironment{SInsetFlow}{\begin{quote}}{\end{quote}}

\newcommand{\SCodePreSkip}{\vskip\abovedisplayskip}
\newcommand{\SCodePostSkip}{\vskip\belowdisplayskip}
\newenvironment{SCodeFlow}{\SCodePreSkip\begin{list}{}{\topsep=0pt\partopsep=0pt%
\listparindent=0pt\itemindent=0pt\labelwidth=0pt\leftmargin=2ex\rightmargin=2ex%
\itemsep=0pt\parsep=0pt}\item}{\end{list}\SCodePostSkip}
\newcommand{\SCodeInsetBox}[1]{\setbox1=\hbox{\hbox{\hspace{2ex}#1\hspace{2ex}}}\vbox{\SCodePreSkip\vtop{\box1\SCodePostSkip}}}

\newcommand{\SVInsetPreSkip}{\vskip\abovedisplayskip}
\newcommand{\SVInsetPostSkip}{\vskip\belowdisplayskip}

\newenvironment{SCentered}{\begin{trivlist}\item \centering}{\end{trivlist}}

\newcommand{\titleAndVersionAndAuthors}[3]{\title{#1\\{\normalsize \SVersionBefore{}#2}}\author{#3}\maketitle}

\newcommand{\titleAndEmptyVersionAndAuthors}[3]{\title{#1}\author{#3}\maketitle}

\newcommand{\SAuthor}[1]{#1}
\newcommand{\SAuthorSep}[1]{\qquad}
\newcommand{\SVersionBefore}[1]{Version }

\newcommand{\SNumberOfAuthors}[1]{}

\let\SOriginalthesubsection\thesubsection
\let\SOriginalthesubsubsection\thesubsubsection

\newcommand{\Ssection}[2]{\section[#1]{#2}\let\thesubsection\SOriginalthesubsection}
\newcommand{\Ssubsection}[2]{\subsection[#1]{#2}\let\thesubsubsection\SOriginalthesubsubsection}

\newcommand{\Ssectionstar}[1]{\section*{#1}\renewcommand*\thesubsection{\arabic{subsection}}\setcounter{subsection}{0}}

\newcommand{\Ssubsubsectionstar}[1]{\subsubsection*{#1}}

\newcommand{\Ssectionstarx}[2]{\Ssectionstar{#2}\phantomsection\addcontentsline{toc}{section}{#1}}

\newcommand{\Ssubsubsectionstarx}[2]{\Ssubsubsectionstar{#2}\phantomsection\addcontentsline{toc}{subsubsection}{#1}}

\newcounter{GrouperTemp}

\newenvironment{SVerbatim}{}{}

\newcommand{\Snolinkurl}[1]{\nolinkurl{#1}}

\newcommand{\SAuthorinfo}[4]{#1}
\newcommand{\SAuthorPlace}[1]{#1}
\newcommand{\SAuthorEmail}[1]{#1}

\newcommand{\SConferenceInfo}[2]{}
\newcommand{\SCopyrightYear}[1]{}
\newcommand{\SCopyrightData}[1]{}
\newcommand{\Sdoi}[1]{}

\newcommand{\SCategory}[3]{}
\newcommand{\SCategoryPlus}[4]{}
\newcommand{\STerms}[1]{}
\newcommand{\SKeywords}[1]{}

\newcommand{\AutobibLink}[1]{\color{ACMPurple}{#1}}

\newcommand{\NoteBox}[1]{\footnote{#1}}
\newcommand{\NoteContent}[1]{#1}

\newcommand{\FootnoteRef}[1]{}
\newcommand{\FootnoteTarget}[1]{}

\newcommand{\FootnoteBlockContent}[1]{}

\newcommand{\SColorize}[2]{\color{#1}{#2}}

\newcommand{\SHyphen}[1]{#1}

\newcommand{\inColor}[2]{{\SHyphen{\Scribtexttt{\SColorize{#1}{#2}}}}}
\definecolor{PaleBlue}{rgb}{0.90,0.90,1.0}
\definecolor{LightGray}{rgb}{0.90,0.90,0.90}
\definecolor{CommentColor}{rgb}{0.76,0.45,0.12}
\definecolor{ParenColor}{rgb}{0.52,0.24,0.14}
\definecolor{IdentifierColor}{rgb}{0.15,0.15,0.50}
\definecolor{ResultColor}{rgb}{0.0,0.0,0.69}
\definecolor{ValueColor}{rgb}{0.13,0.55,0.13}
\definecolor{OutputColor}{rgb}{0.59,0.00,0.59}

\newcommand{\RktCmt}[1]{\inColor{CommentColor}{#1}}
\newcommand{\RktPn}[1]{\inColor{ParenColor}{#1}}
\newcommand{\RktInBG}[1]{\inColor{ParenColor}{#1}}
\newcommand{\RktSym}[1]{\inColor{IdentifierColor}{#1}}

\newcommand{\RktVal}[1]{\inColor{ValueColor}{#1}}

\newcommand{\RktModLink}[1]{\inColor{blue}{#1}}

\newcommand{\RktMeta}[1]{\inColor{IdentifierColor}{#1}}
\newcommand{\RktMod}[1]{\inColor{black}{#1}}
\newcommand{\RktRdr}[1]{\inColor{black}{#1}}

\newcommand{\RktErrCol}[1]{\inColor{red}{#1}}
\newcommand{\RktErr}[1]{{\RktErrCol{\textrm{\textit{#1}}}}}

\newcommand{\highlighted}[1]{\colorbox{PaleBlue}{\hspace{-0.5ex}\RktInBG{#1}\hspace{-0.5ex}}}

\newenvironment{RktBlk}{}{}

\newcommand{\Rfilebox}[2]{\begin{list}{}{\topsep=0pt\partopsep=0pt%
\listparindent=0pt\itemindent=0pt\labelwidth=0pt\leftmargin=0ex\rightmargin=0ex%
\itemsep=0pt\parsep=0pt}\item #1

#2\end{list}}

\newcommand{\Rfiletitle}[1]{\hfill \fbox{#1}}

\newcommand{\Rfilename}[1]{#1}
\newenvironment{Rfilecontent}{}{}

\newcommand{\RBackgroundLabel}[1]{}

\usepackage{ccaption}

\makeatletter
\@ifundefined{belowcaptionskip}{}{}
\makeatother

\newcommand{\Legend}[1]{~

                        \hrule width \hsize height .33pt
                        \vspace{4pt}
                        \legend{#1}}

\newcommand{\FigureTarget}[2]{#1}
\newcommand{\FigureRef}[2]{#1}

\newlength{\FigOrigskip}
\FigOrigskip=\parskip

\newcommand{\FigureSetRef}{\refstepcounter{figure}}

\newenvironment{Figure}{\begin{figure}\FigureSetRef}{\end{figure}}
\newenvironment{FigureMulti}{\begin{figure*}[t!p]\FigureSetRef}{\end{figure*}}

\newenvironment{Herefigure}{\begin{figure}[ht!]\FigureSetRef\centering}{\end{figure}}

\newenvironment{Centerfigure}{\begin{Xfigure}\centering\item}{\end{Xfigure}}

\newenvironment{Xfigure}{\begin{list}{}{\leftmargin=0pt\topsep=0pt\parsep=\FigOrigskip\partopsep=0pt}}{\end{list}}

\newenvironment{FigureInside}{}{}

\newcommand{\Centertext}[1]{\begin{center}#1\end{center}}

\newenvironment{AutoBibliography}{\begin{small}}{\end{small}}
\newcommand{\Autobibentry}[1]{\hspace{0.05\linewidth}\parbox[t]{0.95\linewidth}{\parindent=-0.05\linewidth#1\vspace{1.0ex}}}

\usepackage{calc}
\newlength{\ABcollength}

\newcommand{\Autobibref}[1]{#1}

\providecommand{\AutobibLink}[1]{#1}

\renewcommand{\titleAndVersionAndAuthors}[3]{\title{#1}#3\maketitle}
\renewcommand{\titleAndEmptyVersionAndAuthors}[3]{\titleAndVersionAndAuthors{#1}{#2}{#3}}

\def\SAuthor#1{\SAutoAuthor#1\SAutoAuthorDone{#1}}
\def\SAutoAuthorDone#1{}
\def\SAutoAuthor{\futurelet\next\SAutoAuthorX}
\def\SAutoAuthorX{\ifx\next\SAuthorinfo \let\Snext\relax \else \let\Snext\SToAuthorDone \fi \Snext}
\def\SToAuthorDone{\futurelet\next\SToAuthorDoneX}
\def\SToAuthorDoneX#1{\ifx\next\SAutoAuthorDone \let\Snext\SAddAuthorInfo \else \let\Snext\SToAuthorDone \fi \Snext}
\newcommand{\SAddAuthorInfo}[1]{\SAuthorinfo{#1}{}{}}

\renewcommand{\SAuthorinfo}[4]{\author{#1}{#2}{#3}{#4}}
\renewcommand{\SAuthorSep}[1]{}

\renewcommand{\SAuthorPlace}[1]{\affiliation{#1}}
\renewcommand{\SAuthorEmail}[1]{\email{#1}}

\renewcommand{\SConferenceInfo}[2]{\conferenceinfo{#1}{#2}}
\renewcommand{\SCopyrightYear}[1]{\copyrightyear{#1}}
\renewcommand{\SCopyrightData}[1]{\copyrightdata{#1}}

\renewcommand{\SCategory}[3]{\category{#1}{#2}{#3}}
\renewcommand{\SCategoryPlus}[4]{\category{#1}{#2}{#3}[#4]}
\renewcommand{\STerms}[1]{\terms{#1}}
\renewcommand{\SKeywords}[1]{\keywords{#1}}

\usepackage{tcolorbox}
\usepackage{wrapfig}
\usepackage{tikz}
\usepackage{accsupp}

\usetikzlibrary{tikzmark}

\newcommand{\identity}[1]{#1}
\newcommand{\goAway}[1]{}
\newcommand{\Thyperref}[2]{\hyperref[#2]{#1}}
\setcopyright{rightsretained}
\acmPrice{}
\acmDOI{10.1145/3428290}
\acmYear{2020}
\copyrightyear{2020}
\acmSubmissionID{oopsla20main-p510-p}
\acmJournal{PACMPL}
\acmVolume{4}
\acmNumber{OOPSLA}
\acmArticle{222}
\acmMonth{11}

\ccsdesc[500]{Software and its engineering~Visual languages}
\ccsdesc[500]{Software and its engineering~Macro languages}
\ccsdesc[500]{Software and its engineering~Domain specific languages}
\ccsdesc[300]{Software and its engineering~Graphical user interface languages}
\ccsdesc[500]{Software and its engineering~Integrated and visual development environments}
\ccsdesc[300]{Human-centered computing~Human computer interaction (HCI)}
\ccsdesc[100]{Human-centered computing~Accessibility technologies}

\definecolor{gold}{RGB}{255,215,0}
\begin{document}
\preDoc

\begin{abstract}Many programming problems call for turning geometrical
thoughts into code: tables, hierarchical structures, nests of objects, trees,
forests, graphs, and so on. Linear text does not do justice to such
thoughts. But, it has been the dominant programming medium for the
past and will remain so for the foreseeable future.

This paper proposes a novel mechanism for conveniently extending
textual programming languages with problem{-}specific visual
syntax. It argues the necessity of this language feature,
demonstrates the feasibility with a robust prototype, and
sketches a design plan for adapting the idea to other
languages.\end{abstract}

\keywords{Domain{-}Specific Language}\titleAndEmptyVersionAndAuthors{Adding Interactive Visual Syntax to Textual Code}{}{\SNumberOfAuthors{1}\SAuthor{\SAuthorinfo{Leif Andersen, Michael Ballantyne, Matthias Felleisen}{}{\SAuthorPlace{\institution{PLT}}\SAuthorPlace{\department{Khoury College of Computer Sciences}\institution{Northeastern University}\streetaddress{440 Huntington Ave.}\city{Boston}\state{Massachusetts}\postcode{02115}\country{United States of America}}}{\SAuthorEmail{leif@leifandersen.net}}}}
\label{t:x28part_x22Addingx5fInteractivex5fVisualx5fSyntaxx5ftox5fTextualx5fCodex22x29}

\sectionNewpage

\Ssection{Text is Not Enough, Pictures are Too Useful}{Text is Not Enough, Pictures are Too Useful}\label{t:x28part_x22Textx5fisx5fNotx5fEnoughx5fx5fPicturesx5farex5fToox5fUsefulx22x29}

Code is a message from a developer in the present to a developer in
the future, possibly the same person but aged. This future developer
must comprehend the code and reconstruct the thoughts that went into
it.  Hence, writing code means articulating thoughts as precisely as
possible.

Often these thoughts involve geometrical relationships: tables, nests
of objects, graphs, etc. Furthermore, the geometry differs from problem
domain to problem domain. To this day, though, programmers articulate
their thoughts as linear text. Unsurprisingly, code maintainers
have a hard time reconstructing geometrical thoughts from linear text,
which reduces their productivity.

At first glance, visual languages\Autobibref{~[\hyperref[t:x28autobib_x22Marat_Boshernitsan_and_Michael_Sx2e_DownesVisual_Programming_Languagesx3a_a_SurveyEECS_Departmentx2c_University_of_Californiax2c_Berkeleyx2c_UCBx2fCSDx2d04x2d13682004httpx3ax2fx2fwww2x2eeecsx2eberkeleyx2eedux2fPubsx2fTechRptsx2f2004x2f6201x2ehtmlx22x29]{\AutobibLink{Boshernitsan and Downes}} \hyperref[t:x28autobib_x22Marat_Boshernitsan_and_Michael_Sx2e_DownesVisual_Programming_Languagesx3a_a_SurveyEECS_Departmentx2c_University_of_Californiax2c_Berkeleyx2c_UCBx2fCSDx2d04x2d13682004httpx3ax2fx2fwww2x2eeecsx2eberkeleyx2eedux2fPubsx2fTechRptsx2f2004x2f6201x2ehtmlx22x29]{\AutobibLink{\Thyperref{2004}{autobiblab:7}}}]} eliminate
this problem, but they actually don{'}t. They too offer only a
fixed set of constructs, though visual ones{---}meaning a
visual language fails to address the problem{-}specific nature
of geometric thought. And, as history shows, developers
clearly prefer textual languages over visual ones{---}meaning
graphical syntax should aim to \emph{supplement}, not
displace, textual syntax.

This paper presents a mechanism for extending textual languages with
\emph{interactive and visual} programming constructs tailored to specific
problem domains. It demonstrates the feasibility of this idea with a prototype
implementation. The design space description suggests the architecture could be adapted to a
broad spectrum of programming languages.

To make this idea precise, consider a software
system that implements a game such as Tsuro.\NoteBox{\NoteContent{https://en.wikipedia.org/wiki/Tsuro}} In this
game, players take turns growing a graph from square tiles, each of
which displays four path segments.  A player places one avatar at an
entry point on the periphery of the grid{-}shaped board.  New tiles are
added next to a player{'}s avatar, and all avatars bordering this new
tile are moved as far as possible along the newly extended paths until
they face an empty place again.  If an avatar exits the board, its
owner{-}player is eliminated. The last surviving player wins.

\identity{\begin{wrapfigure}{r}{1.7in}\vspace{-0.4cm}}\raisebox{-0.3999999999999915bp}{\makebox[121.19999999999999bp][l]{\includegraphics[trim=2.4000000000000004 2.4000000000000004 2.4000000000000004 2.4000000000000004]{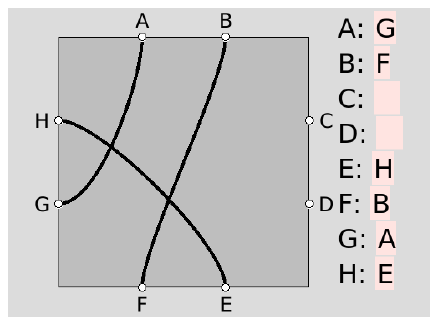}}}\identity{\vspace{-0.5cm} \end{wrapfigure}}

Now imagine a programmer wishing to articulate unit tests in the context of a
Tsuro implementation. When a mechanism for creating interactive and visual
syntax is a available, the tester may add a Tsuro tile as a new language
construct. This developer creates an instance of this syntax via UI actions,
i.e., key strokes or menu selection, and simultaneously inserts it into the
IDE. An instance is referred to as \emph{editor}. Consider the image on the
right, which shows a tile editor. It displays a graphical representation of
the tile (left) and manual entry text fields (right). These text fields and
graphical representation are linked. The graphic updates
whenever a user updates the text; the text fields update when
the programmer connects two nodes graphically via GUI actions. What is shown
here is the state of this syntax just as the developer is about to connect the
nodes labeled \RktVal{"C"} and \RktVal{"D"}.

Every tile editor compiles to code that evaluates to a bidirectional
hash{-}table representation of its connections. For the above example, the
hash{-}table connects the \RktVal{"A"} node with \RktVal{"G"}, as the following lookup operations confirm:

\noindent \begin{SCentered}\identity{~\\\begin{minipage}[c]{0.1\textwidth}%
\end{minipage}%
\begin{minipage}[c]{0.5\textwidth}}

\noindent \begin{Subflow}\Scribtexttt{{\Stttextmore} }\RktPn{(}\RktSym{hash{-}ref}\mbox{\hphantom{\Scribtexttt{x}}}\raisebox{-0.7999999999999972bp}{\makebox[59.64bp][l]{\includegraphics[trim=2.4000000000000004 2.4000000000000004 2.4000000000000004 2.4000000000000004]{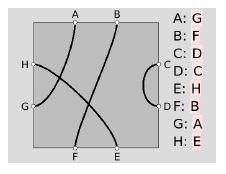}}}\mbox{\hphantom{\Scribtexttt{x}}}\RktVal{{\textquotesingle}}\RktVal{A}\RktPn{)}

\noindent \RktVal{{\textquotesingle}}\RktVal{G}\end{Subflow}

\noindent \identity{\end{minipage}\begin{minipage}[c]{0.4\textwidth}}

\noindent \begin{Subflow}\Scribtexttt{{\Stttextmore} }\RktPn{(}\RktSym{hash{-}ref}\mbox{\hphantom{\Scribtexttt{x}}}\raisebox{-0.7999999999999972bp}{\makebox[59.64bp][l]{\includegraphics[trim=2.4000000000000004 2.4000000000000004 2.4000000000000004 2.4000000000000004]{pict_2.pdf}}}\mbox{\hphantom{\Scribtexttt{x}}}\RktVal{{\textquotesingle}}\RktVal{G}\RktPn{)}

\noindent \RktVal{{\textquotesingle}}\RktVal{A}\end{Subflow}

\noindent \identity{\end{minipage}}\end{SCentered}

\noindent Note how the editor itself assumes the role of a hash{-}table here.

\identity{\begin{wrapfigure}{r}{1.3in}\vspace{-0.4cm}}\raisebox{-0.7999999999999972bp}{\makebox[86.39999999999999bp][l]{\includegraphics[trim=2.4000000000000004 2.4000000000000004 2.4000000000000004 2.4000000000000004]{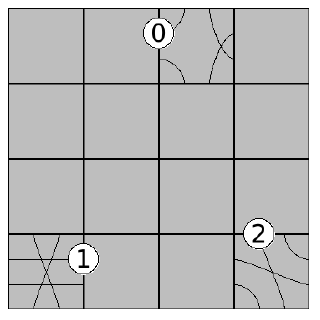}}}\identity{\vspace{-0.3cm} \end{wrapfigure}}
A Tsuro developer will also create interactive syntax for Tsuro boards. The
board syntax supplies a grid of slots. Each slot is initially empty, but the
programmer may place Tsuro tiles there to mimic players{'} actions. Take a look
at the nearby example.  In this image, three players have clearly placed one
tile each (the bottom row extreme left and extreme right plus the third slot
in the top row), and each tile is occupied by an avatar. As far as run time is
concerned, the Tsuro board editor evaluates to a matrix of tiles.

Once a programmer has extended the language with these two Tsuro{-}specific
language constructs, a unit test using these graphical editors looks
as follows in code:

\begin{SCodeFlow}\RktPn{(}\RktSym{check{-}equal{\hbox{\texttt{?}}}}\mbox{\hphantom{\Scribtexttt{x}}}\RktPn{(}\RktSym{send}\mbox{\hphantom{\Scribtexttt{x}}}\raisebox{-0.03999999999999204bp}{\makebox[56.16000000000001bp][l]{\includegraphics[trim=2.4000000000000004 2.4000000000000004 2.4000000000000004 2.4000000000000004]{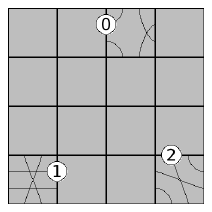}}}\mbox{\hphantom{\Scribtexttt{x}}}\RktSym{addTile}\mbox{\hphantom{\Scribtexttt{x}}}\raisebox{-0.23999999999999488bp}{\makebox[53.676bp][l]{\includegraphics[trim=2.4000000000000004 2.4000000000000004 2.4000000000000004 2.4000000000000004]{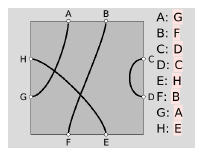}}}\mbox{\hphantom{\Scribtexttt{x}}}\RktSym{player0}\RktPn{)}\mbox{\hphantom{\Scribtexttt{x}}}\raisebox{-0.03999999999999204bp}{\makebox[56.16000000000001bp][l]{\includegraphics[trim=2.4000000000000004 2.4000000000000004 2.4000000000000004 2.4000000000000004]{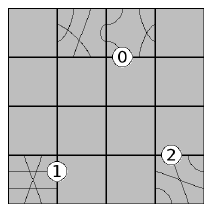}}}\RktPn{)}\end{SCodeFlow}

\noindent This one{-}line unit test checks whether the \RktSym{addTile} method
works properly. The method expects an initial board state, a tile, and a
player. Its result is a new board state with the tile placed in the slot
that the player{'}s avatar faces in the given board state and with the avatar
moved as far as possible so that it again faces an empty spot on the grid.

\begin{Figure}\begin{Centerfigure}\begin{FigureInside}\raisebox{-2.6179687499999886bp}{\makebox[272.80000000000007bp][l]{\includegraphics[trim=2.4000000000000004 2.4000000000000004 2.4000000000000004 2.4000000000000004]{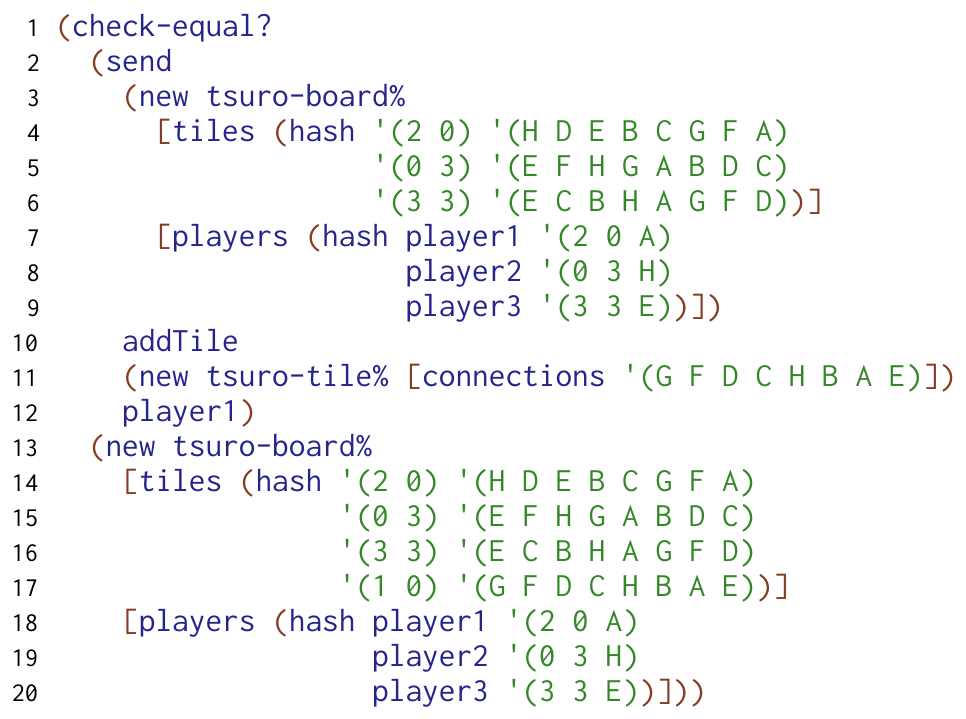}}}\end{FigureInside}\end{Centerfigure}

\Centertext{\Legend{\FigureTarget{\label{t:x28counter_x28x22figurex22_x22tsurox2dtextx22x29x29}\textsf{Fig.}~\textsf{1}. }{t:x28counter_x28x22figurex22_x22tsurox2dtextx22x29x29}\textsf{Textual Test Case}}}\end{Figure}

For comparison, figure~\hyperref[t:x28counter_x28x22figurex22_x22tsurox2dtextx22x29x29]{\FigureRef{1}{t:x28counter_x28x22figurex22_x22tsurox2dtextx22x29x29}} articulates the
same unit test with plain textual code. As with the
graphical unit test, \RktSym{addTile} expects a board, tile,
and player. The board is constructed with tiles and player
start locations. Each tile is a list of eight letters, each
representing its connecting node. The authors invite the
reader to improve on this notational choice and compare
their invention with the visual syntax above.

Interactive visual syntax is just syntax, and syntax composes. An
editor may appear within textual syntax, as shown above. And textual
syntax may appear within interactive syntax.  Let{'}s return to our
Tsuro developer who may wish to write helper functions for unit tests
that produce lists of board configurations for exploring moves. Here
is such a function, again extracted from the authors{'} code base.
\identity{\begin{wrapfigure}{r}{2.8in}\vspace{-0.4cm}}

\noindent \begin{bigtabular}{@{\bigtableleftpad}@{\SBoxedLeft}l@{}}
\SCodeInsetBox{\begin{RktBlk}\begin{tabular}[c]{@{}l@{}}
\hbox{\RktCmt{;}\RktCmt{~}\RktCmt{Tile {-}{\Stttextmore} [Listof Board]}} \\
\hbox{\RktPn{(}\RktSym{define}\mbox{\hphantom{\Scribtexttt{x}}}\RktPn{(}\RktSym{all{-}possible{-}configurations}\mbox{\hphantom{\Scribtexttt{x}}}\RktSym{t}\RktPn{)}} \\
\hbox{\mbox{\hphantom{\Scribtexttt{xx}}}\RktPn{(}\RktSym{for/list}\mbox{\hphantom{\Scribtexttt{x}}}\RktPn{(}\RktPn{[}\RktSym{d}\mbox{\hphantom{\Scribtexttt{x}}}\RktSym{DEGREES}\RktPn{]}\RktPn{)}} \\
\hbox{\mbox{\hphantom{\Scribtexttt{xxxx}}}\RktSym{{\hbox{\texttt{.}}}{\hbox{\texttt{.}}}{\hbox{\texttt{.}}}}\mbox{\hphantom{\Scribtexttt{x}}}\RktPn{(}\RktSym{send}\mbox{\hphantom{\Scribtexttt{x}}}\RktSym{t}\mbox{\hphantom{\Scribtexttt{x}}}\RktSym{rotate}\mbox{\hphantom{\Scribtexttt{x}}}\RktSym{d}\RktPn{)}\mbox{\hphantom{\Scribtexttt{x}}}\RktSym{{\hbox{\texttt{.}}}{\hbox{\texttt{.}}}{\hbox{\texttt{.}}}}\RktPn{)}\RktPn{)}}\end{tabular}\end{RktBlk}}\end{bigtabular}

\noindent \identity{\vspace{-0.5cm} \end{wrapfigure}}
As the type signature says, this function consumes a tile and generates a
list of boards. Specifically, it (\RktSym{for}) loops over a list of
\RktSym{DEGREES}, with each iteration generating an element of the resulting
list (hence \RktSym{for/list}). Each iteration generates a board by rotating
the given tile \RktSym{t} by \RktSym{d} degrees and placing it in a fixed board
context. The dots surrounding the method call are supposed to suggest this fixed
context.

Once again, the developer can either express this context as a map like that of
figure~\hyperref[t:x28counter_x28x22figurex22_x22tsurox2dtextx22x29x29]{\FigureRef{1}{t:x28counter_x28x22figurex22_x22tsurox2dtextx22x29x29}} or use an instance of interactive visual
syntax. Figure~\hyperref[t:x28counter_x28x22figurex22_x22figx3atsurox2dtestx22x29x29]{\FigureRef{2}{t:x28counter_x28x22figurex22_x22figx3atsurox2dtestx22x29x29}} shows the second scenario.
The spot on the board where the tile is to be inserted is a piece of
interactive syntax for editing code. The
zoomed image on the right indicates how a developer manipulates this
code. Clicking on this tile pops up a separate text editor. The developer
manipulates code in this editor and closes the editor when the code is
completed. Creating such an interactive Tsuro board
is only slightly more work than creating the one used for the unit test
above{---}but, in the opinion of the authors, the message it sends to the future maintainer of the code is
infinitely clearer than plain text could ever be.

\begin{Herefigure}\begin{Centerfigure}\begin{FigureInside}\begin{SCentered}\identity{~\\\begin{minipage}[c]{0\textwidth}%
\end{minipage}%
\begin{minipage}[c]{0.6\textwidth}}\identity{~\\\begin{minipage}[c]{0.9\textwidth}}

\noindent \begin{framed}\begin{RktBlk}\begin{SingleColumn}\RktCmt{;}\RktCmt{~}\RktCmt{Tile {-}{\Stttextmore} [Listof Board]}

\RktPn{(}\RktSym{define}\mbox{\hphantom{\Scribtexttt{x}}}\RktPn{(}\RktSym{all{-}possible{-}configurations}\mbox{\hphantom{\Scribtexttt{x}}}\RktSym{t}\RktPn{)}

\mbox{\hphantom{\Scribtexttt{xx}}}\RktPn{(}\RktSym{for/list}\mbox{\hphantom{\Scribtexttt{x}}}\RktPn{(}\RktPn{[}\RktSym{d}\mbox{\hphantom{\Scribtexttt{x}}}\RktSym{DEGREES}\RktPn{]}\RktPn{)}

\mbox{\hphantom{\Scribtexttt{xxxx}}}\identity{\tikzmark{board}}\raisebox{-0.7199999999999989bp}{\makebox[60.480000000000004bp][l]{\includegraphics[trim=2.4000000000000004 2.4000000000000004 2.4000000000000004 2.4000000000000004]{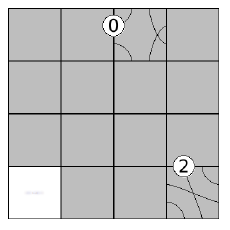}}}\RktPn{)}\RktPn{)}\end{SingleColumn}\end{RktBlk}\end{framed}

\noindent \identity{\end{minipage}\\[.2cm]}\identity{\end{minipage}\begin{minipage}[c]{0.4\textwidth}}\identity{\tikzmark{tile}}\raisebox{-0.1296874999999993bp}{\makebox[97.6bp][l]{\includegraphics[trim=2.4000000000000004 2.4000000000000004 2.4000000000000004 2.4000000000000004]{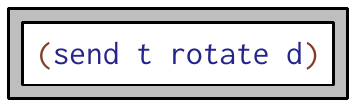}}}\identity{\end{minipage}}\end{SCentered}

\identity{\tikz[remember picture,overlay,baseline=0pt]{
  \draw[dashed] ([shift={(1.5em,0)}]pic cs:board) -- (pic cs:tile);
  \draw[dashed] ([shift={(1.5em,1.5em)}]pic cs:board) -- ([shift={(0,2.5em)}]pic cs:tile);
}}\end{FigureInside}\end{Centerfigure}

\Centertext{\Legend{\FigureTarget{\label{t:x28counter_x28x22figurex22_x22figx3atsurox2dtestx22x29x29}\textsf{Fig.}~\textsf{2}. }{t:x28counter_x28x22figurex22_x22figx3atsurox2dtestx22x29x29}\textsf{Code Inside Interactive Visual Syntax}}}\end{Herefigure}

The preceding examples are code from the authors{'} code base,
using a prototype implementation of the interactive{-}syntax
extension mechanism. In fact, this paper itself is written
using this prototype for live rendering and figure
manipulation. While the prototype is implemented in Racket\Autobibref{~[\hyperref[t:x28autobib_x22Matthew_Flatt_and_PLTReferencex3a_RacketPLT_Design_Incx2ex2c_PLTx2dTRx2d2010x2d12010httpsx3ax2fx2fracketx2dlangx2eorgx2ftr1x2fx22x29]{\AutobibLink{Flatt and PLT}} \hyperref[t:x28autobib_x22Matthew_Flatt_and_PLTReferencex3a_RacketPLT_Design_Incx2ex2c_PLTx2dTRx2d2010x2d12010httpsx3ax2fx2fracketx2dlangx2eorgx2ftr1x2fx22x29]{\AutobibLink{\Thyperref{2010}{autobiblab:25}}}]},
the next section discusses a general design space, which
should help language implementers to adapt this idea to
their world. Following that, the paper demonstrates the
expressive power of interactive syntax, explains the Racket
implementation, contrasts it with other attempts at
combining code with images, and concludes with a concise
explanation of how the prototype falls short and what is
needed to explore this idea from a user{-}facing
perspective.

\sectionNewpage

\Ssection{The Design Space}{The Design Space}\label{t:x28part_x22secx3adesignx22x29}

The project starts from an acknowledgment of the dominance of linear text. Its
goal is to enable developers to supplement linear text with visual syntax as
soon as they are tempted to document any code with some form of diagram.
Furthermore, the transition from linear text to visual and interactive syntax must become
as smooth as possible. Two prior projects \Autobibref{~[\hyperref[t:x28autobib_x22Andrew_Dx2e_Eisenberg_and_Gregor_KiczalesExpressive_Programs_Through_Presentation_ExtensionIn_Procx2e_International_Conference_on_Aspectx2dOriented_Software_Developmentx2c_ppx2e_73x2dx2d842007httpsx3ax2fx2fdoix2eorgx2f10x2e1145x2f1218563x2e1218573x22x29]{\AutobibLink{Eisenberg and Kiczales}} \hyperref[t:x28autobib_x22Andrew_Dx2e_Eisenberg_and_Gregor_KiczalesExpressive_Programs_Through_Presentation_ExtensionIn_Procx2e_International_Conference_on_Aspectx2dOriented_Software_Developmentx2c_ppx2e_73x2dx2d842007httpsx3ax2fx2fdoix2eorgx2f10x2e1145x2f1218563x2e1218573x22x29]{\AutobibLink{\Thyperref{2007}{autobiblab:14}}}; \hyperref[t:x28autobib_x22Gx2e_Wx2e_Frenchx2c_Jx2e_Rx2e_Kennawayx2c_and_Ax2e_Mx2e_DayPrograms_as_Visualx2c_Interactive_DocumentsJournal_of_Softwarex3a_Practice_and_Experience_44x288x29x2c_ppx2e_911x2dx2d9302014httpsx3ax2fx2fdoix2eorgx2f10x2e1002x2fspex2e2182x22x29]{\AutobibLink{French et al\Sendabbrev{.}}} \hyperref[t:x28autobib_x22Gx2e_Wx2e_Frenchx2c_Jx2e_Rx2e_Kennawayx2c_and_Ax2e_Mx2e_DayPrograms_as_Visualx2c_Interactive_DocumentsJournal_of_Softwarex3a_Practice_and_Experience_44x288x29x2c_ppx2e_911x2dx2d9302014httpsx3ax2fx2fdoix2eorgx2f10x2e1002x2fspex2e2182x22x29]{\AutobibLink{\Thyperref{2014}{autobiblab:27}}}]}
share the fundamental assumption and the ultimate goal of this project. Their
shortcomings, however, expose the serious challenges of this undertaking.

In order to gain wide spread adoption, interactive syntax must satisfy three
primary criteria. First, it is imperative to demonstrate the feasibility of
mixing textual and interactive syntax in the context of an ordinary and existing
programming language. Ordinary here refers to the kind of languages developers
already use: with classes, objects, functions, and side effects. Existing does
not mean {``}widely used,{''} though in use by a non{-}trivial community. Adding a
visual{-}syntax mechanism to such a language is the simplest ways to tempt
programmers into its use and to convince them of its usefulness as well as its
usability.\NoteBox{\NoteContent{The popularity of notebooks, REPLs, prompts, and virtual
machines demonstrate the usefulness of visual organizations for experimentation
and data analysis. But, they also show how quickly such an approach becomes
unwieldy as software systems grow.  Techniques such as snapshots and
caf\'{e}s\Autobibref{~[\hyperref[t:x28autobib_x22Rx2e_Kent_DybvigThe_Development_of_Chez_SchemeIn_Procx2e_International_Conference_on_Functional_Programmingx2c_ppx2e_1x2dx2d122006httpsx3ax2fx2fdoix2eorgx2f10x2e1145x2f1160074x2e1159805x22x29]{\AutobibLink{Dybvig}} \hyperref[t:x28autobib_x22Rx2e_Kent_DybvigThe_Development_of_Chez_SchemeIn_Procx2e_International_Conference_on_Functional_Programmingx2c_ppx2e_1x2dx2d122006httpsx3ax2fx2fdoix2eorgx2f10x2e1145x2f1160074x2e1159805x22x29]{\AutobibLink{\Thyperref{2006}{autobiblab:13}}}]} reduce the problem, but do not eliminate it.}}

Second, more than one specific interactive development environment (IDE) must
accommodate interactive syntax. Indeed, programs in the revised language must
not even preclude editing in purely textual IDEs ones such as Vim. Developers
frequently have strong preferences concerning the IDE they use. They are much less
likely to allow teammates to use interactive syntax if it means everyone on the
team must migrate from their favorite IDEs to a specific one.

Finally, developers must be able to amortize the investment in graphical user
interfaces. The construction of interactive{-}syntax extensions demands code that
implements simple GUIs, a potentially time{-}consuming task compared to, say,
drawing ASCII diagrams. If these GUIs can share code with the actual interface
of the software system, though, the cost of creating interactive syntax
extensions may look quite reasonable. The implication is that an
interactive{-}syntax extension mechanism must use the existing GUI libraries of
the chosen language as much as possible.

With these criteria in mind, we can now state specific design desiderata:

\noindent \begin{enumerate}\atItemizeStart

\item \emph{An interactive visual syntax is just syntax. It merely articulates
an idea better than textual syntax.} If the underlying grammar distinguishes
among definitions, expressions, patterns, and other syntactic categories, it
should be possible to use visual syntax in all of them.

\item \emph{Interactive syntax is persistent.} The point of interactive syntax
is that it permits developers to send a visual message across time. In
contrast to wizards and code generators, it is not a GUI that pops up so that a
developer can create textual code. Hence an editor must continuously serialize
and save its state. One developer can then quit the IDE, and another can open
this same file later, at which point the interactive{-}syntax editor can render itself after
deserializing the saved state.

\item \emph{Interactive visual syntax constructs must compose with textual
syntax according to the grammatical productions.} As already demonstrated,
this implies that textual syntax may contain interactive visual syntax and
vice versa. In principle, developers should be able to nest visual and textual
syntax arbitrarily deep.

\item \emph{The ideal mechanism implements a low{-}friction model for the
definition and use of interactive visual syntax.} A developer should be able
to define and use an interactive syntax extension in the same file. Indeed,
this principle can be further extended to lexical scope.  As with traditional
syntax extension mechanisms, a developer should be able to define and use an
interactive{-}syntax extension within a function, method, class, module, or any
other form of code block that sets up a lexical scope.

\item \emph{The creator of interactive visual syntax must be able to exploit
the entire language, including the extension mechanism itself.}  If the
underlying language permits abstraction over syntax, say like Rust, then a
developer must be able to abstract over definitions of interactive visual
syntax; in the same vein, an instance of interactive visual syntax may
create new forms of visual and textual syntax abstractions.

\item \emph{Interactive visual syntax demands \textbf{sandboxing for the IDE}.} The instantiation
of interactive visual syntax into an editor runs code. When developers manipulate editors,
code is run again.  In an ordinary language, such code may have side effects.
Hence, the extension mechanism must ensure that the code does not adversely
affect the functioning of the IDE.

\item \emph{Interactive visual syntax demands \textbf{sandboxing for code composition}.}
Experience with syntax extension mechanisms\Autobibref{~[\hyperref[t:x28autobib_x22Matthew_FlattComposable_and_Compilable_Macrosx2c_You_Want_It_Whenx3fIn_Procx2e_International_Conference_on_Functional_Programmingx2c_ppx2e_72x2dx2d832002x22x29]{\AutobibLink{Flatt}} \hyperref[t:x28autobib_x22Matthew_FlattComposable_and_Compilable_Macrosx2c_You_Want_It_Whenx3fIn_Procx2e_International_Conference_on_Functional_Programmingx2c_ppx2e_72x2dx2d832002x22x29]{\AutobibLink{\Thyperref{2002}{autobiblab:21}}}]} suggests that
it is also desirable to isolate the execution of the edit{-}time code from
other phases, say, the compilation phase and the runtime phase.
This sandboxing greatly facilitates co{-}mingling code from different phases.\end{enumerate}

\noindent The Racket prototype realizes the design principles, and its use is illustrated next.

\sectionNewpage

\Ssection{Constructing Interactive Syntax}{Constructing Interactive Syntax}\label{t:x28part_x22secx3abasicx2dsyntaxx22x29}

Writing interactive{-}syntax extensions parallels the process
of writing traditional syntax extensions. Both traditional
and interactive syntax extensions allow compile{-}time code
and run{-}time code to co{-}exist, in the same program,
the same file, and even the same expression. Interactive syntax
adds the additional notion of an \emph{edit{-}time} phase,
code that runs while the programmer edits the code.

This section describes interactive{-}syntax extensions, its
parallels to traditional syntax extensions\Autobibref{~[\hyperref[t:x28autobib_x22Matthias_Felleisenx2c_Robert_Bruce_Findlerx2c_Matthew_Flattx2c_Shriram_Krishnamurthix2c_Eli_Barzilayx2c_Jay_McCarthyx2c_and_Sam_Tobinx2dHochstadtA_Programmable_Programming_LanguageCommunications_of_the_ACM_61x283x29x2c_ppx2e_62x2dx2d712018httpsx3ax2fx2fdoix2eorgx2f10x2e1145x2f3127323x22x29]{\AutobibLink{Felleisen et al\Sendabbrev{.}}} \hyperref[t:x28autobib_x22Matthias_Felleisenx2c_Robert_Bruce_Findlerx2c_Matthew_Flattx2c_Shriram_Krishnamurthix2c_Eli_Barzilayx2c_Jay_McCarthyx2c_and_Sam_Tobinx2dHochstadtA_Programmable_Programming_LanguageCommunications_of_the_ACM_61x283x29x2c_ppx2e_62x2dx2d712018httpsx3ax2fx2fdoix2eorgx2f10x2e1145x2f3127323x22x29]{\AutobibLink{\Thyperref{2018}{autobiblab:18}}}]},
and how these extensions can implement the Tsuro tiles from
the previous section. The key novelty is
\RktSym{define{-}interactive{-}syntax}, a construct for creating
new interactive{-}syntax forms.

\Ssubsection{Some Basic Background on Syntax Extensions}{Some Basic Background on Syntax Extensions}\label{t:x28part_x22subx3abasicx2dsyntaxx22x29}

Racket comes with a highly expressive sub{-}language of macros
that enable programmers to extend the language. To process a
program, Racket{'}s reader creates a syntax
tree, stored as a compile{-}time object. Next the macro expander traverses this tree
and discovers Racket{'}s core forms while rewriting
instances of macros into new syntax sub{-}trees. In order to
realize this rewriting process, the expander partially
expands all syntax trees in a module and then adds macro definitions to a table of macro
rewriting rules for the second, full expansion pass.

Macros are functions from syntax trees to syntax trees. Instead of

\begin{SCodeFlow}\RktPn{(}\RktSym{define}\mbox{\hphantom{\Scribtexttt{x}}}\RktPn{(}\RktSym{f}\mbox{\hphantom{\Scribtexttt{x}}}\RktSym{x}\RktPn{)}\mbox{\hphantom{\Scribtexttt{x}}} $\_$ $\_$ $\_$ elided $\_$ $\_$ $\_$\RktPn{)}\end{SCodeFlow}

\noindent a programmer writes

\begin{SCodeFlow}\RktPn{(}\RktSym{define{-}syntax}\mbox{\hphantom{\Scribtexttt{x}}}\RktPn{(}\RktSym{m}\mbox{\hphantom{\Scribtexttt{x}}}\RktSym{x}\RktPn{)}\mbox{\hphantom{\Scribtexttt{x}}} $\_$ $\_$ $\_$ elided $\_$ $\_$ $\_$\RktPn{)}\end{SCodeFlow}

\noindent to define the syntax transformer \RktSym{m}. When the
expander encounters \RktPn{(}\RktSym{m}\Scribtexttt{ } $\_$ $\_$ $\_$ elided $\_$ $\_$ $\_$\RktPn{)}, it applies
the transformer to this entire syntax tree. The transformer
is expected to return a new syntax tree, and once this
happens, the expander starts over with the traversal for
this new one. While macros are often used to produce
expressions and definitions, \RktPn{(}\RktSym{m}\Scribtexttt{ } $\_$ $\_$ $\_$ elided $\_$ $\_$ $\_$\RktPn{)} may
also expand to \RktSym{define{-}syntax} and thus introduce new
syntax definitions. The {``}on{-}the{-}fly{''} definitions are why
the macro expander takes the one{-}and{-}a{-}half pass approach,
described above, to elaborating modules
into the Racket core language.

A programmer may specify a macro either as a declarative rewriting rule or
as a procedural process. For the second variant, the macro may wish to rely on
functions that are available at compile time. In Racket a module may import
ordinary libraries \RktSym{for{-}syntax} or it may locally define functions
to be available at compile time with \RktSym{begin{-}for{-}syntax}. Thus,

\begin{SCodeFlow}\begin{RktBlk}\begin{SingleColumn}\RktPn{(}\RktSym{begin{-}for{-}syntax}

\mbox{\hphantom{\Scribtexttt{xx}}}\RktPn{(}\RktSym{define}\mbox{\hphantom{\Scribtexttt{x}}}\RktPn{(}\RktSym{g}\mbox{\hphantom{\Scribtexttt{x}}}\RktSym{a}\mbox{\hphantom{\Scribtexttt{x}}}\RktSym{b}\mbox{\hphantom{\Scribtexttt{x}}}\RktSym{c}\RktPn{)}\mbox{\hphantom{\Scribtexttt{x}}} $\_$ $\_$ $\_$ elided $\_$ $\_$ $\_$\RktPn{)}\RktPn{)}\end{SingleColumn}\end{RktBlk}\end{SCodeFlow}

\noindent makes the ternary function \RktSym{g} available to procedural macros.

Racketeers speak of the compile{-}time phase and the run{-}time
phase. Here a \emph{phase} is a syntactic separation that
determines when code runs and provides a semantic separation of
its effects\Autobibref{~[\hyperref[t:x28autobib_x22Matthew_FlattComposable_and_Compilable_Macrosx2c_You_Want_It_Whenx3fIn_Procx2e_International_Conference_on_Functional_Programmingx2c_ppx2e_72x2dx2d832002x22x29]{\AutobibLink{Flatt}} \hyperref[t:x28autobib_x22Matthew_FlattComposable_and_Compilable_Macrosx2c_You_Want_It_Whenx3fIn_Procx2e_International_Conference_on_Functional_Programmingx2c_ppx2e_72x2dx2d832002x22x29]{\AutobibLink{\Thyperref{2002}{autobiblab:21}}}]}. Naturally,
function definitions for the compile{-}time phase may call
macros defined for the compile{-}time phase, which are in
turned defined in the compile{-}time phase{'}s compile{-}time
phase. Programmatically a module may thus look as follows:
\identity{~\\\begin{minipage}[c]{0.9\textwidth}}

\begin{SCodeFlow}\begin{RktBlk}\begin{SingleColumn}\Smaller{\Smaller{\Scribtexttt{1}\mbox{\hphantom{\Scribtexttt{x}}}}}\RktModLink{\RktMod{\#lang}}\RktMeta{}\mbox{\hphantom{\Scribtexttt{x}}}\RktMeta{}\RktModLink{\RktSym{racket}}\RktMeta{}

\Smaller{\Smaller{\Scribtexttt{2}\mbox{\hphantom{\Scribtexttt{x}}}}}\RktMeta{}\RktPn{(}\RktSym{define{-}syntax}\RktMeta{}\mbox{\hphantom{\Scribtexttt{x}}}\RktMeta{}\RktPn{(}\RktSym{m}\RktMeta{}\mbox{\hphantom{\Scribtexttt{x}}}\RktMeta{}\RktSym{x}\RktPn{)}\RktMeta{}\mbox{\hphantom{\Scribtexttt{x}}}\RktMeta{}\RktSym{{\char`\_}}\RktMeta{}\mbox{\hphantom{\Scribtexttt{x}}}\RktMeta{}\RktSym{{\char`\_}}\RktMeta{}\mbox{\hphantom{\Scribtexttt{x}}}\RktMeta{}\RktSym{{\char`\_}}\RktMeta{}\mbox{\hphantom{\Scribtexttt{x}}}\RktMeta{}\RktPn{(}\RktSym{f}\RktMeta{}\mbox{\hphantom{\Scribtexttt{x}}}\RktMeta{}\RktSym{a}\RktMeta{}\mbox{\hphantom{\Scribtexttt{x}}}\RktMeta{}\RktSym{b}\RktPn{)}\RktMeta{}\mbox{\hphantom{\Scribtexttt{x}}}\RktMeta{}\RktSym{{\char`\_}}\RktMeta{}\mbox{\hphantom{\Scribtexttt{x}}}\RktMeta{}\RktSym{{\char`\_}}\RktMeta{}\mbox{\hphantom{\Scribtexttt{x}}}\RktMeta{}\RktSym{{\char`\_}}\RktPn{)}\RktMeta{}

\Smaller{\Smaller{\Scribtexttt{3}\mbox{\hphantom{\Scribtexttt{x}}}}}\RktMeta{}\RktPn{(}\RktSym{begin{-}for{-}syntax}\RktMeta{}

\Smaller{\Smaller{\Scribtexttt{4}\mbox{\hphantom{\Scribtexttt{x}}}}}\RktMeta{}\mbox{\hphantom{\Scribtexttt{xx}}}\RktMeta{}\RktPn{(}\RktSym{define}\RktMeta{}\mbox{\hphantom{\Scribtexttt{x}}}\RktMeta{}\RktPn{(}\RktSym{f}\RktMeta{}\mbox{\hphantom{\Scribtexttt{x}}}\RktMeta{}\RktSym{y}\RktMeta{}\mbox{\hphantom{\Scribtexttt{x}}}\RktMeta{}\RktSym{z}\RktPn{)}\RktMeta{}\mbox{\hphantom{\Scribtexttt{x}}}\RktMeta{}\RktSym{{\char`\_}}\RktMeta{}\mbox{\hphantom{\Scribtexttt{x}}}\RktMeta{}\RktSym{{\char`\_}}\RktMeta{}\mbox{\hphantom{\Scribtexttt{x}}}\RktMeta{}\RktSym{{\char`\_}}\RktMeta{}\mbox{\hphantom{\Scribtexttt{x}}}\RktMeta{}\RktPn{(}\RktSym{k}\RktMeta{}\mbox{\hphantom{\Scribtexttt{x}}}\RktMeta{}\RktSym{c}\RktMeta{}\mbox{\hphantom{\Scribtexttt{x}}}\RktMeta{}\RktSym{d}\RktMeta{}\mbox{\hphantom{\Scribtexttt{x}}}\RktMeta{}\RktSym{e}\RktPn{)}\RktMeta{}\mbox{\hphantom{\Scribtexttt{x}}}\RktMeta{}\RktSym{{\char`\_}}\RktMeta{}\mbox{\hphantom{\Scribtexttt{x}}}\RktMeta{}\RktSym{{\char`\_}}\RktMeta{}\mbox{\hphantom{\Scribtexttt{x}}}\RktMeta{}\RktSym{{\char`\_}}\RktPn{)}\RktMeta{}

\Smaller{\Smaller{\Scribtexttt{5}\mbox{\hphantom{\Scribtexttt{x}}}}}\RktMeta{}\mbox{\hphantom{\Scribtexttt{xx}}}\RktMeta{}\RktPn{(}\RktSym{define{-}syntax}\RktMeta{}\mbox{\hphantom{\Scribtexttt{x}}}\RktMeta{}\RktPn{(}\RktSym{k}\RktMeta{}\mbox{\hphantom{\Scribtexttt{x}}}\RktMeta{}\RktSym{w}\RktPn{)}\RktMeta{}\mbox{\hphantom{\Scribtexttt{x}}}\RktMeta{}\RktSym{{\char`\_}}\RktMeta{}\mbox{\hphantom{\Scribtexttt{x}}}\RktMeta{}\RktSym{{\char`\_}}\RktMeta{}\mbox{\hphantom{\Scribtexttt{x}}}\RktMeta{}\RktSym{{\char`\_}}\RktMeta{}\mbox{\hphantom{\Scribtexttt{x}}}\RktMeta{}\RktPn{(}\RktSym{g}\RktPn{)}\RktMeta{}\mbox{\hphantom{\Scribtexttt{x}}}\RktMeta{}\RktSym{{\char`\_}}\RktMeta{}\mbox{\hphantom{\Scribtexttt{x}}}\RktMeta{}\RktSym{{\char`\_}}\RktMeta{}\mbox{\hphantom{\Scribtexttt{x}}}\RktMeta{}\RktSym{{\char`\_}}\RktPn{)}\RktMeta{}

\Smaller{\Smaller{\Scribtexttt{6}\mbox{\hphantom{\Scribtexttt{x}}}}}\RktMeta{}\mbox{\hphantom{\Scribtexttt{xx}}}\RktMeta{}\RktPn{(}\RktSym{begin{-}for{-}syntax}\RktMeta{}

\Smaller{\Smaller{\Scribtexttt{7}\mbox{\hphantom{\Scribtexttt{x}}}}}\RktMeta{}\mbox{\hphantom{\Scribtexttt{xxxx}}}\RktMeta{}\RktPn{(}\RktSym{define}\RktMeta{}\mbox{\hphantom{\Scribtexttt{x}}}\RktMeta{}\RktPn{(}\RktSym{g}\RktPn{)}\RktMeta{}\mbox{\hphantom{\Scribtexttt{x}}}\RktMeta{}\RktSym{{\char`\_}}\RktMeta{}\mbox{\hphantom{\Scribtexttt{x}}}\RktMeta{}\RktSym{{\char`\_}}\RktMeta{}\mbox{\hphantom{\Scribtexttt{x}}}\RktMeta{}\RktSym{{\char`\_}}\RktMeta{}\mbox{\hphantom{\Scribtexttt{x}}}\RktMeta{}\RktSym{elided}\RktMeta{}\mbox{\hphantom{\Scribtexttt{x}}}\RktMeta{}\RktSym{{\char`\_}}\RktMeta{}\mbox{\hphantom{\Scribtexttt{x}}}\RktMeta{}\RktSym{{\char`\_}}\RktMeta{}\mbox{\hphantom{\Scribtexttt{x}}}\RktMeta{}\RktSym{{\char`\_}}\RktPn{)}\RktPn{)}\RktPn{)}\RktMeta{}\end{SingleColumn}\end{RktBlk}\end{SCodeFlow}

\noindent \identity{\end{minipage}\\[.2cm]}
Phases in Racket programs are
nested arbitrarily deep, because programmers
appreciate this power.

\Ssubsection{Editing As a Phase}{Editing As a Phase}\label{t:x28part_x22Editingx5fAsx5fax5fPhasex22x29}

Our technical idea is to support the
\emph{editing phase}, linguistically.
The extension to Racket consists of two
linguistic constructs, analogous to \RktSym{define{-}syntax}
and \RktSym{begin{-}for{-}syntax}:

\noindent \begin{itemize}\atItemizeStart

\item \RktSym{define{-}interactive{-}syntax} creates and names
an interactive{-}syntax extension. Roughly speaking, an
interactive{-}syntax extension is a specialized graphical user
interface defined as a class{-}like entity. It comes with one
method of rendering itself in a graphical context, such as
an IDE; a second for reacting to events (keyboard, mouse,
etc.); a third for tracking local state; and a final one for
turning the state into running code.

\item \RktSym{begin{-}for{-}interactive{-}syntax} specifies code
that exists for the editing phase in an interactive{-}syntax
extension. It can only appear at the module top level.\end{itemize}

\noindent That is, an interactive{-}syntax extension executes during
edit time yet denotes compile{-}time and run{-}time code.

The code in a \RktSym{begin{-}for{-}interactive{-}syntax} block
runs when the editor for this module is opened, or
its content is modified.
Like \RktSym{begin{-}for{-}syntax},
\RktSym{begin{-}for{-}interactive{-}syntax} is mostly used as a
mechanism for definitions that are needed to define
interactive{-}syntax.

In Tsuro, for example, a \RktSym{trace{-}player} function
calculates the path of a player{'}s token from the start
position to the end position.
The Tsuro{-}board editors use it to
calculate where they should draw the tokens after the placement of a tile,
meaning the function is needed at edit{-}time, too:

\noindent \begin{SCentered}\identity{~\\\begin{minipage}[c]{0\textwidth}%
\end{minipage}%
\begin{minipage}[c]{1\textwidth}}

\noindent \begin{SCentered}\identity{~\\\begin{minipage}[c]{0\textwidth}%
\end{minipage}%
\begin{minipage}[c]{0.48\textwidth}}\identity{~\\\begin{minipage}[c]{0.99\textwidth}}

\noindent \Rfilebox{\Rfiletitle{\Rfilename{interactive syntax module}}}%
{\begin{Rfilecontent}\begin{framed}\begin{RktBlk}\begin{SingleColumn}\RktPn{(}\RktSym{begin{-}for{-}interactive{-}syntax}

\mbox{\hphantom{\Scribtexttt{xx}}}\RktPn{(}\RktSym{require}\mbox{\hphantom{\Scribtexttt{x}}}\RktVal{"Model/player{\hbox{\texttt{.}}}rkt"}\RktPn{)}\RktPn{)}\end{SingleColumn}\end{RktBlk}\end{framed}\end{Rfilecontent}}

\noindent \identity{\end{minipage}\\[.2cm]}\identity{\end{minipage}\begin{minipage}[c]{0.52\textwidth}}\identity{~\\\begin{minipage}[c]{0.99\textwidth}}

\noindent \Rfilebox{\Rfiletitle{\Rfilename{\Scribtexttt{"Model/player{\hbox{\texttt{.}}}rkt"} module}}}%
{\begin{Rfilecontent}\begin{framed}\begin{RktBlk}\begin{SingleColumn}\RktPn{(}\RktSym{define}\mbox{\hphantom{\Scribtexttt{x}}}\RktPn{(}\RktSym{trace{-}player}\mbox{\hphantom{\Scribtexttt{x}}}\RktSym{board}\mbox{\hphantom{\Scribtexttt{x}}}\RktSym{player}\RktPn{)}

\mbox{\hphantom{\Scribtexttt{xx}}} $\_$ $\_$ $\_$ elided $\_$ $\_$ $\_$\RktPn{)}\end{SingleColumn}\end{RktBlk}\end{framed}\end{Rfilecontent}}

\noindent \identity{\end{minipage}\\[.2cm]}\identity{\end{minipage}}\end{SCentered}

\noindent \identity{\end{minipage}\begin{minipage}[c]{0\textwidth}}\identity{\vspace{1in}}\identity{\end{minipage}}\end{SCentered}

\noindent Requiring an ordinary module at edit time imports its definitions into the desired scope at edit time.

\Ssubsection{Bridging the Gap Between Edit{-}Time and Run{-}Time}{Bridging the Gap Between Edit{-}Time and Run{-}Time}\label{t:x28part_x22Bridgingx5fthex5fGapx5fBetweenx5fEditx2dTimex5fandx5fRunx2dTimex22x29}

The \RktSym{define{-}interactive{-}syntax} form bridges the gap
between run{-}time and edit{-}time code; i.e., it
is analogous to \RktSym{define{-}syntax}. While
\RktSym{define{-}syntax} allows run{-}time code and compile{-}time
code to interact{---}the latter referencing and generating the
former{---}\RktSym{define{-}interactive{-}syntax} connects
edit{-}time code with run{-}time code, generating the latter from
the former.

A Racket syntax extension consists of a new grammar
production and a translation into
existing syntax. By contrast, interactive{-}syntax extensions
consist of four different pieces:

\noindent \begin{enumerate}\atItemizeStart

\item a \emph{presentation}, meaning a method for rendering its current state, runs at edit{-}time;

\item an \emph{interaction}, that is, a method for reacting to direct
manipulation actions, runs at edit{-}time;

\item a \emph{semantics}, which is compile{-}time code that generates run{-}time code; and

\item \emph{persistent storage}, i.e., a specification of persisted data and its
external representation, which exists at both edit{-}time and compile{-}time.\end{enumerate}

Once an interactive{-}syntax definition exists, a developer
may insert an instance into an IDE buffer by a UI action,
such as a mouse click or button press.
DrRacket\Autobibref{~[\hyperref[t:x28autobib_x22Robert_Bruce_Findler_and_PLTDrRacketx3a_Programming_EnvironmentPLT_Design_Incx2ex2c_PLTx2dTRx2d2010x2d22010httpsx3ax2fx2fracketx2dlangx2eorgx2ftr2x2fx22x29]{\AutobibLink{Findler and PLT}} \hyperref[t:x28autobib_x22Robert_Bruce_Findler_and_PLTDrRacketx3a_Programming_EnvironmentPLT_Design_Incx2ex2c_PLTx2dTRx2d2010x2d22010httpsx3ax2fx2fracketx2dlangx2eorgx2ftr2x2fx22x29]{\AutobibLink{\Thyperref{2010}{autobiblab:20}}}]}, the predominant Racket IDE,
implicitly places this editor within a
\RktSym{begin{-}for{-}interactive{-}syntax} block so that its
edit{-}time code runs continuously during program development.
When the programmer requests the execution of the module,
the extension{'}s semantics turns the current state into
run{-}time code, properly integrated into the lexical scope of
its location.

Concretely the \RktSym{define{-}interactive{-}syntax} form consists of
four sub{-}forms:
\identity{~\\\begin{minipage}[c]{0.9\textwidth}}

\begin{SCodeFlow}\begin{RktBlk}\begin{SingleColumn}\Smaller{\Smaller{\Scribtexttt{1}\mbox{\hphantom{\Scribtexttt{xx}}}}}\RktPn{(}\RktSym{define{-}interactive{-}syntax}\RktMeta{}\mbox{\hphantom{\Scribtexttt{x}}}\RktMeta{}\RktSym{name\$}\RktMeta{}\mbox{\hphantom{\Scribtexttt{x}}}\RktMeta{}\RktSym{base\$}\RktMeta{}

\Smaller{\Smaller{\Scribtexttt{2}\mbox{\hphantom{\Scribtexttt{xx}}}}}\RktMeta{}\mbox{\hphantom{\Scribtexttt{xx}}}\RktMeta{}\RktPn{(}\RktSym{define/public}\RktMeta{}\mbox{\hphantom{\Scribtexttt{x}}}\RktMeta{}\RktPn{(}\RktSym{draw}\RktMeta{}\mbox{\hphantom{\Scribtexttt{x}}}\RktMeta{}\RktSym{ctx}\RktPn{)}\RktMeta{}

\Smaller{\Smaller{\Scribtexttt{3}\mbox{\hphantom{\Scribtexttt{xx}}}}}\RktMeta{}\mbox{\hphantom{\Scribtexttt{xxxxx}}}\RktMeta{}\RktSym{{\char`\_}}\RktMeta{}\mbox{\hphantom{\Scribtexttt{x}}}\RktMeta{}\RktSym{{\char`\_}}\RktMeta{}\mbox{\hphantom{\Scribtexttt{x}}}\RktMeta{}\RktSym{{\char`\_}}\RktMeta{}\mbox{\hphantom{\Scribtexttt{x}}}\RktMeta{}\RktSym{visually}\RktMeta{}\mbox{\hphantom{\Scribtexttt{x}}}\RktMeta{}\RktSym{rendering}\RktMeta{}\mbox{\hphantom{\Scribtexttt{x}}}\RktMeta{}\RktSym{{\char`\_}}\RktMeta{}\mbox{\hphantom{\Scribtexttt{x}}}\RktMeta{}\RktSym{{\char`\_}}\RktMeta{}\mbox{\hphantom{\Scribtexttt{x}}}\RktMeta{}\RktSym{{\char`\_}}\RktMeta{}\mbox{\hphantom{\Scribtexttt{x}}}\RktMeta{}\RktPn{)}\RktMeta{}

\Smaller{\Smaller{\Scribtexttt{4}\mbox{\hphantom{\Scribtexttt{xx}}}}}\RktMeta{}\mbox{\hphantom{\Scribtexttt{xx}}}\RktMeta{}\RktPn{(}\RktSym{define/public}\RktMeta{}\mbox{\hphantom{\Scribtexttt{x}}}\RktMeta{}\RktPn{(}\RktSym{on{-}event}\RktMeta{}\mbox{\hphantom{\Scribtexttt{x}}}\RktMeta{}\RktSym{event}\RktPn{)}\RktMeta{}

\Smaller{\Smaller{\Scribtexttt{5}\mbox{\hphantom{\Scribtexttt{xx}}}}}\RktMeta{}\mbox{\hphantom{\Scribtexttt{xxxxx}}}\RktMeta{}\RktSym{{\char`\_}}\RktMeta{}\mbox{\hphantom{\Scribtexttt{x}}}\RktMeta{}\RktSym{{\char`\_}}\RktMeta{}\mbox{\hphantom{\Scribtexttt{x}}}\RktMeta{}\RktSym{{\char`\_}}\RktMeta{}\mbox{\hphantom{\Scribtexttt{x}}}\RktMeta{}\RktSym{interactions}\RktMeta{}\mbox{\hphantom{\Scribtexttt{x}}}\RktMeta{}\RktSym{code}\RktMeta{}\mbox{\hphantom{\Scribtexttt{x}}}\RktMeta{}\RktSym{{\char`\_}}\RktMeta{}\mbox{\hphantom{\Scribtexttt{x}}}\RktMeta{}\RktSym{{\char`\_}}\RktMeta{}\mbox{\hphantom{\Scribtexttt{x}}}\RktMeta{}\RktSym{{\char`\_}}\RktMeta{}\mbox{\hphantom{\Scribtexttt{x}}}\RktMeta{}\RktPn{)}\RktMeta{}

\Smaller{\Smaller{\Scribtexttt{6}\mbox{\hphantom{\Scribtexttt{xx}}}}}\RktMeta{}\mbox{\hphantom{\Scribtexttt{xx}}}\RktMeta{}\RktPn{(}\RktSym{define{-}elaborator}\RktMeta{}

\Smaller{\Smaller{\Scribtexttt{7}\mbox{\hphantom{\Scribtexttt{xx}}}}}\RktMeta{}\mbox{\hphantom{\Scribtexttt{xxxxx}}}\RktMeta{}\RktSym{{\char`\_}}\RktMeta{}\mbox{\hphantom{\Scribtexttt{x}}}\RktMeta{}\RktSym{{\char`\_}}\RktMeta{}\mbox{\hphantom{\Scribtexttt{x}}}\RktMeta{}\RktSym{{\char`\_}}\RktMeta{}\mbox{\hphantom{\Scribtexttt{x}}}\RktMeta{}\RktSym{generating}\RktMeta{}\mbox{\hphantom{\Scribtexttt{x}}}\RktMeta{}\RktSym{run{-}time}\RktMeta{}\mbox{\hphantom{\Scribtexttt{x}}}\RktMeta{}\RktSym{code}\RktMeta{}\mbox{\hphantom{\Scribtexttt{x}}}\RktMeta{}\RktSym{{\char`\_}}\RktMeta{}\mbox{\hphantom{\Scribtexttt{x}}}\RktMeta{}\RktSym{{\char`\_}}\RktMeta{}\mbox{\hphantom{\Scribtexttt{x}}}\RktMeta{}\RktSym{{\char`\_}}\RktMeta{}\mbox{\hphantom{\Scribtexttt{x}}}\RktMeta{}\RktPn{)}\RktMeta{}

\Smaller{\Smaller{\Scribtexttt{8}\mbox{\hphantom{\Scribtexttt{xx}}}}}\RktMeta{}\mbox{\hphantom{\Scribtexttt{xx}}}\RktMeta{}\RktPn{(}\RktMeta{}\mbox{\hphantom{\Scribtexttt{x}}}\RktMeta{}\RktSym{{\char`\_}}\RktMeta{}\mbox{\hphantom{\Scribtexttt{x}}}\RktMeta{}\RktSym{{\char`\_}}\RktMeta{}\mbox{\hphantom{\Scribtexttt{x}}}\RktMeta{}\RktSym{{\char`\_}}\RktMeta{}\mbox{\hphantom{\Scribtexttt{xx}}}\RktMeta{}\RktSym{persistent}\RktMeta{}\mbox{\hphantom{\Scribtexttt{x}}}\RktMeta{}\RktSym{data}\RktMeta{}\mbox{\hphantom{\Scribtexttt{x}}}\RktMeta{}\RktSym{{\char`\_}}\RktMeta{}\mbox{\hphantom{\Scribtexttt{x}}}\RktMeta{}\RktSym{{\char`\_}}\RktMeta{}\mbox{\hphantom{\Scribtexttt{x}}}\RktMeta{}\RktSym{{\char`\_}}\RktMeta{}\mbox{\hphantom{\Scribtexttt{x}}}\RktMeta{}\RktPn{)}\RktPn{)}\RktMeta{}\end{SingleColumn}\end{RktBlk}\end{SCodeFlow}

\noindent \identity{\end{minipage}\\[.2cm]}
The first line specifies the name of the interactive syntax
and from which base class it is derived. Like classes, interactive{-}syntax
definitions benefit from implementation inheritance. The \RktSym{draw} and
\RktSym{on{-}event} methods (lines 2{--}5) make up the forms{'}s
user interface, code that heavily relies on Racket{'}s
platform{-}independent graphical{-}user interfaces\Autobibref{~[\hyperref[t:x28autobib_x22Matthew_Flattx2c_Robert_Bruce_Findlerx2c_and_John_ClementsGUIx3a_Racket_Graphics_ToolkitPLT_Design_Incx2ex2c_PLTx2dTRx2d2010x2d32010httpsx3ax2fx2fracketx2dlangx2eorgx2ftr3x2fx22x29]{\AutobibLink{Flatt et al\Sendabbrev{.}}} \hyperref[t:x28autobib_x22Matthew_Flattx2c_Robert_Bruce_Findlerx2c_and_John_ClementsGUIx3a_Racket_Graphics_ToolkitPLT_Design_Incx2ex2c_PLTx2dTRx2d2010x2d32010httpsx3ax2fx2fracketx2dlangx2eorgx2ftr3x2fx22x29]{\AutobibLink{\Thyperref{2010}{autobiblab:23}}}]}
and tools for the interactive development
environment\Autobibref{~[\hyperref[t:x28autobib_x22Gregory_Cooper_and_Shriram_KrishnamurthiEmbedding_Dynamic_Dataflow_in_a_Callx2dbyx2dValue_LanguageIn_Procx2e_European_Symposium_on_Programmingx2c_ppx2e_294x2dx2d3082004httpsx3ax2fx2fdoix2eorgx2f10x2e1007x2f11693024x5f20x22x29]{\AutobibLink{Cooper and Krishnamurthi}} \hyperref[t:x28autobib_x22Gregory_Cooper_and_Shriram_KrishnamurthiEmbedding_Dynamic_Dataflow_in_a_Callx2dbyx2dValue_LanguageIn_Procx2e_European_Symposium_on_Programmingx2c_ppx2e_294x2dx2d3082004httpsx3ax2fx2fdoix2eorgx2f10x2e1007x2f11693024x5f20x22x29]{\AutobibLink{\Thyperref{2004}{autobiblab:10}}}; \hyperref[t:x28autobib_x22Robert_Bruce_Findler_and_PLTDrRacketx3a_Programming_EnvironmentPLT_Design_Incx2ex2c_PLTx2dTRx2d2010x2d22010httpsx3ax2fx2fracketx2dlangx2eorgx2ftr2x2fx22x29]{\AutobibLink{Findler and PLT}} \hyperref[t:x28autobib_x22Robert_Bruce_Findler_and_PLTDrRacketx3a_Programming_EnvironmentPLT_Design_Incx2ex2c_PLTx2dTRx2d2010x2d22010httpsx3ax2fx2fracketx2dlangx2eorgx2ftr2x2fx22x29]{\AutobibLink{\Thyperref{2010}{autobiblab:20}}}]}. For specifying the semantics
of the new construct, a developer uses the meta{-}DSL for
specifying text{-}based language extensions (lines 6 and 7).
The remaining pieces (lines 8 and below) make up the persistent storage of
the syntax form.

To support the specification and management of persistent state, our design
also supplies a custom language extension. With this extension, a
developer can articulate how to react to changes of the
data, how to serialize it for future use, and how to
deserialize data (if it exists) to resume the use of an
interactive{-}syntax. The extension is called
\RktSym{define{-}state} and its syntax is as follows:
\identity{~\\\begin{minipage}[c]{0.9\textwidth}}

\begin{SCodeFlow}\begin{RktBlk}\begin{SingleColumn}\Smaller{\Smaller{\Scribtexttt{1}\mbox{\hphantom{\Scribtexttt{xx}}}}}\RktPn{(}\RktSym{define{-}state}\RktMeta{}\mbox{\hphantom{\Scribtexttt{x}}}\RktMeta{}\RktSym{name}\RktMeta{}\mbox{\hphantom{\Scribtexttt{x}}}\RktMeta{}\RktSym{default}\RktMeta{}

\Smaller{\Smaller{\Scribtexttt{2}\mbox{\hphantom{\Scribtexttt{xx}}}}}\RktMeta{}\mbox{\hphantom{\Scribtexttt{xxx}}}\RktMeta{}\RktSym{state{-}properties}\RktMeta{}\mbox{\hphantom{\Scribtexttt{x}}}\RktMeta{}\RktSym{{\hbox{\texttt{.}}}{\hbox{\texttt{.}}}{\hbox{\texttt{.}}}}\RktPn{)}\RktMeta{}\end{SingleColumn}\end{RktBlk}\end{SCodeFlow}

\noindent \identity{\end{minipage}\\[.2cm]}
The properties describe these aspects of the state variables.
First, they provide traditional getter/setter methods.
Second, they provide an optional mechanism
for users to marshal seemingly unserializable values into serializable
ones. Finally, they describe how long a value must persist.

Consider the state for an extension that adds word
processors to the language. That state might include the
prose the user typed as well as the cursor{'}s position. The
document{'}s getter/setter methods, as well as its
serialization would be fairly pedestrian and the
\RktSym{define{-}state} form would handle this automatically.
However, persistence is less trivial. The user may expect the
text to be saved in the file, but not the current
cursor position. However, the user does expect the cursor
to remain in place while the document is open.

Semantically, an interactive{-}syntax definition is a
class that runs code both during edit time and compile time.
Definitions using \RktSym{define/public} are methods
and capitalize on object inheritance. The
\RktSym{define{-}interactive{-}syntax} form ensures that
\RktSym{draw} and \RktSym{on{-}event} are among the defined
methods. The \RktSym{define{-}state} form is mapped to the
class fields. Finally, the \RktSym{define{-}elaborator} form
acts as a macro that generates run{-}time code. See \ChapRef{\SectionNumberLink{t:x28part_x22secx3aimplementationx22x29}{5}}{The Implementation of an Interactive{-}Syntax Extension Mechanism} for details.

\Ssubsection{Edit{-}Time Programming, a Tsuro Example}{Edit{-}Time Programming, a Tsuro Example}\label{t:x28part_x22Editx2dTimex5fProgrammingx5fx5fax5fTsurox5fExamplex22x29}

Implementing a dedicated state, renderer, event handler, and
semantics for each interactive{-}syntax extensions is somewhat
labor intensive. To reduce the work load, our prototype
implementation supports standard GUI creation techniques
available in Racket. These techniques fit into three
categories: inheritance, container editors, and graphical
editors.

\begin{Figure}\begin{Centerfigure}\begin{FigureInside}\begin{SCodeFlow}\begin{RktBlk}\begin{SingleColumn}\Smaller{\Smaller{\mbox{\hphantom{\Scribtexttt{x}}}\Scribtexttt{1}\mbox{\hphantom{\Scribtexttt{xx}}}}}\RktPn{(}\RktSym{begin{-}for{-}interactive{-}syntax}\RktMeta{}

\Smaller{\Smaller{\mbox{\hphantom{\Scribtexttt{x}}}\Scribtexttt{2}\mbox{\hphantom{\Scribtexttt{xx}}}}}\RktMeta{}\mbox{\hphantom{\Scribtexttt{xx}}}\RktMeta{}\RktPn{(}\RktSym{define}\RktMeta{}\mbox{\hphantom{\Scribtexttt{x}}}\RktMeta{}\RktSym{TILE{-}NODES}\RktMeta{}\mbox{\hphantom{\Scribtexttt{x}}}\RktMeta{}\RktPn{(}\RktSym{list}\RktMeta{}\mbox{\hphantom{\Scribtexttt{x}}}\RktMeta{}\RktVal{"A"}\RktMeta{}\mbox{\hphantom{\Scribtexttt{x}}}\RktMeta{}\RktVal{"B"}\RktMeta{}\mbox{\hphantom{\Scribtexttt{x}}}\RktMeta{}\RktVal{"C"}\RktMeta{}\mbox{\hphantom{\Scribtexttt{x}}}\RktMeta{}\RktVal{"D"}\RktMeta{}\mbox{\hphantom{\Scribtexttt{x}}}\RktMeta{}\RktVal{"E"}\RktMeta{}\mbox{\hphantom{\Scribtexttt{x}}}\RktMeta{}\RktVal{"F"}\RktMeta{}\mbox{\hphantom{\Scribtexttt{x}}}\RktMeta{}\RktVal{"G"}\RktMeta{}\mbox{\hphantom{\Scribtexttt{x}}}\RktMeta{}\RktVal{"H"}\RktPn{)}\RktPn{)}\RktPn{)}\RktMeta{}

\Smaller{\Smaller{\mbox{\hphantom{\Scribtexttt{x}}}\Scribtexttt{3}\mbox{\hphantom{\Scribtexttt{xx}}}}}\RktMeta{~}

\Smaller{\Smaller{\mbox{\hphantom{\Scribtexttt{x}}}\Scribtexttt{4}\mbox{\hphantom{\Scribtexttt{xx}}}}}\RktMeta{}\RktPn{(}\RktSym{define{-}interactive{-}syntax}\RktMeta{}\mbox{\hphantom{\Scribtexttt{x}}}\RktMeta{}\RktSym{tsuro{-}tile\$}\RktMeta{}\mbox{\hphantom{\Scribtexttt{x}}}\RktMeta{}\RktSym{horizontal{-}block\$}\RktMeta{}\mbox{\hphantom{\Scribtexttt{x}}}\RktMeta{}\RktPn{(}\RktSym{super{-}new}\RktPn{)}\RktMeta{}

\Smaller{\Smaller{\mbox{\hphantom{\Scribtexttt{x}}}\Scribtexttt{5}\mbox{\hphantom{\Scribtexttt{xx}}}}}\RktMeta{}\mbox{\hphantom{\Scribtexttt{xx}}}\RktMeta{}\RktCmt{;;}\mbox{\hphantom{\Scribtexttt{x}}}\RktCmt{STATE}\RktMeta{}

\Smaller{\Smaller{\mbox{\hphantom{\Scribtexttt{x}}}\Scribtexttt{6}\mbox{\hphantom{\Scribtexttt{xx}}}}}\RktMeta{}\mbox{\hphantom{\Scribtexttt{xx}}}\RktMeta{}\RktPn{(}\RktSym{define{-}state}\RktMeta{}\mbox{\hphantom{\Scribtexttt{x}}}\RktMeta{}\RktSym{pairs}\RktMeta{}\mbox{\hphantom{\Scribtexttt{x}}}\RktMeta{}\RktPn{(}\RktSym{hash}\RktPn{)}\RktMeta{}

\Smaller{\Smaller{\mbox{\hphantom{\Scribtexttt{x}}}\Scribtexttt{7}\mbox{\hphantom{\Scribtexttt{xx}}}}}\RktMeta{}\mbox{\hphantom{\Scribtexttt{xxxx}}}\RktMeta{}\RktPn{\#{\hbox{\texttt{:}}}elaborator}\RktMeta{}\mbox{\hphantom{\Scribtexttt{x}}}\RktMeta{}\RktVal{\#t}\RktMeta{}

\Smaller{\Smaller{\mbox{\hphantom{\Scribtexttt{x}}}\Scribtexttt{8}\mbox{\hphantom{\Scribtexttt{xx}}}}}\RktMeta{}\mbox{\hphantom{\Scribtexttt{xxxx}}}\RktMeta{}\RktPn{\#{\hbox{\texttt{:}}}getter}\RktMeta{}\mbox{\hphantom{\Scribtexttt{x}}}\RktMeta{}\RktVal{\#t}\RktPn{)}\RktMeta{}

\Smaller{\Smaller{\mbox{\hphantom{\Scribtexttt{x}}}\Scribtexttt{9}\mbox{\hphantom{\Scribtexttt{xx}}}}}\RktMeta{~}

\Smaller{\Smaller{\Scribtexttt{10}\mbox{\hphantom{\Scribtexttt{xx}}}}}\RktMeta{}\mbox{\hphantom{\Scribtexttt{xx}}}\RktMeta{}\RktCmt{;;}\mbox{\hphantom{\Scribtexttt{x}}}\RktCmt{Char}\mbox{\hphantom{\Scribtexttt{x}}}\RktCmt{Char}\mbox{\hphantom{\Scribtexttt{x}}}\RktCmt{{-}{\Stttextmore}}\mbox{\hphantom{\Scribtexttt{x}}}\RktCmt{Void}\RktMeta{}

\Smaller{\Smaller{\Scribtexttt{11}\mbox{\hphantom{\Scribtexttt{xx}}}}}\RktMeta{}\mbox{\hphantom{\Scribtexttt{xx}}}\RktMeta{}\RktCmt{;;}\mbox{\hphantom{\Scribtexttt{x}}}\RktCmt{EFFECT}\mbox{\hphantom{\Scribtexttt{x}}}\RktCmt{connects}\mbox{\hphantom{\Scribtexttt{x}}}\RktCmt{letter}\mbox{\hphantom{\Scribtexttt{x}}}\RktCmt{to}\mbox{\hphantom{\Scribtexttt{x}}}\RktCmt{other}\mbox{\hphantom{\Scribtexttt{x}}}\RktCmt{and}\mbox{\hphantom{\Scribtexttt{x}}}\RktCmt{vice}\mbox{\hphantom{\Scribtexttt{x}}}\RktCmt{versa}\mbox{\hphantom{\Scribtexttt{x}}}\RktCmt{in}\mbox{\hphantom{\Scribtexttt{x}}}\RktCmt{pairs}\RktMeta{}

\Smaller{\Smaller{\Scribtexttt{12}\mbox{\hphantom{\Scribtexttt{xx}}}}}\RktMeta{}\mbox{\hphantom{\Scribtexttt{xx}}}\RktMeta{}\RktPn{(}\RktSym{define/public}\RktMeta{}\mbox{\hphantom{\Scribtexttt{x}}}\RktMeta{}\RktPn{(}\RktSym{connect{\hbox{\texttt{!}}}}\RktMeta{}\mbox{\hphantom{\Scribtexttt{x}}}\RktMeta{}\RktSym{letter}\RktMeta{}\mbox{\hphantom{\Scribtexttt{x}}}\RktMeta{}\RktSym{other}\RktPn{)}\RktMeta{}

\Smaller{\Smaller{\Scribtexttt{13}\mbox{\hphantom{\Scribtexttt{xx}}}}}\RktMeta{}\mbox{\hphantom{\Scribtexttt{xxxx}}}\RktMeta{}\RktPn{(}\RktSym{send}\RktMeta{}\mbox{\hphantom{\Scribtexttt{x}}}\RktMeta{}\RktPn{(}\RktSym{hash{-}ref}\RktMeta{}\mbox{\hphantom{\Scribtexttt{x}}}\RktMeta{}\RktSym{field{-}gui}\RktMeta{}\mbox{\hphantom{\Scribtexttt{x}}}\RktMeta{}\RktSym{letter}\RktPn{)}\RktMeta{}\mbox{\hphantom{\Scribtexttt{x}}}\RktMeta{}\RktSym{set{-}text{\hbox{\texttt{!}}}}\RktMeta{}\mbox{\hphantom{\Scribtexttt{x}}}\RktMeta{}\RktSym{other}\RktPn{)}\RktMeta{}

\Smaller{\Smaller{\Scribtexttt{14}\mbox{\hphantom{\Scribtexttt{xx}}}}}\RktMeta{}\mbox{\hphantom{\Scribtexttt{xxxx}}}\RktMeta{}\RktPn{(}\RktSym{send}\RktMeta{}\mbox{\hphantom{\Scribtexttt{x}}}\RktMeta{}\RktPn{(}\RktSym{hash{-}ref}\RktMeta{}\mbox{\hphantom{\Scribtexttt{x}}}\RktMeta{}\RktSym{field{-}gui}\RktMeta{}\mbox{\hphantom{\Scribtexttt{x}}}\RktMeta{}\RktSym{other}\RktPn{)}\RktMeta{}\mbox{\hphantom{\Scribtexttt{x}}}\RktMeta{}\RktSym{set{-}text{\hbox{\texttt{!}}}}\RktMeta{}\mbox{\hphantom{\Scribtexttt{x}}}\RktMeta{}\RktSym{letter}\RktPn{)}\RktMeta{}

\Smaller{\Smaller{\Scribtexttt{15}\mbox{\hphantom{\Scribtexttt{xx}}}}}\RktMeta{}\mbox{\hphantom{\Scribtexttt{xxxx}}}\RktMeta{}\RktPn{(}\RktSym{set{\hbox{\texttt{!}}}}\RktMeta{}\mbox{\hphantom{\Scribtexttt{x}}}\RktMeta{}\RktSym{pairs}\RktMeta{}\mbox{\hphantom{\Scribtexttt{x}}}\RktMeta{}\RktPn{(}\RktSym{hash{-}set*}\RktMeta{}\mbox{\hphantom{\Scribtexttt{x}}}\RktMeta{}\RktSym{pairs}\RktMeta{}\mbox{\hphantom{\Scribtexttt{x}}}\RktMeta{}\RktSym{letter}\RktMeta{}\mbox{\hphantom{\Scribtexttt{x}}}\RktMeta{}\RktSym{other}\RktMeta{}\mbox{\hphantom{\Scribtexttt{x}}}\RktMeta{}\RktSym{other}\RktMeta{}\mbox{\hphantom{\Scribtexttt{x}}}\RktMeta{}\RktSym{letter}\RktPn{)}\RktPn{)}\RktMeta{}

\Smaller{\Smaller{\Scribtexttt{16}\mbox{\hphantom{\Scribtexttt{xx}}}}}\RktMeta{}\mbox{\hphantom{\Scribtexttt{xxxx}}}\RktMeta{}\RktPn{(}\RktSym{send}\RktMeta{}\mbox{\hphantom{\Scribtexttt{x}}}\RktMeta{}\RktSym{picture}\RktMeta{}\mbox{\hphantom{\Scribtexttt{x}}}\RktMeta{}\RktSym{set{-}tile{\hbox{\texttt{!}}}}\RktMeta{}\mbox{\hphantom{\Scribtexttt{x}}}\RktMeta{}\RktPn{(}\RktSym{draw{-}tile}\RktMeta{}\mbox{\hphantom{\Scribtexttt{x}}}\RktMeta{}\RktSym{pairs}\RktPn{)}\RktPn{)}\RktPn{)}\RktMeta{}

\Smaller{\Smaller{\Scribtexttt{17}\mbox{\hphantom{\Scribtexttt{xx}}}}}\RktMeta{~}

\Smaller{\Smaller{\Scribtexttt{18}\mbox{\hphantom{\Scribtexttt{xx}}}}}\RktMeta{}\mbox{\hphantom{\Scribtexttt{xx}}}\RktMeta{}\RktCmt{;;}\mbox{\hphantom{\Scribtexttt{x}}}\RktCmt{VIEW}\mbox{\hphantom{\Scribtexttt{x}}}\RktCmt{{\hbox{\texttt{:}}}}\mbox{\hphantom{\Scribtexttt{x}}}\RktCmt{two}\mbox{\hphantom{\Scribtexttt{x}}}\RktCmt{horizontally}\mbox{\hphantom{\Scribtexttt{x}}}\RktCmt{aligned}\mbox{\hphantom{\Scribtexttt{x}}}\RktCmt{elements}\RktMeta{}

\Smaller{\Smaller{\Scribtexttt{19}\mbox{\hphantom{\Scribtexttt{xx}}}}}\RktMeta{}\mbox{\hphantom{\Scribtexttt{xx}}}\RktMeta{}\RktPn{(}\RktSym{define}\RktMeta{}\mbox{\hphantom{\Scribtexttt{x}}}\RktMeta{}\RktSym{picture}\RktMeta{}\mbox{\hphantom{\Scribtexttt{x}}}\RktMeta{}\RktPn{(}\RktSym{new}\RktMeta{}\mbox{\hphantom{\Scribtexttt{x}}}\RktMeta{}\RktSym{tsuro{-}picture\$}\RktMeta{}\mbox{\hphantom{\Scribtexttt{x}}}\RktMeta{}\RktPn{[}\RktSym{parent}\RktMeta{}\mbox{\hphantom{\Scribtexttt{x}}}\RktMeta{}\RktSym{this}\RktPn{]}\RktMeta{}

\Smaller{\Smaller{\Scribtexttt{20}\mbox{\hphantom{\Scribtexttt{xx}}}}}\RktMeta{}\mbox{\hphantom{\Scribtexttt{xxxxxxxxx}}}\RktMeta{}\RktPn{[}\RktSym{connections}\RktMeta{}\mbox{\hphantom{\Scribtexttt{x}}}\RktMeta{}\RktSym{pairs}\RktPn{]}\RktPn{)}\RktPn{)}\RktMeta{}

\Smaller{\Smaller{\Scribtexttt{21}\mbox{\hphantom{\Scribtexttt{xx}}}}}\RktMeta{~}

\Smaller{\Smaller{\Scribtexttt{22}\mbox{\hphantom{\Scribtexttt{xx}}}}}\RktMeta{}\mbox{\hphantom{\Scribtexttt{xx}}}\RktMeta{}\RktPn{(}\RktSym{define}\RktMeta{}\mbox{\hphantom{\Scribtexttt{x}}}\RktMeta{}\RktSym{fields}\RktMeta{}\mbox{\hphantom{\Scribtexttt{xx}}}\RktMeta{}\RktPn{(}\RktSym{new}\RktMeta{}\mbox{\hphantom{\Scribtexttt{x}}}\RktMeta{}\RktSym{vertical{-}block\$}\RktMeta{}\mbox{\hphantom{\Scribtexttt{x}}}\RktMeta{}\RktPn{[}\RktSym{parent}\RktMeta{}\mbox{\hphantom{\Scribtexttt{x}}}\RktMeta{}\RktSym{this}\RktPn{]}\RktPn{)}\RktPn{)}\RktMeta{}

\Smaller{\Smaller{\Scribtexttt{23}\mbox{\hphantom{\Scribtexttt{xx}}}}}\RktMeta{}\mbox{\hphantom{\Scribtexttt{xx}}}\RktMeta{}\RktCmt{;;}\mbox{\hphantom{\Scribtexttt{x}}}\RktCmt{Char}\mbox{\hphantom{\Scribtexttt{x}}}\RktCmt{{-}{\Stttextmore}}\mbox{\hphantom{\Scribtexttt{x}}}\RktCmt{Void}\RktMeta{}

\Smaller{\Smaller{\Scribtexttt{24}\mbox{\hphantom{\Scribtexttt{xx}}}}}\RktMeta{}\mbox{\hphantom{\Scribtexttt{xx}}}\RktMeta{}\RktCmt{;;}\mbox{\hphantom{\Scribtexttt{x}}}\RktCmt{EFFECT}\mbox{\hphantom{\Scribtexttt{x}}}\RktCmt{creates}\mbox{\hphantom{\Scribtexttt{x}}}\RktCmt{a}\mbox{\hphantom{\Scribtexttt{x}}}\RktCmt{text}\mbox{\hphantom{\Scribtexttt{x}}}\RktCmt{field}\mbox{\hphantom{\Scribtexttt{x}}}\RktCmt{as}\mbox{\hphantom{\Scribtexttt{x}}}\RktCmt{a}\mbox{\hphantom{\Scribtexttt{x}}}\RktCmt{child}\mbox{\hphantom{\Scribtexttt{x}}}\RktCmt{of}\mbox{\hphantom{\Scribtexttt{x}}}\RktCmt{fields}\RktMeta{}

\Smaller{\Smaller{\Scribtexttt{25}\mbox{\hphantom{\Scribtexttt{xx}}}}}\RktMeta{}\mbox{\hphantom{\Scribtexttt{xx}}}\RktMeta{}\RktPn{(}\RktSym{define}\RktMeta{}\mbox{\hphantom{\Scribtexttt{x}}}\RktMeta{}\RktPn{(}\RktSym{add{-}tsuro{-}field{\hbox{\texttt{!}}}}\RktMeta{}\mbox{\hphantom{\Scribtexttt{x}}}\RktMeta{}\RktSym{letter}\RktPn{)}\RktMeta{}

\Smaller{\Smaller{\Scribtexttt{26}\mbox{\hphantom{\Scribtexttt{xx}}}}}\RktMeta{}\mbox{\hphantom{\Scribtexttt{xxxx}}}\RktMeta{}\RktCmt{;;}\mbox{\hphantom{\Scribtexttt{x}}}\RktCmt{TextField}\mbox{\hphantom{\Scribtexttt{x}}}\RktCmt{Event}\mbox{\hphantom{\Scribtexttt{x}}}\RktCmt{{-}{\Stttextmore}}\mbox{\hphantom{\Scribtexttt{x}}}\RktCmt{Void}\RktMeta{}

\Smaller{\Smaller{\Scribtexttt{27}\mbox{\hphantom{\Scribtexttt{xx}}}}}\RktMeta{}\mbox{\hphantom{\Scribtexttt{xxxx}}}\RktMeta{}\RktCmt{;;}\mbox{\hphantom{\Scribtexttt{x}}}\RktCmt{EFFECT}\mbox{\hphantom{\Scribtexttt{x}}}\RktCmt{connect}\mbox{\hphantom{\Scribtexttt{x}}}\RktCmt{the}\mbox{\hphantom{\Scribtexttt{x}}}\RktCmt{specified}\mbox{\hphantom{\Scribtexttt{x}}}\RktCmt{char}\mbox{\hphantom{\Scribtexttt{x}}}\RktCmt{in}\mbox{\hphantom{\Scribtexttt{x}}}\RktCmt{f}\mbox{\hphantom{\Scribtexttt{x}}}\RktCmt{with}\mbox{\hphantom{\Scribtexttt{x}}}\RktCmt{this}\mbox{\hphantom{\Scribtexttt{x}}}\RktCmt{letter}\RktMeta{}

\Smaller{\Smaller{\Scribtexttt{28}\mbox{\hphantom{\Scribtexttt{xx}}}}}\RktMeta{}\mbox{\hphantom{\Scribtexttt{xxxx}}}\RktMeta{}\RktPn{(}\RktSym{define}\RktMeta{}\mbox{\hphantom{\Scribtexttt{x}}}\RktMeta{}\RktPn{(}\RktSym{letter{-}callback}\RktMeta{}\mbox{\hphantom{\Scribtexttt{x}}}\RktMeta{}\RktSym{f}\RktMeta{}\mbox{\hphantom{\Scribtexttt{x}}}\RktMeta{}\RktSym{e}\RktPn{)}\RktMeta{}

\Smaller{\Smaller{\Scribtexttt{29}\mbox{\hphantom{\Scribtexttt{xx}}}}}\RktMeta{}\mbox{\hphantom{\Scribtexttt{xxxxxx}}}\RktMeta{}\RktPn{(}\RktSym{connect{\hbox{\texttt{!}}}}\RktMeta{}\mbox{\hphantom{\Scribtexttt{x}}}\RktMeta{}\RktPn{(}\RktSym{send}\RktMeta{}\mbox{\hphantom{\Scribtexttt{x}}}\RktMeta{}\RktSym{f}\RktMeta{}\mbox{\hphantom{\Scribtexttt{x}}}\RktMeta{}\RktSym{get{-}text}\RktPn{)}\RktMeta{}\mbox{\hphantom{\Scribtexttt{x}}}\RktMeta{}\RktSym{letter}\RktPn{)}\RktPn{)}\RktMeta{}

\Smaller{\Smaller{\Scribtexttt{30}\mbox{\hphantom{\Scribtexttt{xx}}}}}\RktMeta{~}

\Smaller{\Smaller{\Scribtexttt{31}\mbox{\hphantom{\Scribtexttt{xx}}}}}\RktMeta{}\mbox{\hphantom{\Scribtexttt{xxxx}}}\RktMeta{}\RktCmt{;;}\mbox{\hphantom{\Scribtexttt{x}}}\RktCmt{Container}\mbox{\hphantom{\Scribtexttt{x}}}\RktCmt{{-}{\Stttextmore}}\mbox{\hphantom{\Scribtexttt{x}}}\RktCmt{Void}\RktMeta{}

\Smaller{\Smaller{\Scribtexttt{32}\mbox{\hphantom{\Scribtexttt{xx}}}}}\RktMeta{}\mbox{\hphantom{\Scribtexttt{xxxx}}}\RktMeta{}\RktCmt{;;}\mbox{\hphantom{\Scribtexttt{x}}}\RktCmt{EFFECT}\mbox{\hphantom{\Scribtexttt{x}}}\RktCmt{create}\mbox{\hphantom{\Scribtexttt{x}}}\RktCmt{an}\mbox{\hphantom{\Scribtexttt{x}}}\RktCmt{option}\mbox{\hphantom{\Scribtexttt{x}}}\RktCmt{field}\mbox{\hphantom{\Scribtexttt{x}}}\RktCmt{as}\mbox{\hphantom{\Scribtexttt{x}}}\RktCmt{a}\mbox{\hphantom{\Scribtexttt{x}}}\RktCmt{child}\mbox{\hphantom{\Scribtexttt{x}}}\RktCmt{of}\mbox{\hphantom{\Scribtexttt{x}}}\RktCmt{p}\RktMeta{}

\Smaller{\Smaller{\Scribtexttt{33}\mbox{\hphantom{\Scribtexttt{xx}}}}}\RktMeta{}\mbox{\hphantom{\Scribtexttt{xxxx}}}\RktMeta{}\RktPn{(}\RktSym{define}\RktMeta{}\mbox{\hphantom{\Scribtexttt{x}}}\RktMeta{}\RktPn{(}\RktSym{option{-}maker}\RktMeta{}\mbox{\hphantom{\Scribtexttt{x}}}\RktMeta{}\RktSym{p}\RktPn{)}\RktMeta{}

\Smaller{\Smaller{\Scribtexttt{34}\mbox{\hphantom{\Scribtexttt{xx}}}}}\RktMeta{}\mbox{\hphantom{\Scribtexttt{xxxxxx}}}\RktMeta{}\RktPn{(}\RktSym{new}\RktMeta{}\mbox{\hphantom{\Scribtexttt{x}}}\RktMeta{}\RktSym{text{-}field\$}\RktMeta{}\mbox{\hphantom{\Scribtexttt{x}}}\RktMeta{}\RktPn{[}\RktSym{parent}\RktMeta{}\mbox{\hphantom{\Scribtexttt{x}}}\RktMeta{}\RktSym{p}\RktPn{]}\RktMeta{}\mbox{\hphantom{\Scribtexttt{x}}}\RktMeta{}\RktPn{[}\RktSym{callback}\RktMeta{}\mbox{\hphantom{\Scribtexttt{x}}}\RktMeta{}\RktSym{letter{-}callback}\RktPn{]}\RktPn{)}\RktPn{)}\RktMeta{}

\Smaller{\Smaller{\Scribtexttt{35}\mbox{\hphantom{\Scribtexttt{xx}}}}}\RktMeta{~}

\Smaller{\Smaller{\Scribtexttt{36}\mbox{\hphantom{\Scribtexttt{xx}}}}}\RktMeta{}\mbox{\hphantom{\Scribtexttt{xxxx}}}\RktMeta{}\RktPn{(}\RktSym{new}\RktMeta{}\mbox{\hphantom{\Scribtexttt{x}}}\RktMeta{}\RktSym{labeled{-}option\$}\RktMeta{}\mbox{\hphantom{\Scribtexttt{x}}}\RktMeta{}\RktPn{[}\RktSym{parent}\RktMeta{}\mbox{\hphantom{\Scribtexttt{x}}}\RktMeta{}\RktSym{fields}\RktPn{]}\RktMeta{}

\Smaller{\Smaller{\Scribtexttt{37}\mbox{\hphantom{\Scribtexttt{xx}}}}}\RktMeta{}\mbox{\hphantom{\Scribtexttt{xxxxxxxxx}}}\RktMeta{}\RktPn{[}\RktSym{label}\RktMeta{}\mbox{\hphantom{\Scribtexttt{xx}}}\RktMeta{}\RktPn{(}\RktSym{format}\RktMeta{}\mbox{\hphantom{\Scribtexttt{x}}}\RktMeta{}\RktVal{"$\sim$a{\hbox{\texttt{:}}}}\mbox{\hphantom{\Scribtexttt{x}}}\RktVal{"}\RktMeta{}\mbox{\hphantom{\Scribtexttt{x}}}\RktMeta{}\RktSym{letter}\RktPn{)}\RktPn{]}\RktMeta{}

\Smaller{\Smaller{\Scribtexttt{38}\mbox{\hphantom{\Scribtexttt{xx}}}}}\RktMeta{}\mbox{\hphantom{\Scribtexttt{xxxxxxxxx}}}\RktMeta{}\RktPn{[}\RktSym{option}\RktMeta{}\mbox{\hphantom{\Scribtexttt{x}}}\RktMeta{}\RktSym{option{-}maker}\RktPn{]}\RktPn{)}\RktPn{)}\RktMeta{}

\Smaller{\Smaller{\Scribtexttt{39}\mbox{\hphantom{\Scribtexttt{xx}}}}}\RktMeta{~}

\Smaller{\Smaller{\Scribtexttt{40}\mbox{\hphantom{\Scribtexttt{xx}}}}}\RktMeta{}\mbox{\hphantom{\Scribtexttt{xx}}}\RktMeta{}\RktPn{(}\RktSym{define}\RktMeta{}\mbox{\hphantom{\Scribtexttt{x}}}\RktMeta{}\RktSym{field{-}gui}\RktMeta{}\mbox{\hphantom{\Scribtexttt{x}}}\RktMeta{}\RktCmt{;;}\mbox{\hphantom{\Scribtexttt{x}}}\RktCmt{create}\mbox{\hphantom{\Scribtexttt{x}}}\RktCmt{all}\mbox{\hphantom{\Scribtexttt{x}}}\RktCmt{text}\mbox{\hphantom{\Scribtexttt{x}}}\RktCmt{entry}\mbox{\hphantom{\Scribtexttt{x}}}\RktCmt{fields}\RktMeta{}

\Smaller{\Smaller{\Scribtexttt{41}\mbox{\hphantom{\Scribtexttt{xx}}}}}\RktMeta{}\mbox{\hphantom{\Scribtexttt{xxxx}}}\RktMeta{}\RktPn{(}\RktSym{for/hash}\RktMeta{}\mbox{\hphantom{\Scribtexttt{x}}}\RktMeta{}\RktPn{(}\RktPn{[}\RktSym{a}\RktMeta{}\mbox{\hphantom{\Scribtexttt{x}}}\RktMeta{}\RktSym{TILE{-}NODES}\RktPn{]}\RktPn{)}\RktMeta{}

\Smaller{\Smaller{\Scribtexttt{42}\mbox{\hphantom{\Scribtexttt{xx}}}}}\RktMeta{}\mbox{\hphantom{\Scribtexttt{xxxxxx}}}\RktMeta{}\RktPn{(}\RktSym{values}\RktMeta{}\mbox{\hphantom{\Scribtexttt{x}}}\RktMeta{}\RktSym{a}\RktMeta{}\mbox{\hphantom{\Scribtexttt{x}}}\RktMeta{}\RktPn{(}\RktSym{send}\RktMeta{}\mbox{\hphantom{\Scribtexttt{x}}}\RktMeta{}\RktPn{(}\RktSym{add{-}tsuro{-}field{\hbox{\texttt{!}}}}\RktMeta{}\mbox{\hphantom{\Scribtexttt{x}}}\RktMeta{}\RktSym{a}\RktPn{)}\RktMeta{}\mbox{\hphantom{\Scribtexttt{x}}}\RktMeta{}\RktSym{get{-}option}\RktPn{)}\RktPn{)}\RktPn{)}\RktPn{)}\RktMeta{}

\Smaller{\Smaller{\Scribtexttt{43}\mbox{\hphantom{\Scribtexttt{xx}}}}}\RktMeta{~}

\Smaller{\Smaller{\Scribtexttt{44}\mbox{\hphantom{\Scribtexttt{xx}}}}}\RktMeta{}\mbox{\hphantom{\Scribtexttt{xx}}}\RktMeta{}\RktCmt{;;}\mbox{\hphantom{\Scribtexttt{x}}}\RktCmt{CODE}\mbox{\hphantom{\Scribtexttt{x}}}\RktCmt{GENERATION}\RktMeta{}

\Smaller{\Smaller{\Scribtexttt{45}\mbox{\hphantom{\Scribtexttt{xx}}}}}\RktMeta{}\mbox{\hphantom{\Scribtexttt{xx}}}\RktMeta{}\RktPn{(}\RktSym{define{-}elaborator}\RktMeta{}\mbox{\hphantom{\Scribtexttt{x}}}\RktMeta{}\RktSym{this}\RktMeta{}\mbox{\hphantom{\Scribtexttt{x}}}\RktMeta{}\RktSym{\#{\textasciigrave}}\RktSym{{\textquotesingle}}\RktSym{\#,}\RktPn{(}\RktSym{send}\RktMeta{}\mbox{\hphantom{\Scribtexttt{x}}}\RktMeta{}\RktSym{this}\RktMeta{}\mbox{\hphantom{\Scribtexttt{x}}}\RktMeta{}\RktSym{get{-}pairs}\RktPn{)}\RktPn{)}\RktPn{)}\RktMeta{}\end{SingleColumn}\end{RktBlk}\end{SCodeFlow}\end{FigureInside}\end{Centerfigure}

\Centertext{\Legend{\FigureTarget{\label{t:x28counter_x28x22figurex22_x22figx3atsurox2ddefinitionx22x29x29}\textsf{Fig.}~\textsf{3}. }{t:x28counter_x28x22figurex22_x22figx3atsurox2ddefinitionx22x29x29}\textsf{Example Editor for Tsuro Tile}}}\end{Figure}

Developers use inheritance and mixins\Autobibref{~[\hyperref[t:x28autobib_x22Matthew_Flattx2c_Robert_Bruce_Findlerx2c_and_Matthias_FelleisenScheme_with_Classesx2c_Mixinsx2c_and_TraitsIn_Procx2e_Asian_Symposium_Programming_Languages_and_Systemsx2c_ppx2e_270x2dx2d2892006x22x29]{\AutobibLink{Flatt et al\Sendabbrev{.}}} \hyperref[t:x28autobib_x22Matthew_Flattx2c_Robert_Bruce_Findlerx2c_and_Matthias_FelleisenScheme_with_Classesx2c_Mixinsx2c_and_TraitsIn_Procx2e_Asian_Symposium_Programming_Languages_and_Systemsx2c_ppx2e_270x2dx2d2892006x22x29]{\AutobibLink{\Thyperref{2006}{autobiblab:24}}}]}
to write only absolutely necessary \RktSym{draw} and
\RktSym{on{-}event} methods. Inheritance works just like in
Java. For example, every editor extends a \RktSym{base\$}
class, which supplies basic drawing and event handling.
Mixin functions abstract over inheritance. The
\RktSym{define{-}interactive{-}syntax{-}mixin} form creates new
mixin types. These mixins are added to an editor{'}s code
by applying them to the editor{'}s base class.

Container editors facilitate editor composition, which
almost completely eliminates the need for manually creating
methods for the resulting product. The three most
predominant container blocks are \RktSym{vertical{-}block\$}
for vertical alignment, \RktSym{horizontal{-}block\$} for
horizontal alignment, and \RktSym{pasteboard\$} for
free{-}flowing editors. Each of these containers work with the
\RktSym{widget\$} editor type. Each child has a super class that
supplies its drawing and event{-}handling methods.

To illustrate these abstraction and composition mechanisms,
figure~\hyperref[t:x28counter_x28x22figurex22_x22figx3atsurox2ddefinitionx22x29x29]{\FigureRef{3}{t:x28counter_x28x22figurex22_x22figx3atsurox2ddefinitionx22x29x29}} presents the
implementation of the Tsuro tile extension. The purpose of
this interactive syntax is to permit the programmer to
insert a graphical image of the tile where an expression is expected.
The construction and maintenance of this tile demands a
capability for connecting and re{-}connecting the entry points
of the tile, as well as for displaying
the current state of the connections both graphically and as
text.

Rather than implementing the \RktSym{draw} and
\RktSym{on{-}event} methods directly, the Tsuro syntax relies
on container classes (lines 4, 18{--}38) to manage
layout and events. These labels and fields (lines 22{--}42) are
provided by the standard library and can draw
themselves. The \RktSym{tsuro{-}picture\$} editor (lines 18{--}20)
works with \RktSym{trace{-}player} to provide drawing and event
handling functionality.

The field and label sub{-}editors serve similar purposes; they
both render text. The field editor, however, also handles
user interaction, while the label one does not. These
editors use the \RktSym{text\$\$} and \RktSym{focus\$\$}
interactive{-}syntax mixins from the standard library; see figure~\hyperref[t:x28counter_x28x22figurex22_x22figx3afieldx2dandx2dlabelx22x29x29]{\FigureRef{4}{t:x28counter_x28x22figurex22_x22figx3afieldx2dandx2dlabelx22x29x29}}.
The \RktSym{text\$\$} mixin (lines 1{--}5) handles both the text
portion of the editor{'}s state and drawing directly. The
\RktSym{focus\$\$} mixin (lines 7{--}10) handles user
interaction. Finally, the \RktSym{text{-}field\$} editor (lines
12{--}14) combines the two mixins and applies them to the
\RktSym{widget\$} base.

The code elaborator (lines 44{--}45) in
figure~\hyperref[t:x28counter_x28x22figurex22_x22figx3atsurox2ddefinitionx22x29x29]{\FigureRef{3}{t:x28counter_x28x22figurex22_x22figx3atsurox2ddefinitionx22x29x29}} turns \RktSym{pairs},
the state of the extension, into a
hash table for the run{-}time phase, using the
traditional syntax extension mechanism.

Importantly, developers may compose interactive{-}syntax
extensions. Thus, for example, an instance of
\RktSym{tsuro{-}tile\$} works with the interactive{-}syntax
extension for the full Tsuro board. Each tile in the board
is stored directly in the board editor, renders itself in
the board{'}s GUI context, and reacts to events flowing down
from this container.

\begin{Herefigure}\begin{Centerfigure}\begin{FigureInside}\begin{SCodeFlow}\begin{RktBlk}\begin{SingleColumn}\Smaller{\Smaller{\mbox{\hphantom{\Scribtexttt{x}}}\Scribtexttt{1}\mbox{\hphantom{\Scribtexttt{x}}}}}\RktCmt{;;}\mbox{\hphantom{\Scribtexttt{x}}}\RktCmt{Mixin}\mbox{\hphantom{\Scribtexttt{x}}}\RktCmt{for}\mbox{\hphantom{\Scribtexttt{x}}}\RktCmt{drawing}\mbox{\hphantom{\Scribtexttt{x}}}\RktCmt{text}\mbox{\hphantom{\Scribtexttt{x}}}\RktCmt{in}\mbox{\hphantom{\Scribtexttt{x}}}\RktCmt{an}\mbox{\hphantom{\Scribtexttt{x}}}\RktCmt{editor}\RktMeta{}

\Smaller{\Smaller{\mbox{\hphantom{\Scribtexttt{x}}}\Scribtexttt{2}\mbox{\hphantom{\Scribtexttt{x}}}}}\RktMeta{}\RktPn{(}\RktSym{define{-}interactive{-}syntax{-}mixin}\RktMeta{}\mbox{\hphantom{\Scribtexttt{x}}}\RktMeta{}\RktSym{text\$\$}\RktMeta{}

\Smaller{\Smaller{\mbox{\hphantom{\Scribtexttt{x}}}\Scribtexttt{3}\mbox{\hphantom{\Scribtexttt{x}}}}}\RktMeta{}\mbox{\hphantom{\Scribtexttt{xx}}}\RktMeta{}\RktPn{(}\RktSym{super{-}new}\RktPn{)}\RktMeta{}

\Smaller{\Smaller{\mbox{\hphantom{\Scribtexttt{x}}}\Scribtexttt{4}\mbox{\hphantom{\Scribtexttt{x}}}}}\RktMeta{}\mbox{\hphantom{\Scribtexttt{xx}}}\RktMeta{}\RktPn{(}\RktSym{define{-}state}\RktMeta{}\mbox{\hphantom{\Scribtexttt{x}}}\RktMeta{}\RktSym{text}\RktMeta{}\mbox{\hphantom{\Scribtexttt{x}}}\RktMeta{}\RktVal{""}\RktPn{)}\RktMeta{}

\Smaller{\Smaller{\mbox{\hphantom{\Scribtexttt{x}}}\Scribtexttt{5}\mbox{\hphantom{\Scribtexttt{x}}}}}\RktMeta{}\mbox{\hphantom{\Scribtexttt{xx}}}\RktMeta{}\RktPn{(}\RktSym{define/augment}\RktMeta{}\mbox{\hphantom{\Scribtexttt{x}}}\RktMeta{}\RktPn{(}\RktSym{draw}\RktMeta{}\mbox{\hphantom{\Scribtexttt{x}}}\RktMeta{}\RktSym{dc}\RktPn{)}\RktMeta{}\mbox{\hphantom{\Scribtexttt{x}}}\RktMeta{}\RktSym{{\hbox{\texttt{.}}}{\hbox{\texttt{.}}}{\hbox{\texttt{.}}}}\RktPn{)}\RktPn{)}\RktMeta{}

\Smaller{\Smaller{\mbox{\hphantom{\Scribtexttt{x}}}\Scribtexttt{6}\mbox{\hphantom{\Scribtexttt{x}}}}}\RktMeta{~}

\Smaller{\Smaller{\mbox{\hphantom{\Scribtexttt{x}}}\Scribtexttt{7}\mbox{\hphantom{\Scribtexttt{x}}}}}\RktMeta{}\RktCmt{;;}\mbox{\hphantom{\Scribtexttt{x}}}\RktCmt{Mixin}\mbox{\hphantom{\Scribtexttt{x}}}\RktCmt{for}\mbox{\hphantom{\Scribtexttt{x}}}\RktCmt{basic}\mbox{\hphantom{\Scribtexttt{x}}}\RktCmt{user}\mbox{\hphantom{\Scribtexttt{x}}}\RktCmt{interaction}\RktMeta{}

\Smaller{\Smaller{\mbox{\hphantom{\Scribtexttt{x}}}\Scribtexttt{8}\mbox{\hphantom{\Scribtexttt{x}}}}}\RktMeta{}\RktPn{(}\RktSym{define{-}interactive{-}syntax{-}mixin}\RktMeta{}\mbox{\hphantom{\Scribtexttt{x}}}\RktMeta{}\RktSym{focus\$\$}\RktMeta{}

\Smaller{\Smaller{\mbox{\hphantom{\Scribtexttt{x}}}\Scribtexttt{9}\mbox{\hphantom{\Scribtexttt{x}}}}}\RktMeta{}\mbox{\hphantom{\Scribtexttt{xx}}}\RktMeta{}\RktPn{(}\RktSym{super{-}new}\RktPn{)}\RktMeta{}

\Smaller{\Smaller{\Scribtexttt{10}\mbox{\hphantom{\Scribtexttt{x}}}}}\RktMeta{}\mbox{\hphantom{\Scribtexttt{xx}}}\RktMeta{}\RktPn{(}\RktSym{define/augment}\RktMeta{}\mbox{\hphantom{\Scribtexttt{x}}}\RktMeta{}\RktPn{(}\RktSym{on{-}event}\RktMeta{}\mbox{\hphantom{\Scribtexttt{x}}}\RktMeta{}\RktSym{event}\RktPn{)}\RktMeta{}\mbox{\hphantom{\Scribtexttt{x}}}\RktMeta{}\RktSym{{\hbox{\texttt{.}}}{\hbox{\texttt{.}}}{\hbox{\texttt{.}}}}\RktPn{)}\RktPn{)}\RktMeta{}

\Smaller{\Smaller{\Scribtexttt{11}\mbox{\hphantom{\Scribtexttt{x}}}}}\RktMeta{~}

\Smaller{\Smaller{\Scribtexttt{12}\mbox{\hphantom{\Scribtexttt{x}}}}}\RktMeta{}\RktCmt{;;}\mbox{\hphantom{\Scribtexttt{x}}}\RktCmt{A}\mbox{\hphantom{\Scribtexttt{x}}}\RktCmt{text}\mbox{\hphantom{\Scribtexttt{x}}}\RktCmt{field}\mbox{\hphantom{\Scribtexttt{x}}}\RktCmt{widget}\RktMeta{}

\Smaller{\Smaller{\Scribtexttt{13}\mbox{\hphantom{\Scribtexttt{x}}}}}\RktMeta{}\RktPn{(}\RktSym{define{-}interactive{-}syntax}\RktMeta{}\mbox{\hphantom{\Scribtexttt{x}}}\RktMeta{}\RktSym{text{-}field\$}\RktMeta{}\mbox{\hphantom{\Scribtexttt{x}}}\RktMeta{}\RktPn{(}\RktSym{focus\$\$}\RktMeta{}\mbox{\hphantom{\Scribtexttt{x}}}\RktMeta{}\RktPn{(}\RktSym{text\$\$}\RktMeta{}\mbox{\hphantom{\Scribtexttt{x}}}\RktMeta{}\RktSym{widget\$}\RktPn{)}\RktPn{)}\RktMeta{}

\Smaller{\Smaller{\Scribtexttt{14}\mbox{\hphantom{\Scribtexttt{x}}}}}\RktMeta{}\mbox{\hphantom{\Scribtexttt{xx}}}\RktMeta{}\RktPn{(}\RktSym{super{-}new}\RktPn{)}\RktPn{)}\RktMeta{}\end{SingleColumn}\end{RktBlk}\end{SCodeFlow}\end{FigureInside}\end{Centerfigure}

\Centertext{\Legend{\FigureTarget{\label{t:x28counter_x28x22figurex22_x22figx3afieldx2dandx2dlabelx22x29x29}\textsf{Fig.}~\textsf{4}. }{t:x28counter_x28x22figurex22_x22figx3afieldx2dandx2dlabelx22x29x29}\textsf{Field editor using mixins}}}\end{Herefigure}

\sectionNewpage

\Ssection{Evaluating a Plethora of Examples}{Evaluating a Plethora of Examples}\label{t:x28part_x22secx3aexamplesx22x29}

The Tsuro{-}specific syntax extensions illustrate two aspects
of programming with interactive visual extensions. First,
the interactive composition of visual and textual code can
obviously express ideas better than just text (or just
pictures). In a sense, this first insight is not surprising.
Like English, many natural languages come with the idiom
that {``}a picture is worth a thousand words.{''} What might
surprise readers (as it did the authors) is that there is
barely any support for this idea in the
world of programming languages. Perhaps language designers
could not imagine how widely this idea is applicable or how
to make this idea work easily.

Second, the implementation sketch demonstrates the ease of developing such
interactive extensions. The effort looks eminently reasonable in the context of a
prototype, especially since the essential code of this particular example can be shared
between the GUI interface to Tsuro and the unit test suites. A
continued development of this prototype is likely to reduce the development burden
even more, just like research on syntactic extensions has reduced the work
of macro writers.

Naturally, a single example cannot serve as the basis of an
evaluation. A truly proper evaluation of this new language
feature must demonstrate its expressive power with a number
of distinct cases. Additionally, it must show that the
effort remains reasonable across this spectrum of examples.
This section starts with a list of inspirational sources:
numerous text book illustrations of algorithms with
diagrams, pictorial illustrations in standards such as
RFCs, and ASCII diagrams in code repositories. The second subsection
surveys a range of uses and our implementation of those
uses, with an emphasis on where and how interactive syntax
can be deployed. The final two subsections present two cases
in some depth.

\Ssubsection{Examples of Diagram Documentation}{Examples of Diagram Documentation}\label{t:x28part_x22subx3aexamplesx22x29}

Text books, documentation, source code inspection, and practical experience
all motivate the idea of interactive{-}syntax extensions.

\Ssubsubsectionstarx{Tree Algorithms}{Tree Algorithms}\label{t:x28part_x22Treex5fAlgorithmsx22x29}

 Every standard algorithms book and every tree
automata monograph \Autobibref{~[\hyperref[t:x28autobib_x22Hubert_Comonx2c_Max_Dauchetx2c_Remi_Gilleronx2c_Florent_Jacquemardx2c_Denis_Lugiezx2c_Christof_Lxf6dingx2c_Sophie_Tisonx2c_and_Marc_TommasiTree_Automata_Techniques_and_Applications2007httpx3ax2fx2ftatax2egforgex2einriax2efrx2fx22x29]{\AutobibLink{Comon et al\Sendabbrev{.}}} \hyperref[t:x28autobib_x22Hubert_Comonx2c_Max_Dauchetx2c_Remi_Gilleronx2c_Florent_Jacquemardx2c_Denis_Lugiezx2c_Christof_Lxf6dingx2c_Sophie_Tisonx2c_and_Marc_TommasiTree_Automata_Techniques_and_Applications2007httpx3ax2fx2ftatax2egforgex2einriax2efrx2fx22x29]{\AutobibLink{\Thyperref{2007}{autobiblab:9}}}; \hyperref[t:x28autobib_x22Thomas_Hx2e_Cormenx2c_Charles_Ex2e_Leisersonx2c_Ronald_Lx2e_Rivestx2c_and_Clifford_SteinIntroduction_to_Algorithmsx2c_Third_EditionMIT_Press2009x22x29]{\AutobibLink{Cormen et al\Sendabbrev{.}}} \hyperref[t:x28autobib_x22Thomas_Hx2e_Cormenx2c_Charles_Ex2e_Leisersonx2c_Ronald_Lx2e_Rivestx2c_and_Clifford_SteinIntroduction_to_Algorithmsx2c_Third_EditionMIT_Press2009x22x29]{\AutobibLink{\Thyperref{2009}{autobiblab:11}}}]} comes with many diagrams to
describe tree manipulations. Programmers often include ASCII diagrams of
trees in comments to document complex code.\NoteBox{\NoteContent{https://git.musl{-}libc.org/cgit/musl/tree/src/search/tsearch.c?id=v1.1.21}} These
diagrams contain concrete trees and depict abstract tree transformations.

\Ssubsubsectionstarx{Matrix}{Matrix}\label{t:x28part_x22Matrixx22x29}

 Astute programmers
format matrix{-}manipulation code to reflect literal matrices
when possible.\NoteBox{\NoteContent{http://www.opengl{-}tutorial.org/}} Mathematical
programming books depict matrices as rectangles in otherwise
linear text\Autobibref{~[\hyperref[t:x28autobib_x22Robert_Fourerx2c_David_Mx2e_Gayx2c_and_Brian_Wx2e_KernighanAMPLx3a_A_Modeling_Language_for_Mathematical_Programming2nd_editionx2e_Cengage_Learning2002httpsx3ax2fx2famplx2ecomx2fresourcesx2fthex2damplx2dbookx2fx22x29]{\AutobibLink{Fourer et al\Sendabbrev{.}}} \hyperref[t:x28autobib_x22Robert_Fourerx2c_David_Mx2e_Gayx2c_and_Brian_Wx2e_KernighanAMPLx3a_A_Modeling_Language_for_Mathematical_Programming2nd_editionx2e_Cengage_Learning2002httpsx3ax2fx2famplx2ecomx2fresourcesx2fthex2damplx2dbookx2fx22x29]{\AutobibLink{\Thyperref{2002}{autobiblab:26}}}]}.

\Ssubsubsectionstarx{File System Data Structures}{File System Data Structures}\label{t:x28part_x22Filex5fSystemx5fDatax5fStructuresx22x29}

Any systems course that covers the \Scribtexttt{inode} file
representation describes it with box{-}and{-}pointer
diagrams.\NoteBox{\NoteContent{https://www.youtube.com/watch?v=tMVj22EWg6A}} Likewise, source code for these data
structures frequently include ASCII sketches of these
diagrams.

\Ssubsubsectionstarx{TCP}{TCP}\label{t:x28part_x22TCPx22x29}

RFC{-}793\Autobibref{~[\hyperref[t:x28autobib_x22Jon_PostelTransmission_Control_ProtocolInternet_Engineering_Task_Forcex2c_RFC_7931981httpsx3ax2fx2ftoolsx2eietfx2eorgx2fhtmlx2frfc793x22x29]{\AutobibLink{Postel}} \hyperref[t:x28autobib_x22Jon_PostelTransmission_Control_ProtocolInternet_Engineering_Task_Forcex2c_RFC_7931981httpsx3ax2fx2ftoolsx2eietfx2eorgx2fhtmlx2frfc793x22x29]{\AutobibLink{\Thyperref{1981}{autobiblab:44}}}]} for TCP lays out the format of
messages via a table{-}shaped diagram. Each row represents a
32{-}bit word that is split into four 8{-}bit octets.

\Ssubsubsectionstarx{Pictures as Bindings}{Pictures as Bindings}\label{t:x28part_x22Picturesx5fasx5fBindingsx22x29}

 Many
visual programming environments, such as Game
Maker\Autobibref{~[\hyperref[t:x28autobib_x22Mark_OvermarsTeaching_Computer_Science_Through_Game_DesignComputer_37x284x29x2c_ppx2e_81x2dx2d832004httpsx3ax2fx2fdoix2eorgx2f10x2e1109x2fMCx2e2004x2e1297314x22x29]{\AutobibLink{Overmars}} \hyperref[t:x28autobib_x22Mark_OvermarsTeaching_Computer_Science_Through_Game_DesignComputer_37x284x29x2c_ppx2e_81x2dx2d832004httpsx3ax2fx2fdoix2eorgx2f10x2e1109x2fMCx2e2004x2e1297314x22x29]{\AutobibLink{\Thyperref{2004}{autobiblab:41}}}]}, allow developers to lay out their
programs as actors placed on a spatial grid. Actors are
depicted as pictorial avatars and the code defining each actor{'}s
behavior refers to other actors using avatars. In
other words, pictures act as the variable names referencing
objects in this environment.

\Ssubsubsectionstarx{Video Editors}{Video Editors}\label{t:x28part_x22Videox5fEditorsx22x29}

 Video editing is predominantly done via non{-}linear,
graphical editors. Such purely graphical editors are prone to force people
to perform manually repetitive tasks.

\Ssubsubsectionstarx{Circuits}{Circuits}\label{t:x28part_x22Circuitsx22x29}

 Circuits are naturally described graphically. Reviewers
might be familiar with Tikz and CircuitTikz, two LaTeX libraries for
drawing diagrams and specifically circuit diagrams. Coding diagrams in
these languages is rather painful, though; manipulating them afterwards to
put them into the proper place within a paper can also pose challenges.

Electrical engineers code circuits in the domain{-}specific
SPICE\Autobibref{~[\hyperref[t:x28autobib_x22Holger_Vogtx2c_Marcel_Hendrixx2c_and_Paolo_NenziNgspice_Users_ManualNGSPICEx2c_302019httpx3ax2fx2fngspicex2esourceforgex2enetx2fdocsx2fngspicex2d30x2dmanualx2epdfx22x29]{\AutobibLink{Vogt et al\Sendabbrev{.}}} \hyperref[t:x28autobib_x22Holger_Vogtx2c_Marcel_Hendrixx2c_and_Paolo_NenziNgspice_Users_ManualNGSPICEx2c_302019httpx3ax2fx2fngspicex2esourceforgex2enetx2fdocsx2fngspicex2d30x2dmanualx2epdfx22x29]{\AutobibLink{\Thyperref{2019}{autobiblab:50}}}]} simulation language or hardware
description languages such as Xilinx ISE. While both come
with tools to edit circuits graphically, engineers
cannot mix and match textual and graphical part definitions.

\Ssubsection{The Expressive Power of Interactive{-}Syntax Extensions}{The Expressive Power of Interactive{-}Syntax Extensions}\label{t:x28part_x22subx3aattributesx22x29}

We have implemented the examples from the previous
section with interactive syntax. Doing so yields easily
readable code and several insights on the expressive power
of mixing visual and textual syntax. Here we present a
classification of the linguistic roles that these extensions
play within code. Figure~\hyperref[t:x28counter_x28x22figurex22_x22figx3aworkedx2dexamplesx22x29x29]{\FigureRef{5}{t:x28counter_x28x22figurex22_x22figx3aworkedx2dexamplesx22x29x29}}
provides a concise overview. The first column lists the name
of the example, the second the role that interactive syntax
plays. The third column reports the number of lines of code
needed for these extensions.

\begin{Figure}\begin{Centerfigure}\begin{FigureInside}\raisebox{-0.8968749999999943bp}{\makebox[339.20000000000005bp][l]{\includegraphics[trim=2.4000000000000004 2.4000000000000004 2.4000000000000004 2.4000000000000004]{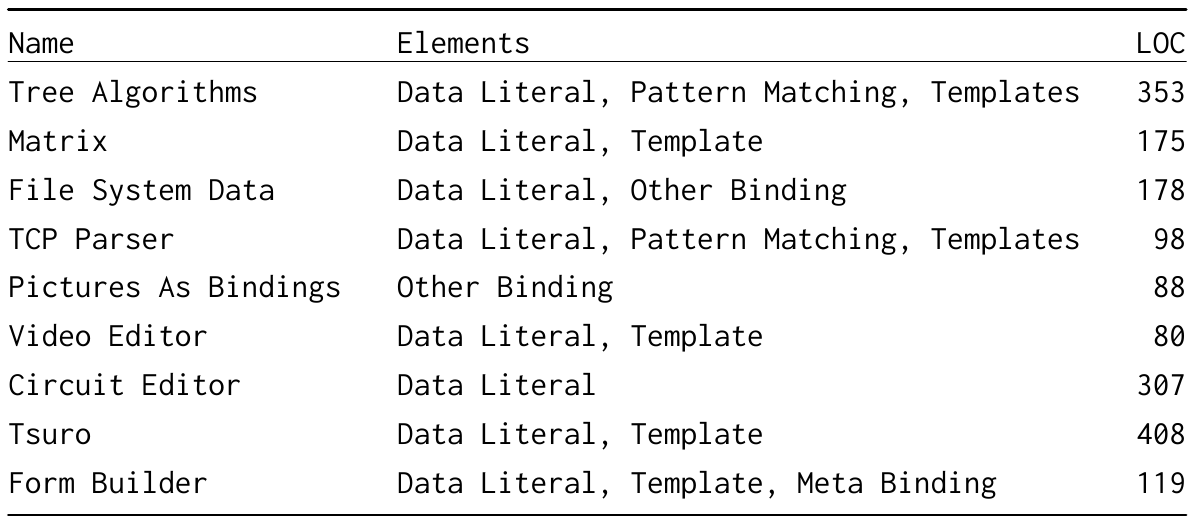}}}\end{FigureInside}\end{Centerfigure}

\Centertext{\Legend{\FigureTarget{\label{t:x28counter_x28x22figurex22_x22figx3aworkedx2dexamplesx22x29x29}\textsf{Fig.}~\textsf{5}. }{t:x28counter_x28x22figurex22_x22figx3aworkedx2dexamplesx22x29x29}\textsf{Attributes of the Worked Language Extensions}}}\end{Figure}

\Ssubsubsectionstarx{Data Literal}{Data Literal}\label{t:x28part_x22Datax5fLiteralx22x29}

 The simplest role is that of a data literal.  In this
role, developers interact with the syntax only to enter plain textual data;
the elaborator tends to translate these editors into structures or
objects.  As the Tsuro examples in the introduction point out,
data{-}literal forms of interactive syntax can be replaced by a lot of
text{---}at the cost of reduced readability.

\Ssubsubsectionstarx{Template}{Template}\label{t:x28part_x22Templatex22x29}

 The template{'}s role
generalizes data literals. Instead of entering plain data
into an editor, a developer inserts code into text fields of these instances.
The Tsuro board in the introduction and the tree in
figure~\hyperref[t:x28counter_x28x22figurex22_x22rbx2dbalancex22x29x29]{\FigureRef{7}{t:x28counter_x28x22figurex22_x22rbx2dbalancex22x29x29}} are examples of such templates.
The templates build a board and tree, respectively, using pattern
variables embedded in an editor.

\Ssubsubsectionstarx{Pattern Matching}{Pattern Matching}\label{t:x28part_x22Patternx5fMatchingx22x29}

 Languages
such as Scala emphasize pattern matching, and interactive
syntax can greatly enhance the message that a pattern
expresses. In this context, a developer fills an editor with
pattern variables, and the code generator synthesizes a
pattern from the visual parts and these pattern variables.
In this role, pattern{-}matching editors serve as binding
constructs.

\Ssubsubsectionstarx{Meta Binding}{Meta Binding}\label{t:x28part_x22Metax5fBindingx22x29}

 Since syntax extension is a form of meta programming,
interactive syntax naturally plays a meta{-}programming role, too. We refer
to this role as meta{-}binding in figure~\hyperref[t:x28counter_x28x22figurex22_x22figx3aworkedx2dexamplesx22x29x29]{\FigureRef{5}{t:x28counter_x28x22figurex22_x22figx3aworkedx2dexamplesx22x29x29}}. Here
editors are used to construct new types of syntax, and because the
prototype is in Racket, they can generate both graphical and textual syntax
extensions. \SecRefUC{\SectionNumberLink{t:x28part_x22secx3aexamplex2dmetax22x29}{4.4}}{In Depth: Form Builders} demonstrates this idea with form
builders.

\Ssubsubsectionstarx{Other Binding}{Other Binding}\label{t:x28part_x22Otherx5fBindingx22x29}

 Finally, editors can play the role of a binding
form. This role allows multiple editors to {``}talk{''} to each other. The
\Scribtexttt{inode} data structure supplies an example of this kind.

\Ssubsection{In Depth: Red{-}Black Trees}{In Depth: Red{-}Black Trees}\label{t:x28part_x22subx3arbtx22x29}

\begin{Figure}\begin{Centerfigure}\begin{FigureInside}\begin{SCodeFlow}\begin{RktBlk}\begin{SingleColumn}\Smaller{\Smaller{\mbox{\hphantom{\Scribtexttt{x}}}\Scribtexttt{1}\mbox{\hphantom{\Scribtexttt{xx}}}}}\RktCmt{;}\mbox{\hphantom{\Scribtexttt{x}}}\RktCmt{Balance}\mbox{\hphantom{\Scribtexttt{x}}}\RktCmt{({-}}\mbox{\hphantom{\Scribtexttt{x}}}\RktCmt{black)}\mbox{\hphantom{\Scribtexttt{x}}}\RktCmt{(=}\mbox{\hphantom{\Scribtexttt{x}}}\RktCmt{red)}\RktMeta{}

\Smaller{\Smaller{\mbox{\hphantom{\Scribtexttt{x}}}\Scribtexttt{2}\mbox{\hphantom{\Scribtexttt{xx}}}}}\RktMeta{}\RktCmt{;}\RktMeta{}

\Smaller{\Smaller{\mbox{\hphantom{\Scribtexttt{x}}}\Scribtexttt{3}\mbox{\hphantom{\Scribtexttt{xx}}}}}\RktMeta{}\RktCmt{;}\mbox{\hphantom{\Scribtexttt{xx}}}\RktCmt{/}\mbox{\hphantom{\Scribtexttt{xxxxxx}}}\RktCmt{{-}z{-}}\mbox{\hphantom{\Scribtexttt{xx}}}\RktCmt{{\Stttextbar}}\mbox{\hphantom{\Scribtexttt{xxxx}}}\RktCmt{{-}z{-}}\mbox{\hphantom{\Scribtexttt{xx}}}\RktCmt{{\Stttextbar}}\mbox{\hphantom{\Scribtexttt{xx}}}\RktCmt{{-}x{-}}\mbox{\hphantom{\Scribtexttt{xxxx}}}\RktCmt{{\Stttextbar}}\mbox{\hphantom{\Scribtexttt{xx}}}\RktCmt{{-}x{-}}\mbox{\hphantom{\Scribtexttt{xxxxxx}}}\RktCmt{{\char`\\}}\RktMeta{}

\Smaller{\Smaller{\mbox{\hphantom{\Scribtexttt{x}}}\Scribtexttt{4}\mbox{\hphantom{\Scribtexttt{xx}}}}}\RktMeta{}\RktCmt{;}\mbox{\hphantom{\Scribtexttt{x}}}\RktCmt{{\Stttextbar}}\mbox{\hphantom{\Scribtexttt{xxxxxxx}}}\RktCmt{/}\mbox{\hphantom{\Scribtexttt{x}}}\RktCmt{{\char`\\}}\mbox{\hphantom{\Scribtexttt{xx}}}\RktCmt{{\Stttextbar}}\mbox{\hphantom{\Scribtexttt{xxxx}}}\RktCmt{/}\mbox{\hphantom{\Scribtexttt{x}}}\RktCmt{{\char`\\}}\mbox{\hphantom{\Scribtexttt{xx}}}\RktCmt{{\Stttextbar}}\mbox{\hphantom{\Scribtexttt{xx}}}\RktCmt{/}\mbox{\hphantom{\Scribtexttt{x}}}\RktCmt{{\char`\\}}\mbox{\hphantom{\Scribtexttt{xxxx}}}\RktCmt{{\Stttextbar}}\mbox{\hphantom{\Scribtexttt{xx}}}\RktCmt{/}\mbox{\hphantom{\Scribtexttt{x}}}\RktCmt{{\char`\\}}\mbox{\hphantom{\Scribtexttt{xxxxxxx}}}\RktCmt{{\Stttextbar}}\mbox{\hphantom{\Scribtexttt{xxxxxxxx}}}\RktCmt{=y=}\RktMeta{}

\Smaller{\Smaller{\mbox{\hphantom{\Scribtexttt{x}}}\Scribtexttt{5}\mbox{\hphantom{\Scribtexttt{xx}}}}}\RktMeta{}\RktCmt{;}\mbox{\hphantom{\Scribtexttt{x}}}\RktCmt{{\Stttextbar}}\mbox{\hphantom{\Scribtexttt{xxxxx}}}\RktCmt{=y=}\mbox{\hphantom{\Scribtexttt{xx}}}\RktCmt{D}\mbox{\hphantom{\Scribtexttt{x}}}\RktCmt{{\Stttextbar}}\mbox{\hphantom{\Scribtexttt{xx}}}\RktCmt{=x=}\mbox{\hphantom{\Scribtexttt{xx}}}\RktCmt{D}\mbox{\hphantom{\Scribtexttt{x}}}\RktCmt{{\Stttextbar}}\mbox{\hphantom{\Scribtexttt{x}}}\RktCmt{A}\mbox{\hphantom{\Scribtexttt{xx}}}\RktCmt{=z=}\mbox{\hphantom{\Scribtexttt{xx}}}\RktCmt{{\Stttextbar}}\mbox{\hphantom{\Scribtexttt{x}}}\RktCmt{A}\mbox{\hphantom{\Scribtexttt{xx}}}\RktCmt{=y=}\mbox{\hphantom{\Scribtexttt{xxxxx}}}\RktCmt{{\Stttextbar}}\mbox{\hphantom{\Scribtexttt{xxxxxxxx}}}\RktCmt{/}\mbox{\hphantom{\Scribtexttt{x}}}\RktCmt{{\char`\\}}\RktMeta{}

\Smaller{\Smaller{\mbox{\hphantom{\Scribtexttt{x}}}\Scribtexttt{6}\mbox{\hphantom{\Scribtexttt{xx}}}}}\RktMeta{}\RktCmt{;}\mbox{\hphantom{\Scribtexttt{x}}}\RktCmt{{\Stttextbar}}\mbox{\hphantom{\Scribtexttt{xxxxx}}}\RktCmt{/}\mbox{\hphantom{\Scribtexttt{x}}}\RktCmt{{\char`\\}}\mbox{\hphantom{\Scribtexttt{xxxx}}}\RktCmt{{\Stttextbar}}\mbox{\hphantom{\Scribtexttt{xx}}}\RktCmt{/}\mbox{\hphantom{\Scribtexttt{x}}}\RktCmt{{\char`\\}}\mbox{\hphantom{\Scribtexttt{xxxx}}}\RktCmt{{\Stttextbar}}\mbox{\hphantom{\Scribtexttt{xxxx}}}\RktCmt{/}\mbox{\hphantom{\Scribtexttt{x}}}\RktCmt{{\char`\\}}\mbox{\hphantom{\Scribtexttt{xx}}}\RktCmt{{\Stttextbar}}\mbox{\hphantom{\Scribtexttt{xxxx}}}\RktCmt{/}\mbox{\hphantom{\Scribtexttt{x}}}\RktCmt{{\char`\\}}\mbox{\hphantom{\Scribtexttt{xxxxx}}}\RktCmt{{\Stttextbar}}\mbox{\hphantom{\Scribtexttt{x}}}\RktCmt{{-}{-}{\Stttextmore}}\mbox{\hphantom{\Scribtexttt{xx}}}\RktCmt{{-}x{-}}\mbox{\hphantom{\Scribtexttt{x}}}\RktCmt{{-}z{-}}\RktMeta{}

\Smaller{\Smaller{\mbox{\hphantom{\Scribtexttt{x}}}\Scribtexttt{7}\mbox{\hphantom{\Scribtexttt{xx}}}}}\RktMeta{}\RktCmt{;}\mbox{\hphantom{\Scribtexttt{x}}}\RktCmt{{\Stttextbar}}\mbox{\hphantom{\Scribtexttt{xxx}}}\RktCmt{=x=}\mbox{\hphantom{\Scribtexttt{xx}}}\RktCmt{C}\mbox{\hphantom{\Scribtexttt{xxx}}}\RktCmt{{\Stttextbar}}\mbox{\hphantom{\Scribtexttt{x}}}\RktCmt{A}\mbox{\hphantom{\Scribtexttt{xx}}}\RktCmt{=y=}\mbox{\hphantom{\Scribtexttt{xx}}}\RktCmt{{\Stttextbar}}\mbox{\hphantom{\Scribtexttt{xx}}}\RktCmt{=y=}\mbox{\hphantom{\Scribtexttt{xx}}}\RktCmt{D}\mbox{\hphantom{\Scribtexttt{x}}}\RktCmt{{\Stttextbar}}\mbox{\hphantom{\Scribtexttt{xxx}}}\RktCmt{B}\mbox{\hphantom{\Scribtexttt{xx}}}\RktCmt{=z=}\mbox{\hphantom{\Scribtexttt{xxx}}}\RktCmt{{\Stttextbar}}\mbox{\hphantom{\Scribtexttt{xxxxxx}}}\RktCmt{/{\Stttextbar}}\mbox{\hphantom{\Scribtexttt{xxx}}}\RktCmt{{\Stttextbar}{\char`\\}}\RktMeta{}

\Smaller{\Smaller{\mbox{\hphantom{\Scribtexttt{x}}}\Scribtexttt{8}\mbox{\hphantom{\Scribtexttt{xx}}}}}\RktMeta{}\RktCmt{;}\mbox{\hphantom{\Scribtexttt{x}}}\RktCmt{{\Stttextbar}}\mbox{\hphantom{\Scribtexttt{xxx}}}\RktCmt{/}\mbox{\hphantom{\Scribtexttt{x}}}\RktCmt{{\char`\\}}\mbox{\hphantom{\Scribtexttt{xxxxxx}}}\RktCmt{{\Stttextbar}}\mbox{\hphantom{\Scribtexttt{xxxx}}}\RktCmt{/}\mbox{\hphantom{\Scribtexttt{x}}}\RktCmt{{\char`\\}}\mbox{\hphantom{\Scribtexttt{xx}}}\RktCmt{{\Stttextbar}}\mbox{\hphantom{\Scribtexttt{xx}}}\RktCmt{/}\mbox{\hphantom{\Scribtexttt{x}}}\RktCmt{{\char`\\}}\mbox{\hphantom{\Scribtexttt{xxxx}}}\RktCmt{{\Stttextbar}}\mbox{\hphantom{\Scribtexttt{xxxxxx}}}\RktCmt{/}\mbox{\hphantom{\Scribtexttt{x}}}\RktCmt{{\char`\\}}\mbox{\hphantom{\Scribtexttt{xxx}}}\RktCmt{{\Stttextbar}}\mbox{\hphantom{\Scribtexttt{xxxxx}}}\RktCmt{A}\mbox{\hphantom{\Scribtexttt{x}}}\RktCmt{B}\mbox{\hphantom{\Scribtexttt{xxx}}}\RktCmt{C}\mbox{\hphantom{\Scribtexttt{x}}}\RktCmt{D}\RktMeta{}

\Smaller{\Smaller{\mbox{\hphantom{\Scribtexttt{x}}}\Scribtexttt{9}\mbox{\hphantom{\Scribtexttt{xx}}}}}\RktMeta{}\RktCmt{;}\mbox{\hphantom{\Scribtexttt{xx}}}\RktCmt{{\char`\\}}\mbox{\hphantom{\Scribtexttt{x}}}\RktCmt{A}\mbox{\hphantom{\Scribtexttt{xxx}}}\RktCmt{B}\mbox{\hphantom{\Scribtexttt{xxxxx}}}\RktCmt{{\Stttextbar}}\mbox{\hphantom{\Scribtexttt{xxx}}}\RktCmt{B}\mbox{\hphantom{\Scribtexttt{xxx}}}\RktCmt{C}\mbox{\hphantom{\Scribtexttt{x}}}\RktCmt{{\Stttextbar}}\mbox{\hphantom{\Scribtexttt{x}}}\RktCmt{B}\mbox{\hphantom{\Scribtexttt{xx}}}\RktCmt{C}\mbox{\hphantom{\Scribtexttt{xxxx}}}\RktCmt{{\Stttextbar}}\mbox{\hphantom{\Scribtexttt{xxxxx}}}\RktCmt{C}\mbox{\hphantom{\Scribtexttt{xxx}}}\RktCmt{D}\mbox{\hphantom{\Scribtexttt{x}}}\RktCmt{/}\RktMeta{}

\Smaller{\Smaller{\Scribtexttt{10}\mbox{\hphantom{\Scribtexttt{xx}}}}}\RktMeta{}\RktPn{(}\RktSym{define/match}\RktMeta{}\mbox{\hphantom{\Scribtexttt{x}}}\RktMeta{}\RktPn{(}\RktSym{balance}\RktMeta{}\mbox{\hphantom{\Scribtexttt{x}}}\RktMeta{}\RktSym{t}\RktPn{)}\RktMeta{}

\Smaller{\Smaller{\Scribtexttt{11}\mbox{\hphantom{\Scribtexttt{xx}}}}}\RktMeta{}\mbox{\hphantom{\Scribtexttt{xx}}}\RktMeta{}\RktPn{[}\RktPn{(}\RktSym{or}\RktMeta{}\mbox{\hphantom{\Scribtexttt{x}}}\RktMeta{}\RktPn{(}\RktSym{tree}\RktMeta{}\mbox{\hphantom{\Scribtexttt{x}}}\RktMeta{}\RktSym{z}\RktMeta{}\mbox{\hphantom{\Scribtexttt{x}}}\RktMeta{}\RktSym{{\textquotesingle}}\RktSym{black}\RktMeta{}\mbox{\hphantom{\Scribtexttt{x}}}\RktMeta{}\RktPn{(}\RktSym{tree}\RktMeta{}\mbox{\hphantom{\Scribtexttt{x}}}\RktMeta{}\RktSym{y}\RktMeta{}\mbox{\hphantom{\Scribtexttt{x}}}\RktMeta{}\RktSym{{\textquotesingle}}\RktSym{red}\RktMeta{}\mbox{\hphantom{\Scribtexttt{x}}}\RktMeta{}\RktPn{(}\RktSym{tree}\RktMeta{}\mbox{\hphantom{\Scribtexttt{x}}}\RktMeta{}\RktSym{x}\RktMeta{}\mbox{\hphantom{\Scribtexttt{x}}}\RktMeta{}\RktSym{{\textquotesingle}}\RktSym{red}\RktMeta{}\mbox{\hphantom{\Scribtexttt{x}}}\RktMeta{}\RktSym{A}\RktMeta{}\mbox{\hphantom{\Scribtexttt{x}}}\RktMeta{}\RktSym{B}\RktPn{)}\RktMeta{}\mbox{\hphantom{\Scribtexttt{x}}}\RktMeta{}\RktSym{C}\RktPn{)}\RktMeta{}\mbox{\hphantom{\Scribtexttt{x}}}\RktMeta{}\RktSym{D}\RktPn{)}\RktMeta{}

\Smaller{\Smaller{\Scribtexttt{12}\mbox{\hphantom{\Scribtexttt{xx}}}}}\RktMeta{}\mbox{\hphantom{\Scribtexttt{xxxxxxx}}}\RktMeta{}\RktPn{(}\RktSym{tree}\RktMeta{}\mbox{\hphantom{\Scribtexttt{x}}}\RktMeta{}\RktSym{z}\RktMeta{}\mbox{\hphantom{\Scribtexttt{x}}}\RktMeta{}\RktSym{{\textquotesingle}}\RktSym{black}\RktMeta{}\mbox{\hphantom{\Scribtexttt{x}}}\RktMeta{}\RktPn{(}\RktSym{tree}\RktMeta{}\mbox{\hphantom{\Scribtexttt{x}}}\RktMeta{}\RktSym{x}\RktMeta{}\mbox{\hphantom{\Scribtexttt{x}}}\RktMeta{}\RktSym{{\textquotesingle}}\RktSym{red}\RktMeta{}\mbox{\hphantom{\Scribtexttt{x}}}\RktMeta{}\RktSym{A}\RktMeta{}\mbox{\hphantom{\Scribtexttt{x}}}\RktMeta{}\RktPn{(}\RktSym{tree}\RktMeta{}\mbox{\hphantom{\Scribtexttt{x}}}\RktMeta{}\RktSym{y}\RktMeta{}\mbox{\hphantom{\Scribtexttt{x}}}\RktMeta{}\RktSym{{\textquotesingle}}\RktSym{red}\RktMeta{}\mbox{\hphantom{\Scribtexttt{x}}}\RktMeta{}\RktSym{B}\RktMeta{}\mbox{\hphantom{\Scribtexttt{x}}}\RktMeta{}\RktSym{C}\RktPn{)}\RktPn{)}\RktMeta{}\mbox{\hphantom{\Scribtexttt{x}}}\RktMeta{}\RktSym{D}\RktPn{)}\RktMeta{}

\Smaller{\Smaller{\Scribtexttt{13}\mbox{\hphantom{\Scribtexttt{xx}}}}}\RktMeta{}\mbox{\hphantom{\Scribtexttt{xxxxxxx}}}\RktMeta{}\RktPn{(}\RktSym{tree}\RktMeta{}\mbox{\hphantom{\Scribtexttt{x}}}\RktMeta{}\RktSym{x}\RktMeta{}\mbox{\hphantom{\Scribtexttt{x}}}\RktMeta{}\RktSym{{\textquotesingle}}\RktSym{black}\RktMeta{}\mbox{\hphantom{\Scribtexttt{x}}}\RktMeta{}\RktSym{A}\RktMeta{}\mbox{\hphantom{\Scribtexttt{x}}}\RktMeta{}\RktPn{(}\RktSym{tree}\RktMeta{}\mbox{\hphantom{\Scribtexttt{x}}}\RktMeta{}\RktSym{z}\RktMeta{}\mbox{\hphantom{\Scribtexttt{x}}}\RktMeta{}\RktSym{{\textquotesingle}}\RktSym{red}\RktMeta{}\mbox{\hphantom{\Scribtexttt{x}}}\RktMeta{}\RktPn{(}\RktSym{tree}\RktMeta{}\mbox{\hphantom{\Scribtexttt{x}}}\RktMeta{}\RktSym{y}\RktMeta{}\mbox{\hphantom{\Scribtexttt{x}}}\RktMeta{}\RktSym{{\textquotesingle}}\RktSym{red}\RktMeta{}\mbox{\hphantom{\Scribtexttt{x}}}\RktMeta{}\RktSym{B}\RktMeta{}\mbox{\hphantom{\Scribtexttt{x}}}\RktMeta{}\RktSym{C}\RktPn{)}\RktMeta{}\mbox{\hphantom{\Scribtexttt{x}}}\RktMeta{}\RktSym{D}\RktPn{)}\RktPn{)}\RktMeta{}

\Smaller{\Smaller{\Scribtexttt{14}\mbox{\hphantom{\Scribtexttt{xx}}}}}\RktMeta{}\mbox{\hphantom{\Scribtexttt{xxxxxxx}}}\RktMeta{}\RktPn{(}\RktSym{tree}\RktMeta{}\mbox{\hphantom{\Scribtexttt{x}}}\RktMeta{}\RktSym{x}\RktMeta{}\mbox{\hphantom{\Scribtexttt{x}}}\RktMeta{}\RktSym{{\textquotesingle}}\RktSym{black}\RktMeta{}\mbox{\hphantom{\Scribtexttt{x}}}\RktMeta{}\RktSym{A}\RktMeta{}\mbox{\hphantom{\Scribtexttt{x}}}\RktMeta{}\RktPn{(}\RktSym{tree}\RktMeta{}\mbox{\hphantom{\Scribtexttt{x}}}\RktMeta{}\RktSym{y}\RktMeta{}\mbox{\hphantom{\Scribtexttt{x}}}\RktMeta{}\RktSym{{\textquotesingle}}\RktSym{red}\RktMeta{}\mbox{\hphantom{\Scribtexttt{x}}}\RktMeta{}\RktSym{B}\RktMeta{}\mbox{\hphantom{\Scribtexttt{x}}}\RktMeta{}\RktPn{(}\RktSym{tree}\RktMeta{}\mbox{\hphantom{\Scribtexttt{x}}}\RktMeta{}\RktSym{z}\RktMeta{}\mbox{\hphantom{\Scribtexttt{x}}}\RktMeta{}\RktSym{{\textquotesingle}}\RktSym{red}\RktMeta{}\mbox{\hphantom{\Scribtexttt{x}}}\RktMeta{}\RktSym{C}\RktMeta{}\mbox{\hphantom{\Scribtexttt{x}}}\RktMeta{}\RktSym{D}\RktPn{)}\RktPn{)}\RktPn{)}\RktPn{)}\RktMeta{}

\Smaller{\Smaller{\Scribtexttt{15}\mbox{\hphantom{\Scribtexttt{xx}}}}}\RktMeta{}\mbox{\hphantom{\Scribtexttt{xxx}}}\RktMeta{}\RktPn{(}\RktSym{tree}\RktMeta{}\mbox{\hphantom{\Scribtexttt{x}}}\RktMeta{}\RktSym{y}\RktMeta{}\mbox{\hphantom{\Scribtexttt{x}}}\RktMeta{}\RktSym{{\textquotesingle}}\RktSym{red}\RktMeta{}\mbox{\hphantom{\Scribtexttt{x}}}\RktMeta{}\RktPn{(}\RktSym{tree}\RktMeta{}\mbox{\hphantom{\Scribtexttt{x}}}\RktMeta{}\RktSym{x}\RktMeta{}\mbox{\hphantom{\Scribtexttt{x}}}\RktMeta{}\RktSym{{\textquotesingle}}\RktSym{black}\RktMeta{}\mbox{\hphantom{\Scribtexttt{x}}}\RktMeta{}\RktSym{A}\RktMeta{}\mbox{\hphantom{\Scribtexttt{x}}}\RktMeta{}\RktSym{B}\RktPn{)}\RktMeta{}\mbox{\hphantom{\Scribtexttt{x}}}\RktMeta{}\RktPn{(}\RktSym{tree}\RktMeta{}\mbox{\hphantom{\Scribtexttt{x}}}\RktMeta{}\RktSym{z}\RktMeta{}\mbox{\hphantom{\Scribtexttt{x}}}\RktMeta{}\RktSym{{\textquotesingle}}\RktSym{black}\RktMeta{}\mbox{\hphantom{\Scribtexttt{x}}}\RktMeta{}\RktSym{C}\RktMeta{}\mbox{\hphantom{\Scribtexttt{x}}}\RktMeta{}\RktSym{D}\RktPn{)}\RktPn{)}\RktPn{]}\RktMeta{}

\Smaller{\Smaller{\Scribtexttt{16}\mbox{\hphantom{\Scribtexttt{xx}}}}}\RktMeta{}\mbox{\hphantom{\Scribtexttt{xx}}}\RktMeta{}\RktPn{[}\RktSym{else}\RktMeta{}\mbox{\hphantom{\Scribtexttt{x}}}\RktMeta{}\RktSym{t}\RktPn{]}\RktPn{)}\RktMeta{}\end{SingleColumn}\end{RktBlk}\end{SCodeFlow}\end{FigureInside}\end{Centerfigure}

\Centertext{\Legend{\FigureTarget{\label{t:x28counter_x28x22figurex22_x22textx2dbalancex22x29x29}\textsf{Fig.}~\textsf{6}. }{t:x28counter_x28x22figurex22_x22textx2dbalancex22x29x29}\textsf{A Textual Balance function for a Red{-}Black Tree}}}\end{Figure}

When programmers explain tree algorithms, they frequently describe the
essential ideas with diagrams. Often  these diagrams make it into the
library documentation and the programmer
who maintains the code has to go back and forth between the code and
the diagrams in the documentation. A poor man{'}s fix is to render such
diagrams as ASCII art.\NoteBox{\NoteContent{https://blog.regehr.org/archives/1653}}

The balancing algorithm for red{-}black trees\Autobibref{~[\hyperref[t:x28autobib_x22Rudolf_BayerSymmetric_Binary_Bx2dTreesx3a_Data_Structure_and_Maintenance_AlgorithmsActa_Informatica_1x284x29x2c_ppx2e_290x2dx2d3061972httpsx3ax2fx2fdoix2eorgx2f10x2e1007x2fBF00289509x22x29]{\AutobibLink{Bayer}} \hyperref[t:x28autobib_x22Rudolf_BayerSymmetric_Binary_Bx2dTreesx3a_Data_Structure_and_Maintenance_AlgorithmsActa_Informatica_1x284x29x2c_ppx2e_290x2dx2d3061972httpsx3ax2fx2fdoix2eorgx2f10x2e1007x2fBF00289509x22x29]{\AutobibLink{\Thyperref{1972}{autobiblab:4}}}]}
illustrates this kind of work particularly well.
Figure~\hyperref[t:x28counter_x28x22figurex22_x22textx2dbalancex22x29x29]{\FigureRef{6}{t:x28counter_x28x22figurex22_x22textx2dbalancex22x29x29}} shows a code snippet from a
tree{-}manipulation library in Racket. The snippet depicts a
function for balancing red{-}black trees using pattern
matching. The comment block (lines 1{--}9) makes up the
internal ASCII{-}art documentation of the functionality, while
the code itself (lines 10{--}16) is written with Racket{'}s
expressive pattern{-}matching construct.

An interactive{-}syntax extension empowers the developers to
express the algorithm directly as a diagram, which
guarantees that the diagram and the code are always in sync.
The key point is that interactive syntax can show up in
the \emph{pattern} part of a \RktSym{match} expression as
well as in the \emph{template} part, both situated within
ordinary program text.

A look at figure~\hyperref[t:x28counter_x28x22figurex22_x22rbx2dbalancex22x29x29]{\FigureRef{7}{t:x28counter_x28x22figurex22_x22rbx2dbalancex22x29x29}} makes this point for the
red{-}black tree balance algorithm. The \RktSym{match} expression consists
of two clauses, but only the first one matters for the current discussion.

\begin{Figure}\begin{Centerfigure}\begin{FigureInside}\begin{SCodeFlow}\begin{RktBlk}\begin{SingleColumn}\RktPn{(}\RktSym{define}\mbox{\hphantom{\Scribtexttt{x}}}\RktPn{(}\RktSym{balance}\mbox{\hphantom{\Scribtexttt{x}}}\RktSym{t}\RktPn{)}

\mbox{\hphantom{\Scribtexttt{xx}}}\RktPn{(}\RktSym{match}\mbox{\hphantom{\Scribtexttt{x}}}\RktSym{t}

\mbox{\hphantom{\Scribtexttt{xxxx}}}\RktPn{[}\RktPn{(}\RktSym{or}

\mbox{\hphantom{\Scribtexttt{xxxxxx}}}\raisebox{-0.23999999999999488bp}{\makebox[85.12bp][l]{\includegraphics[trim=2.4000000000000004 2.4000000000000004 2.4000000000000004 2.4000000000000004]{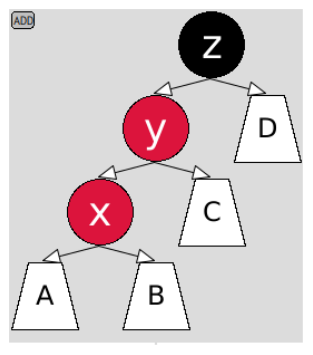}}}\mbox{\hphantom{\Scribtexttt{xx}}}\raisebox{-0.23999999999999488bp}{\makebox[69.11999999999999bp][l]{\includegraphics[trim=2.4000000000000004 2.4000000000000004 2.4000000000000004 2.4000000000000004]{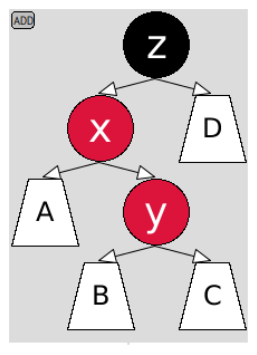}}}\mbox{\hphantom{\Scribtexttt{xx}}}\raisebox{-0.23999999999999488bp}{\makebox[69.11999999999999bp][l]{\includegraphics[trim=2.4000000000000004 2.4000000000000004 2.4000000000000004 2.4000000000000004]{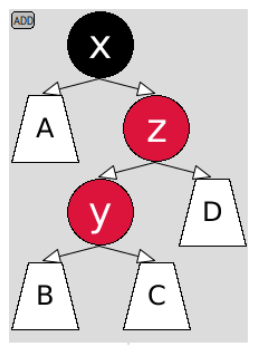}}}\mbox{\hphantom{\Scribtexttt{xx}}}\raisebox{-0.23999999999999488bp}{\makebox[85.12bp][l]{\includegraphics[trim=2.4000000000000004 2.4000000000000004 2.4000000000000004 2.4000000000000004]{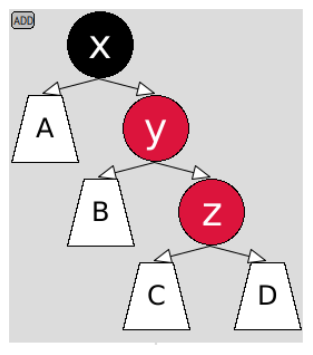}}}\RktPn{)}

\mbox{\hphantom{\Scribtexttt{xxxxx}}}\RktCmt{;}\RktCmt{~}\RktCmt{={\Stttextmore}}

\mbox{\hphantom{\Scribtexttt{xxxxx}}}\raisebox{-0.23999999999999488bp}{\makebox[85.12bp][l]{\includegraphics[trim=2.4000000000000004 2.4000000000000004 2.4000000000000004 2.4000000000000004]{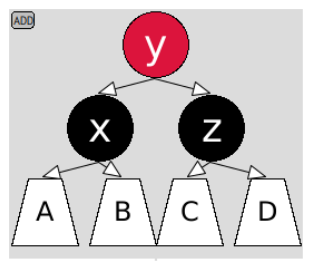}}}\RktPn{]}

\mbox{\hphantom{\Scribtexttt{xxxx}}}\RktPn{[}\RktSym{else}\mbox{\hphantom{\Scribtexttt{x}}}\RktSym{t}\RktPn{]}\RktPn{)}\RktPn{)}\end{SingleColumn}\end{RktBlk}\end{SCodeFlow}\end{FigureInside}\end{Centerfigure}

\Centertext{\Legend{\FigureTarget{\label{t:x28counter_x28x22figurex22_x22rbx2dbalancex22x29x29}\textsf{Fig.}~\textsf{7}. }{t:x28counter_x28x22figurex22_x22rbx2dbalancex22x29x29}\textsf{Visual Red{-}Black Tree Balance}}}\end{Figure}

The \RktSym{or} pattern combines several sub{-}patterns; if any one of them
matches, the pattern matches the given tree \RktSym{t}. This example uses
four sub{-}patterns, each expressed with editors; each
represents one of the four possible situations in which a balance must
take place. The situation should remind the reader of the diagram in
Okasaki{'}s functional implementation\Autobibref{~[\hyperref[t:x28autobib_x22Chris_OkasakiRedx2dblack_Trees_in_a_Functional_SettingJournal_of_Functional_Programming_9x284x29x2c_ppx2e_471x2dx2d4771999httpsx3ax2fx2fdoix2eorgx2f10x2e1017x2fS0956796899003494x22x29]{\AutobibLink{Okasaki}} \hyperref[t:x28autobib_x22Chris_OkasakiRedx2dblack_Trees_in_a_Functional_SettingJournal_of_Functional_Programming_9x284x29x2c_ppx2e_471x2dx2d4771999httpsx3ax2fx2fdoix2eorgx2f10x2e1017x2fS0956796899003494x22x29]{\AutobibLink{\Thyperref{1999}{autobiblab:38}}}]}, which uses the same
four trees on the second page of his paper.
The four sub{-}patterns name nodes{---}\RktSym{x}, \RktSym{y}, and
\RktSym{z}{---}and subtrees{---}\RktSym{A}, \RktSym{B}, \RktSym{C}, and
\RktSym{D}{---}with consistent sets of pattern variables.

The template{---}on the second line{---}refers to these pieces of the
pattern. It is also an editor and shows how the nodes and
sub{-}trees are put into a different position. The resulting tree is
clearly balanced relative to the matched subtrees.

In sum, the code consists of four input patterns that map to the
same output pattern. Any programmer who opens this file in the future will
immediately understand the relationship between the input tree shapes and
the output tree{---}plus the connection to the published paper.

\Ssubsection{In Depth: Form Builders}{In Depth: Form Builders}\label{t:x28part_x22secx3aexamplex2dmetax22x29}

The presented examples
generate only run{-}time code, and they exclusively focus
on patterns and data structures.
Interactive{-}syntax extensions, however, can also generate
compile{-}time and even edit{-}time code. This means that
interactive{-}syntax extensions can impose both domain{-}specific
validation requirements and provide the tools to enforce them.

As an example, consider embedding table{-}like forms into code. To make
this idea truly concrete, imagine  managing a large introductory
course, with several hundred students and a staff of a few dozen teaching
assistants. The course coordinator must log
information for each student and staff member. Furthermore,
each role requires different information, e.g., each
student gets a grade, while staff members have grading assignments. To manage
all this information, the course coordinator can use interactive
syntax to create information forms. Rather than making forms
directly, the coordinator creates a form builder with
interactive syntax to generate each type of form. More generally,
in this scenario interactive syntax is used to make new
types of interactive{-}syntax.

Normally, forms act like the Tsuro editor and
elaborate to tables containing their contents. These tables
can act as dictionaries used directly in source programs, or
as SQL statements that insert data into a form{-}specific database.

Programmers can add additional constraints to fields in those
forms. Forms that do not meet these constraints are
considered syntax errors. For example, here is a small
program with an incorrectly filled grade form:
\identity{~\\\begin{minipage}[c]{0.9\textwidth}}

\noindent \begin{Subflow}\Scribtexttt{{\Stttextmore} }\RktPn{(}\RktSym{dict{-}ref}\mbox{\hphantom{\Scribtexttt{x}}}\raisebox{-0.0bp}{\makebox[158.4bp][l]{\includegraphics[trim=2.4000000000000004 2.4000000000000004 2.4000000000000004 2.4000000000000004]{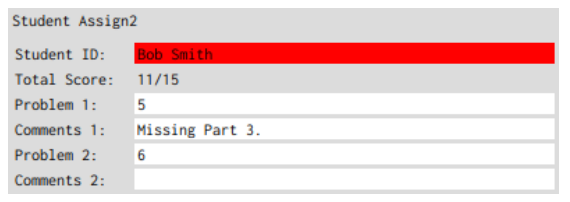}}}\mbox{\hphantom{\Scribtexttt{x}}}\RktVal{{\textquotesingle}}\RktVal{score}\RktPn{)}

\noindent \begin{Subflow}\RktErr{Bad syntax in student{-}form\$: Invalid "Student ID", expected an: id?, got: "Bob Smith"}\end{Subflow}\end{Subflow}

\noindent \identity{\end{minipage}\\[.2cm]}
The {``}Student ID{''} field expects an identification number, but the user put
in a name rather than an ID. Because forms may be created with checks
that ensure correct data is entered, feedback happens during edit time.
As an aside, note the {``}Total Score{''} pseudo{-}field in this form, which displays a
result, but is not a text field. In addition to traditional (text)
fields, forms can contain any arbitrary interactive{-}syntax
extension. For example, the {``}Total Score{''} area sums the result of
\Scribtexttt{"Problem 1"} and \Scribtexttt{"Problem 2"}.

\identity{\begin{wrapfigure}{l}{1.65in}\vspace{-0.4cm}}\raisebox{-0.3999999999999915bp}{\makebox[116.0bp][l]{\includegraphics[trim=2.4000000000000004 2.4000000000000004 2.4000000000000004 2.4000000000000004]{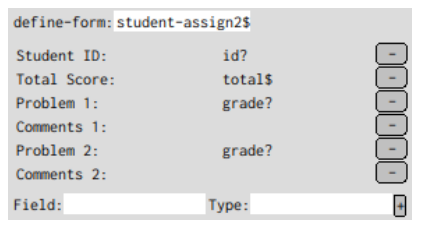}}}\identity{\vspace{-0.3cm} \end{wrapfigure}}

Rather than creating form editors textually, the coordinator can
use a graphical form builder from a library of interactive syntax constructs to create new types of
forms. The editor on the left is an example of this
meta{-}form. Using an IDE action, the coordinator instantiated
this form editor and used the text field labeled
\RktSym{define{-}form} to give a name to all its instances. As
displayed, this meta{-}editor already specifies a list of
fields: \Scribtexttt{"Student ID"}, \Scribtexttt{"Total Score"}, \Scribtexttt{"Problem
1"}, \Scribtexttt{"Problem 1 Comments"}, \Scribtexttt{"Problem 2"}, and \Scribtexttt{"Problem 2 Comments"}. Each field comes with a \BeginAccSupp{method=plain,ActualText={-},space}\raisebox{-0.03999999999999915bp}{\makebox[11.52bp][l]{\includegraphics[trim=2.4000000000000004 2.4000000000000004 2.4000000000000004 2.4000000000000004]{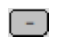}}}\EndAccSupp{}
button, with which the coordinator can remove the field from
the form. Additionally, each field contains a reference to
either an editor or a predicate. If an editor is provided,
say \RktSym{total\$}, than the form uses it instead of the
default text field. This enables custom sub{-}editors to
display information based on other fields. Likewise, if a
predicate is provided, it handles the validation of entries
in the corresponding text field.

At the bottom is a text field plus a \BeginAccSupp{method=plain,ActualText={+},space}\raisebox{-0.03999999999999915bp}{\makebox[4.800000000000001bp][l]{\includegraphics[trim=2.4000000000000004 2.4000000000000004 2.4000000000000004 2.4000000000000004]{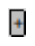}}}\EndAccSupp{} button; it permits
the coordinator to add a new field to
\RktSym{student{-}assign2\$}. Indeed, once a field is deleted
or added, every instance of this \RktSym{student{-}form\%} is
updated to match its meta{-}definition. For example, if the
coordinator were to add a \Scribtexttt{"Problem 3"} label, a
correspondingly labeled, blank text field would show up in
both of the editors in the list on the next line.

In addition to defining new form types in the program, the
\RktSym{define{-}interactive{-}syntax} meta{-}editor introduces a
companion SQL table to the program. The table{'}s schema is
specific to the introduced form type, which allows users to
add form instances into the table. The
\RktSym{student{-}assign2\$} table, for example, expects five
fields: \Scribtexttt{"Student ID"}, \Scribtexttt{"Problem 1"}, \Scribtexttt{"Problem 1
Comments"}, \Scribtexttt{"Problem 2"}, and \Scribtexttt{"Problem 2 Comments"}.
The \Scribtexttt{"Total Score"} field is absent because the
\RktSym{total\$} editor does not provide an SQL schema.

Figure~\hyperref[t:x28counter_x28x22figurex22_x22figx3aformx2ddefinitionx22x29x29]{\FigureRef{8}{t:x28counter_x28x22figurex22_x22figx3aformx2ddefinitionx22x29x29}} presents the
implementation of the form builder syntax. This
interactive{-}syntax extension is like a graphical GUI editor
in that it generates another interactive{-}syntax extension.
The edit{-}time code, such as the state and view, is similar
to the Tsuro editor and hence of little interest. Its
elaborator contains compile{-}time code that generates more
code that runs at compile time, run time, and edit time.
First, at compile time it
generates identifiers used throughout the elaborator (lines
4{-}9, compile time). Second, the elaborator defines an SQL
schema for form instances at run time (lines 11{-}12, run
time). Finally it synthesizes the edit{-}time code for the
form itself (lines 13{-}17, edit time).

\begin{Figure}\begin{Centerfigure}\begin{FigureInside}\identity{\tikzmark{meta}}\identity{\tikz[remember picture,overlay,baseline=0pt]{
\fill[pink] ([shift={(-19.5em,0)}]pic cs:meta) rectangle ([shift={(19.5em,-20.5em)}]pic cs:meta);
\fill[lime] ([shift={(-19.5em,-2.4em)}]pic cs:meta) rectangle ([shift={(19.5em,-20.5em)}]pic cs:meta);
\fill[gold]    ([shift={(-19.5em,-10.7em)}]pic cs:meta) rectangle ([shift={(19.5em,-20.5em)}]pic cs:meta);
\fill[pink] ([shift={(-19.5em,-14.3em)}]pic cs:meta) rectangle ([shift={(19.5em,-20.5em)}]pic cs:meta);
}}

\noindent \begin{SInsetFlow}\begin{bigtabular}{@{\bigtableleftpad}l@{}l@{}l@{}l@{}l@{}l@{}l@{}}
\hbox{ } &
\hbox{ } &
\hbox{ } &
\begin{RktBlk}\begin{tabular}[c]{@{}l@{}}
\hbox{\mbox{\hphantom{\Scribtexttt{x}}}\Scribtexttt{1}} \\
\hbox{\mbox{\hphantom{\Scribtexttt{x}}}\Scribtexttt{2}} \\
\hbox{\mbox{\hphantom{\Scribtexttt{x}}}\Scribtexttt{3}} \\
\hbox{\mbox{\hphantom{\Scribtexttt{x}}}\Scribtexttt{4}} \\
\hbox{\mbox{\hphantom{\Scribtexttt{x}}}\Scribtexttt{5}} \\
\hbox{\mbox{\hphantom{\Scribtexttt{x}}}\Scribtexttt{6}} \\
\hbox{\mbox{\hphantom{\Scribtexttt{x}}}\Scribtexttt{7}} \\
\hbox{\mbox{\hphantom{\Scribtexttt{x}}}\Scribtexttt{8}} \\
\hbox{\mbox{\hphantom{\Scribtexttt{x}}}\Scribtexttt{9}} \\
\hbox{\Scribtexttt{10}} \\
\hbox{\Scribtexttt{11}} \\
\hbox{\Scribtexttt{12}} \\
\hbox{\Scribtexttt{13}} \\
\hbox{\Scribtexttt{14}} \\
\hbox{\Scribtexttt{15}} \\
\hbox{\Scribtexttt{16}} \\
\hbox{\Scribtexttt{17}}\end{tabular}\end{RktBlk} &
\hbox{ } &
\hbox{ } &
\begin{RktBlk}\begin{tabular}[c]{@{}l@{}}
\hbox{\RktPn{(}\RktSym{define{-}interactive{-}syntax}\mbox{\hphantom{\Scribtexttt{x}}}\RktSym{form{-}builder\$}\mbox{\hphantom{\Scribtexttt{x}}}\RktSym{vertical{-}block\$}} \\
\hbox{\mbox{\hphantom{\Scribtexttt{xx}}}\RktSym{\mbox{{-}{-}{-}e}lided{-}{-}{-}}} \\
\hbox{\mbox{\hphantom{\Scribtexttt{xx}}}\RktPn{(}\RktSym{define{-}elaborate}\mbox{\hphantom{\Scribtexttt{x}}}\RktSym{this}} \\
\hbox{\mbox{\hphantom{\Scribtexttt{xxxx}}}\RktPn{\#{\hbox{\texttt{:}}}with}\mbox{\hphantom{\Scribtexttt{x}}}\RktSym{name/sql}\mbox{\hphantom{\Scribtexttt{xxxxxxxxxxxxxx}}}\RktPn{(}\RktSym{retrieve{-}sql{-}name}\mbox{\hphantom{\Scribtexttt{xxxxxxxxxxxxxx}}}\RktSym{this}\RktPn{)}} \\
\hbox{\mbox{\hphantom{\Scribtexttt{xxxx}}}\RktPn{\#{\hbox{\texttt{:}}}with}\mbox{\hphantom{\Scribtexttt{x}}}\RktSym{fields/sql}\mbox{\hphantom{\Scribtexttt{xxxxxxxxxxxx}}}\RktPn{(}\RktSym{retrieve{-}sql{-}field}\mbox{\hphantom{\Scribtexttt{xxxxxxxxxxxxx}}}\RktSym{this}\RktPn{)}} \\
\hbox{\mbox{\hphantom{\Scribtexttt{xxxx}}}\RktPn{\#{\hbox{\texttt{:}}}with}\mbox{\hphantom{\Scribtexttt{x}}}\RktSym{name\$}\mbox{\hphantom{\Scribtexttt{xxxxxxxxxxxxxxxxx}}}\RktPn{(}\RktSym{retrieve{-}name}\mbox{\hphantom{\Scribtexttt{xxxxxxxxxxxxxxxxxx}}}\RktSym{this}\RktPn{)}} \\
\hbox{\mbox{\hphantom{\Scribtexttt{xxxx}}}\RktPn{\#{\hbox{\texttt{:}}}with}\mbox{\hphantom{\Scribtexttt{x}}}\RktPn{(}\RktSym{field{-}name}\mbox{\hphantom{\Scribtexttt{x}}}\RktSym{{\hbox{\texttt{.}}}{\hbox{\texttt{.}}}{\hbox{\texttt{.}}}}\RktPn{)}\mbox{\hphantom{\Scribtexttt{xxxxxx}}}\RktPn{(}\RktSym{retrieve{-}field{-}names}\mbox{\hphantom{\Scribtexttt{xxxxxxxxxxx}}}\RktSym{this}\RktPn{)}} \\
\hbox{\mbox{\hphantom{\Scribtexttt{xxxx}}}\RktPn{\#{\hbox{\texttt{:}}}with}\mbox{\hphantom{\Scribtexttt{x}}}\RktPn{(}\RktSym{field{-}validator}\mbox{\hphantom{\Scribtexttt{x}}}\RktSym{{\hbox{\texttt{.}}}{\hbox{\texttt{.}}}{\hbox{\texttt{.}}}}\RktPn{)}\mbox{\hphantom{\Scribtexttt{x}}}\RktPn{(}\RktSym{retrieve{-}field{-}validation{-}code}\mbox{\hphantom{\Scribtexttt{x}}}\RktSym{this}\RktPn{)}} \\
\hbox{\mbox{\hphantom{\Scribtexttt{xxxx}}}\RktSym{\mbox{{-}{-}{-}e}lided{-}{-}{-}}} \\
\hbox{\mbox{\hphantom{\Scribtexttt{xxxx}}}\RktRdr{\#{\textasciigrave}}\RktPn{(}\RktSym{begin}} \\
\hbox{\mbox{\hphantom{\Scribtexttt{xxxxxxxx}}}\RktPn{(}\RktSym{query{-}exec}\mbox{\hphantom{\Scribtexttt{x}}}\RktSym{db}\mbox{\hphantom{\Scribtexttt{x}}}\RktPn{(}\RktSym{create{-}table}\mbox{\hphantom{\Scribtexttt{x}}}\RktSym{name/sql}} \\
\hbox{\mbox{\hphantom{\Scribtexttt{xxxxxxxxxxxxxxxxxxxxxxxxxxxxxxxxxxxxx}}}\RktPn{\#{\hbox{\texttt{:}}}columns}\mbox{\hphantom{\Scribtexttt{x}}}\RktPn{[}\RktSym{fields/sql}\mbox{\hphantom{\Scribtexttt{x}}}\RktSym{text}\RktPn{]}\mbox{\hphantom{\Scribtexttt{x}}}\RktSym{{\hbox{\texttt{.}}}{\hbox{\texttt{.}}}{\hbox{\texttt{.}}}}\RktPn{)}\RktPn{)}} \\
\hbox{\mbox{\hphantom{\Scribtexttt{xxxxxxxx}}}\RktPn{(}\RktSym{define{-}interactive{-}syntax}\mbox{\hphantom{\Scribtexttt{x}}}\RktSym{name\$}\mbox{\hphantom{\Scribtexttt{x}}}\RktSym{table{-}base\$}} \\
\hbox{\mbox{\hphantom{\Scribtexttt{xxxxxxxxxx}}}\RktPn{(}\RktSym{new}\mbox{\hphantom{\Scribtexttt{x}}}\RktSym{field\$}\mbox{\hphantom{\Scribtexttt{x}}}\RktSym{\mbox{{-}{-}{-}e}lided{-}{-}{-}}} \\
\hbox{\mbox{\hphantom{\Scribtexttt{xxxxxxxxxxxxxxx}}}\RktPn{[}\RktSym{name}\mbox{\hphantom{\Scribtexttt{x}}}\RktSym{field{-}name}\RktPn{]}} \\
\hbox{\mbox{\hphantom{\Scribtexttt{xxxxxxxxxxxxxxx}}}\RktPn{[}\RktSym{on{-}change}\mbox{\hphantom{\Scribtexttt{x}}}\RktSym{field{-}validator}\RktPn{]}\RktPn{)}\mbox{\hphantom{\Scribtexttt{x}}}\RktSym{{\hbox{\texttt{.}}}{\hbox{\texttt{.}}}{\hbox{\texttt{.}}}}} \\
\hbox{\mbox{\hphantom{\Scribtexttt{xxxxxxxxxx}}}\RktSym{\mbox{{-}{-}{-}e}lided{-}{-}{-}}\RktPn{)}\RktPn{)}\RktPn{)}\RktPn{)}}\end{tabular}\end{RktBlk}\end{bigtabular}\end{SInsetFlow}

\noindent \begin{SCentered}\identity{~\\\begin{minipage}[c]{0.33\textwidth}}\raisebox{-0.1999999999999993bp}{\makebox[12.0bp][l]{\includegraphics[trim=2.4000000000000004 2.4000000000000004 2.4000000000000004 2.4000000000000004]{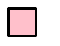}}}Edit Time\identity{\end{minipage}\begin{minipage}[c]{0.33\textwidth}}\raisebox{-0.1999999999999993bp}{\makebox[12.0bp][l]{\includegraphics[trim=2.4000000000000004 2.4000000000000004 2.4000000000000004 2.4000000000000004]{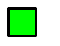}}}Compile Time\identity{\end{minipage}\begin{minipage}[c]{0.3\textwidth}}\raisebox{-0.1999999999999993bp}{\makebox[12.0bp][l]{\includegraphics[trim=2.4000000000000004 2.4000000000000004 2.4000000000000004 2.4000000000000004]{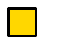}}}Run Time\identity{\end{minipage}}\end{SCentered}\end{FigureInside}\end{Centerfigure}

\Centertext{\Legend{\FigureTarget{\label{t:x28counter_x28x22figurex22_x22figx3aformx2ddefinitionx22x29x29}\textsf{Fig.}~\textsf{8}. }{t:x28counter_x28x22figurex22_x22figx3aformx2ddefinitionx22x29x29}\textsf{Example Editor for a Form Builder}}}\end{Figure}

The generated interactive{-}syntax extension looks deceivingly simple:

\noindent \begin{itemize}\atItemizeStart

\item The new interactive{-}syntax class, named \RktSym{name} as specified in
the state of \RktSym{form{-}builder\$}, inherits from the \RktSym{table{-}base\$}
superclass (lines 13). This use of inheritance illustrates a common use of
generated interactive{-}syntax definitions. Rather than putting the entire
implementation for an editor in the template of the elaborator, we factor
most of it out into an external class. Besides separating common code from
elaborator{-}specific one, it also generates significantly smaller code than
a full{-}fledged implementation class, which significantly improves
compile{-}time and edit{-}time performance. The \RktSym{table{-}base\$} class itself
is rather pedestrian GUI code.

\item The elaborator of the generated interactive{-}syntax simply
expands to a dictionary with the form{'}s field names and
values.\end{itemize}

\noindent While generating interactive{-}syntax extensions from
interactive{-}syntax extensions sounds complicated, the
decades{-}old precedent of  syntax
extension design should help the reader understand the incredible
power in such meta{-}capabilities.

In short, interactive visual syntax can be used to introduce
new interactive visual syntax. Here a general
form{-}generating interactive syntax construct is instantiated
for creating assignment specific forms. This meta{-}editor
adds another editor type to the program. At some point, of
course, the process has to switch back to text, which is
where the form extension itself is defined.

\sectionNewpage

\Ssection{The Implementation of an Interactive{-}Syntax Extension Mechanism}{The Implementation of an Interactive{-}Syntax Extension Mechanism}\label{t:x28part_x22secx3aimplementationx22x29}

The implementation of an interactive{-}syntax extension
mechanism poses four challenges: editor syntax, edit{-}time
semantics, editor elaboration, and IDE integration. This
section first describes a general implementation approach
that realizes our design desiderata and suggests abstract
solutions to these four challenges. It then explains how the
application of this approach to Racket yields a reasonably
robust prototype\NoteBox{\NoteContent{https://github.com/videolang/interactive{-}syntax}} including a
connection to the DrRacket IDE. We conjecture that, with
significantly more labor, this approach would also work for
other languages with macro systems such as Clojure, Elixir,
Julia, Rust, Scala, and even C++, as well as alternative
IDEs.

\Ssubsection{Ingredients for an Interactive{-}Syntax System}{Ingredients for an Interactive{-}Syntax System}\label{t:x28part_x22secx3apartsx22x29}

As mentioned above, four efforts are required to implement support for
interactive{-}syntax extensions to a programming language.
First, the language{'}s syntax must be augmented with a
textual \emph{editor} form to house these extensions.
Second, the language{'}s semantics must be extended to support
the execution of code at edit time. Third, the language must
include some mechanism for interactive{-}syntax elaboration.
Finally, it demands the construction of plugins for
graphical IDEs so that they use these extended semantics to
interpret the textual \emph{editor} form as a visual and
interactive graphical user interface. Any language tool
chain that can accommodate these four components can also
support interactive{-}syntax extensions.

\Ssubsubsectionstarx{The Editor Form}{The Editor Form}\label{t:x28part_x22edit1x22x29}

For a language to support interactive{-}syntax extensions,
implementers of that language must extend its syntax to
include a textual representation for editors. This textual
representation is how an editor is stored on the developer{'}s
file system. Additionally, using a purely textual representation allows
developers to edit their code in any environment, such as
plain text editors, not just interactive{-}syntax specific
ones.

This textual extension must contain three
parts: an editor{'}s state, a pointer to a means of converting that state
into a graphical editor, and a pointer to a means of converting that
state into elaborated code.
It can be added either to the language{'}s
core implementation or as an external language extension.
Many common languages, such as C, already allow for small
language extensions like this. Languages that don{'}t, such as
Java, may still have tools that can emulate language extensions to a
limited degree.

As an example, a form builder from the previous
section may have this textual representation:

\begin{SCodeFlow}\begin{RktBlk}\begin{SingleColumn}\RktMeta{}\RktSym{\$editor}\RktMeta{}\mbox{\hphantom{\Scribtexttt{x}}}\RktMeta{}\RktPn{{\char`\{}}\RktMeta{}

\RktMeta{}\mbox{\hphantom{\Scribtexttt{xx}}}\RktMeta{}\RktSym{binding{\hbox{\texttt{:}}}}\RktMeta{}\mbox{\hphantom{\Scribtexttt{x}}}\RktMeta{}\RktPn{[}\RktVal{"lib/form{-}builder{\hbox{\texttt{.}}}rkt"}\RktSym{,}\RktMeta{}\mbox{\hphantom{\Scribtexttt{x}}}\RktMeta{}\RktVal{"form{-}builder\$"}\RktPn{]}\RktSym{,}\RktMeta{}

\RktMeta{}\mbox{\hphantom{\Scribtexttt{xx}}}\RktMeta{}\RktSym{state{\hbox{\texttt{:}}}}\RktMeta{}\mbox{\hphantom{\Scribtexttt{x}}}\RktMeta{}\RktPn{{\char`\{}}\RktMeta{}

\RktMeta{}\mbox{\hphantom{\Scribtexttt{xxxx}}}\RktMeta{}\RktSym{name{\hbox{\texttt{:}}}}\RktMeta{}\mbox{\hphantom{\Scribtexttt{x}}}\RktMeta{}\RktVal{"person\$"}\RktSym{,}\RktMeta{}

\RktMeta{}\mbox{\hphantom{\Scribtexttt{xxxx}}}\RktMeta{}\RktSym{keys{\hbox{\texttt{:}}}}\RktMeta{}\mbox{\hphantom{\Scribtexttt{x}}}\RktMeta{}\RktPn{[}\RktVal{"Name"}\RktMeta{}\mbox{\hphantom{\Scribtexttt{x}}}\RktMeta{}\RktSym{,}\RktVal{"Age"}\RktPn{]}\RktMeta{}

\RktMeta{}\mbox{\hphantom{\Scribtexttt{xx}}}\RktMeta{}\RktPn{{\char`\}}}\RktMeta{}

\RktMeta{}\RktPn{{\char`\}}}\RktMeta{}\end{SingleColumn}\end{RktBlk}\end{SCodeFlow}

\noindent Here, the \Scribtexttt{binding} tag refers to the module and name of
an interactive{-}syntax binding, which serves the dual purpose of
converting state into a graphical editor and into elaborated
code. The rest of the syntax contains the editor{'}s state as a hash table.
Due to this design choice, a plain text editor, such as Emacs or Vim, simply
displays this text when a developer opens a module that contains interactive
syntax. IDEs with support for interactive syntax can display the editors as mini
GUIs embedded in program text, using the information stored in the \Scribtexttt{binding} field.

The interpretation of editor syntax is analogous to
closures. Like a closure, an editor combines a code pointer
and its current state into a new kind of value. The code
pointer, found in the \RktSym{binding} keyword in the
example above, refers to the
\RktSym{define{-}interactive{-}syntax} definition that the
editor instantiates. The state component records those
aspects of the editor{'}s state that this definition specifies
as persistent. Together these two pieces suffice to fully
re{-}instantiate the editor as a graphical element in IDEs
that can interpret these {``}closures{''} at edit time.

\Ssubsubsectionstarx{Edit{-}time semantics}{Edit{-}time semantics}\label{t:x28part_x22sem1x22x29}

The second component required for interactive{-}syntax
extensions is semantics for edit{-}time code. This semantics
is distinct from already existing semantics for compile{-}time
and run{-}time code. Furthermore, it must be possible
to interleave edit{-}time code with other forms of code in a program{'}s
body and to formulate such edit{-}time code in the same language
as run{-}time code or
compile{-}time code. Finally, while not strictly necessary,
the language should enforce a level of isolation on effects
among these phases to allow a clean separate compilation of
modules in the presence of co{-}mingled
elements\Autobibref{~[\hyperref[t:x28autobib_x22Matthew_FlattComposable_and_Compilable_Macrosx2c_You_Want_It_Whenx3fIn_Procx2e_International_Conference_on_Functional_Programmingx2c_ppx2e_72x2dx2d832002x22x29]{\AutobibLink{Flatt}} \hyperref[t:x28autobib_x22Matthew_FlattComposable_and_Compilable_Macrosx2c_You_Want_It_Whenx3fIn_Procx2e_International_Conference_on_Functional_Programmingx2c_ppx2e_72x2dx2d832002x22x29]{\AutobibLink{\Thyperref{2002}{autobiblab:21}}}]}.

Two syntactic forms must link run{-}time code and edit{-}time
code. The first form simply allows edit{-}time code to be
spliced into run time code. The second form must create
bindings in run{-}time code, that refer to extensions defined
at edit{-}time.

As with the syntax extensions, this semantics can be either
built into a language{'}s core, or it can be added with an
external preprocessor. If done externally, only one
preprocessor is needed to support all interactive{-}syntax extensions.

\Ssubsubsectionstarx{Editor Elaboration}{Editor Elaboration}\label{t:x28part_x22elab1x22x29}

The language also needs an expander to turn each editor form
into its elaborated code. This expander must additionally be
able to evaluate user code including the definition of new
types of interactive{-}syntax. An expander can use its host
language{'}s normal meta{-}programming facilities. However,
these facilities must implement the one{-}and{-}a{-}half pass
approach discussed in \ChapRef{\SectionNumberLink{t:x28part_x22secx3abasicx2dsyntaxx22x29}{3}}{Constructing Interactive Syntax}.

Expanding an editor into code happens in five steps. First,
the expander must recognize editor syntax. Second, it must
deserialize an editor{'}s state. Third, the expander must
locate the editor{'}s elaborator. Fourth, the expander
invokes the elaborator with the editor{'}s state. Finally, it
splices the generated code into the program body.

\Ssubsubsectionstarx{Cooperating With an IDE}{Cooperating With an IDE}\label{t:x28part_x22ide1x22x29}

An IDE must cooperate with
the language to run the edit{-}time code of interactive{-}syntax
extensions and to connect this code to its own graphical context.
An implementation can either modify the IDE directly or via a
plugin. If an IDE supports interactive{-}syntax extensions through a
plugin, only one plugin is required to support all extensions.

Enabling an IDE to interpret editors demands two kinds of extensions.
First, the IDE must recognize interactive{-}syntax extensions so that it
supports UI actions for the insertion of instances into code and for
updating existing editors if their underlying definition
changes. Second, the IDE must supply a graphical context to editors so
that they can render themselves and receive relevant UI events.

The code that comprises an interactive{-}syntax extension is
fundamentally user code. As such, the IDE must
sandbox these extensions. This sandbox is similar to the
traditional environment that an operating system may use or a web
browser for its tabs. For example, a sandbox may prevent an
extension from creating or modifying files without some file system
permission. This sandbox environment must also allow an editor to
gracefully fail, reverting back to some default rendering or even a variant of its textual
representation.

\Ssubsection{A prototype using Racket and DrRacket}{A prototype using Racket and DrRacket}\label{t:x28part_x22secx3aracketx2dpartsx22x29}

As a proof of concept, we have constructed a prototype of interactive{-}syntax
extensions for the Racket language and DrRacket IDE\Autobibref{~[\hyperref[t:x28autobib_x22Robert_Bruce_Findlerx2c_John_Clementsx2c_Cormac_Flanaganx2c_Matthew_Flattx2c_Shriram_Krishnamurthix2c_Paul_Stecklerx2c_and_Matthias_FelleisenDrSchemex3a_A_Programming_Environment_for_SchemeJournal_of_Functional_Programming_12x282x29x2c_ppx2e_159x2dx2d1822002httpsx3ax2fx2fdoix2eorgx2f10x2e1017x2fS0956796801004208x22x29]{\AutobibLink{Findler et al\Sendabbrev{.}}} \hyperref[t:x28autobib_x22Robert_Bruce_Findlerx2c_John_Clementsx2c_Cormac_Flanaganx2c_Matthew_Flattx2c_Shriram_Krishnamurthix2c_Paul_Stecklerx2c_and_Matthias_FelleisenDrSchemex3a_A_Programming_Environment_for_SchemeJournal_of_Functional_Programming_12x282x29x2c_ppx2e_159x2dx2d1822002httpsx3ax2fx2fdoix2eorgx2f10x2e1017x2fS0956796801004208x22x29]{\AutobibLink{\Thyperref{2002}{autobiblab:19}}}]}.

\Ssubsubsectionstarx{The Editor Form}{The Editor Form}\label{t:x28part_x22edit2x22x29}

First, an editor{'}s textual representation is
composed of binding information and state syntax.
Concretely, an editor is represented in text like the form in the proceeding subsection.

Making this form valid syntax requires a change to Racket{'}s
reader. The extended reader generates a valid syntax object
with a known (macro) interpretation. Racket{'}s reader is
extensible, meaning that interactive{-}syntax can, in
principle, work with any Racket{-}based language\Autobibref{~[\hyperref[t:x28autobib_x22Matthias_Felleisenx2c_Robert_Bruce_Findlerx2c_Matthew_Flattx2c_Shriram_Krishnamurthix2c_Eli_Barzilayx2c_Jay_McCarthyx2c_and_Sam_Tobinx2dHochstadtA_Programmable_Programming_LanguageCommunications_of_the_ACM_61x283x29x2c_ppx2e_62x2dx2d712018httpsx3ax2fx2fdoix2eorgx2f10x2e1145x2f3127323x22x29]{\AutobibLink{Felleisen et al\Sendabbrev{.}}} \hyperref[t:x28autobib_x22Matthias_Felleisenx2c_Robert_Bruce_Findlerx2c_Matthew_Flattx2c_Shriram_Krishnamurthix2c_Eli_Barzilayx2c_Jay_McCarthyx2c_and_Sam_Tobinx2dHochstadtA_Programmable_Programming_LanguageCommunications_of_the_ACM_61x283x29x2c_ppx2e_62x2dx2d712018httpsx3ax2fx2fdoix2eorgx2f10x2e1145x2f3127323x22x29]{\AutobibLink{\Thyperref{2018}{autobiblab:18}}}]}
that also uses Racket{'}s reader extensions.

\Ssubsubsectionstarx{Edit{-}time semantics}{Edit{-}time semantics}\label{t:x28part_x22sem2x22x29}

Second, while Racket already supports a hierarchy of compile{-}time phases (for
syntax extensions that generate syntax extensions), it has no mechanism for
adding a new phase. Since we wish to demonstrate that interactive{-}syntax
extensions can be added to a language without changing the underlying
virtual machine or interpreter, the prototype employs a surprisingly robust
work{-}around. We conjecture that such work{-}arounds exist for other languages
too, though it is also likely that in many cases, an implementer would have
to explicitly add an edit phase.

Interactive{-}syntax extensions are implemented as syntax extensions. They
elaborate constructs such as \RktSym{define{-}interactive{-}syntax}, editors, and
\RktSym{begin{-}for{-}interactive{-}syntax} into a mix of further (plain) syntax
extensions and submodules.

Figure~\hyperref[t:x28counter_x28x22figurex22_x22figx3aeditx2delabx22x29x29]{\FigureRef{9}{t:x28counter_x28x22figurex22_x22figx3aeditx2delabx22x29x29}} shows an example of such an elaboration. The
module in the top half consists of three pieces: the definition of an
interactive{-}syntax extension, an editor of this extension (the \RktVal{\#f}
denotes {``}locally defined,{''} the \RktSym{simple\$} points to the definition;
the editor has no state), and a simplistic edit{-}time test of this definition. The
module at the bottom is (approximately) the transformation of the module at
the top into Racket code.

In the expanded program, all edit{-}time code is placed into a
single \RktModLink{\RktSym{edit}} submodule\Autobibref{~[\hyperref[t:x28autobib_x22Matthew_FlattSubmodules_in_Racketx2c_You_Want_it_Whenx2c_Againx3fIn_Procx2e_Generative_Programmingx3a_Concepts_x26_Experiencesx2c_ppx2e_13x2dx2d222013httpsx3ax2fx2fdoix2eorgx2f10x2e1145x2f2517208x2e2517211x22x29]{\AutobibLink{Flatt}} \hyperref[t:x28autobib_x22Matthew_FlattSubmodules_in_Racketx2c_You_Want_it_Whenx2c_Againx3fIn_Procx2e_Generative_Programmingx3a_Concepts_x26_Experiencesx2c_ppx2e_13x2dx2d222013httpsx3ax2fx2fdoix2eorgx2f10x2e1145x2f2517208x2e2517211x22x29]{\AutobibLink{\Thyperref{2013}{autobiblab:22}}}]}.\NoteBox{\NoteContent{Appendix A contains a brief introduction to
submodules in Racket.}} This new submodule is inserted at the
bottom of the expansion module. The definition of the
interactive{-}syntax extension is separated into two pieces
(as indicated with code highlighting and indicies): the
elaborator called \RktSym{simple\${\hbox{\texttt{:}}}elaborator}, which exists
at compile time, and the interactive{-}syntax class
called \RktSym{simple\$}, which exists at edit time. Recall
that the elaborator translates the state of an editor into
run{-}time code; the interactive{-}syntax class
inherits and implements the edit{-}time interaction
functionality for the syntax extension. As for the textual
editor form, its reference to the \RktSym{simple\$}
interactive{-}syntax extension is refined to a reference to
the elaborator; for edit time execution, the IDE plugin
performs a separate name resolution. Finally, the test code
in the \RktSym{begin{-}for{-}interactive{-}syntax} block is also
moved into the \RktSym{edit} submodule.

\begin{Figure}\begin{Centerfigure}\begin{FigureInside}\identity{~\\\begin{minipage}[c]{0.9\textwidth}}

\noindent \begin{framed}\begin{RktBlk}\begin{SingleColumn}\RktModLink{\RktMod{\#lang}}\mbox{\hphantom{\Scribtexttt{x}}}\RktModLink{\RktSym{editor{-}racket}}

\mbox{\hphantom{\Scribtexttt{x}}}

\RktPn{(}\RktSym{define{-}interactive{-}syntax}\mbox{\hphantom{\Scribtexttt{x}}}\RktSym{simple\$}\mbox{\hphantom{\Scribtexttt{x}}}\RktSym{base\$}

\mbox{\hphantom{\Scribtexttt{xx}}}\highlighted{\RktPn{(}\RktSym{super{-}new}\RktPn{)}}\textit{}\textsub{\textit{}1\textit{}}\textit{}

\mbox{\hphantom{\Scribtexttt{xx}}}\RktPn{(}\RktSym{define{-}elaborate}\mbox{\hphantom{\Scribtexttt{x}}}\RktSym{this}

\mbox{\hphantom{\Scribtexttt{xxxx}}}\highlighted{\RktRdr{\#{\textquotesingle}}\RktPn{(}\RktSym{void}\RktPn{)}}\textit{}\textsub{\textit{}2\textit{}}\textit{}\RktPn{)}\RktPn{)}

\mbox{\hphantom{\Scribtexttt{x}}}

\RktMeta{}\RktPn{\#editor(\#f}\mbox{\hphantom{\Scribtexttt{x}}}\RktPn{{\hbox{\texttt{.}}}}\mbox{\hphantom{\Scribtexttt{x}}}\RktPn{simple\$)()}\RktMeta{}

\mbox{\hphantom{\Scribtexttt{x}}}

\RktPn{(}\RktSym{begin{-}for{-}interactive{-}syntax}

\mbox{\hphantom{\Scribtexttt{xx}}}\RktPn{(}\RktSym{require}\mbox{\hphantom{\Scribtexttt{x}}}\RktSym{editor/test}\RktPn{)}

\mbox{\hphantom{\Scribtexttt{xx}}}\RktPn{(}\RktSym{test{-}window}\mbox{\hphantom{\Scribtexttt{x}}}\RktPn{(}\RktSym{new}\mbox{\hphantom{\Scribtexttt{x}}}\RktSym{simple\$}\RktPn{)}\RktPn{)}\RktPn{)}\end{SingleColumn}\end{RktBlk}\end{framed}

\noindent \identity{\end{minipage}\\[.2cm]}

\identity{~\\\begin{minipage}[c]{0.9\textwidth}}

\noindent elaborates to

\noindent \identity{\end{minipage}\\[.2cm]}

\identity{~\\\begin{minipage}[c]{0.9\textwidth}}

\noindent \begin{framed}\begin{RktBlk}\begin{SingleColumn}\RktModLink{\RktMod{\#lang}}\mbox{\hphantom{\Scribtexttt{x}}}\RktModLink{\RktSym{racket}}

\mbox{\hphantom{\Scribtexttt{x}}}

\RktPn{(}\RktSym{provide}\mbox{\hphantom{\Scribtexttt{x}}}\RktSym{simple\${\hbox{\texttt{:}}}elaborator}\RktPn{)}

\mbox{\hphantom{\Scribtexttt{x}}}

\RktPn{(}\RktSym{define{-}syntax}\mbox{\hphantom{\Scribtexttt{x}}}\RktPn{(}\RktSym{simple\${\hbox{\texttt{:}}}elaborator}\mbox{\hphantom{\Scribtexttt{x}}}\RktSym{stx}\RktPn{)}

\mbox{\hphantom{\Scribtexttt{xx}}}\RktPn{(}\RktSym{class/syntax}\mbox{\hphantom{\Scribtexttt{x}}}\RktSym{base\${\hbox{\texttt{:}}}elaborator}

\mbox{\hphantom{\Scribtexttt{xxxxxxxxxxxxxxxx}}}\highlighted{\RktRdr{\#{\textquotesingle}}\RktPn{(}\RktSym{void}\RktPn{)}}\textit{}\textsub{\textit{}2\textit{}}\textit{}\RktPn{)}\RktPn{)}

\mbox{\hphantom{\Scribtexttt{x}}}

\RktMeta{}\RktPn{\#editor(\#f}\mbox{\hphantom{\Scribtexttt{x}}}\RktPn{{\hbox{\texttt{.}}}}\mbox{\hphantom{\Scribtexttt{x}}}\RktPn{simple\${\hbox{\texttt{:}}}elaborator)()}\RktMeta{}

\mbox{\hphantom{\Scribtexttt{x}}}

\RktPn{(}\RktSym{module+}\mbox{\hphantom{\Scribtexttt{x}}}\RktSym{edit}

\mbox{\hphantom{\Scribtexttt{xx}}}\RktPn{(}\RktSym{provide}\mbox{\hphantom{\Scribtexttt{x}}}\RktSym{simple\$}\RktPn{)}

\mbox{\hphantom{\Scribtexttt{x}}}

\mbox{\hphantom{\Scribtexttt{xx}}}\RktPn{(}\RktSym{define}\mbox{\hphantom{\Scribtexttt{x}}}\RktSym{simple\$}

\mbox{\hphantom{\Scribtexttt{xxxx}}}\RktPn{(}\RktSym{class/interactive{-}syntax}\mbox{\hphantom{\Scribtexttt{x}}}\RktSym{base\$}

\mbox{\hphantom{\Scribtexttt{xxxxxxxxxxxxxxxxxxxxxxxxxxxxxx}}}\highlighted{\RktPn{(}\RktSym{super{-}new}\RktPn{)}}\textit{}\textsub{\textit{}1\textit{}}\textit{}\RktPn{)}\RktPn{)}

\mbox{\hphantom{\Scribtexttt{x}}}

\mbox{\hphantom{\Scribtexttt{xx}}}\RktPn{(}\RktSym{require}\mbox{\hphantom{\Scribtexttt{x}}}\RktSym{editor/test}\RktPn{)}

\mbox{\hphantom{\Scribtexttt{xx}}}\RktPn{(}\RktSym{test{-}window}\mbox{\hphantom{\Scribtexttt{x}}}\RktPn{(}\RktSym{new}\mbox{\hphantom{\Scribtexttt{x}}}\RktSym{simple\$}\RktPn{)}\RktPn{)}\RktPn{)}\end{SingleColumn}\end{RktBlk}\end{framed}

\noindent \identity{\end{minipage}\\[.2cm]}
\identity{\vspace{0.1em}}\end{FigureInside}\end{Centerfigure}

\Centertext{\Legend{\FigureTarget{\label{t:x28counter_x28x22figurex22_x22figx3aeditx2delabx22x29x29}\textsf{Fig.}~\textsf{9}. }{t:x28counter_x28x22figurex22_x22figx3aeditx2delabx22x29x29}\textsf{Elaboration of an Interactive{-}Syntax Extension}}}\end{Figure}

Placing the edit{-}time code into a separate submodule permits the runtime
system to distinguish between editor{-}specific code and general{-}purpose
program code. In particular, it ensures that the runtime system can execute
the editor portion of an interactive{-}syntax extension independently of its
host module. Indeed, the runtime system realizes this goal by merely requiring the
\RktSym{edit} submodule and thus obtaining the provided \RktSym{simple\$}
interactive{-}syntax class, which implements the GUI interactions.  By
contrast, the generated run{-}time code of an editor must remain subject to
the host module{'}s scope.

\Ssubsubsectionstarx{Editor Elaboration}{Editor Elaboration}\label{t:x28part_x22elab2x22x29}

Third, editor elaboration is a straightforward use of
Racket{'}s macro expander. The Racket parser converts the
\RktMeta{}\RktPn{\#editor()()}\RktMeta{} form in the source into
traditional Racket syntax containing a \RktSym{\#\%editor}
macro. The \RktSym{\#\%editor} macro
(figure~\hyperref[t:x28counter_x28x22figurex22_x22figx3aeditorx2dmacrox22x29x29]{\FigureRef{10}{t:x28counter_x28x22figurex22_x22figx3aeditorx2dmacrox22x29x29}}) finds the elaborator
(lines 3{--}5), deserializes the state (lines 6{--}7), and
expands to the elaborator with the deserialized state (line
8). From here, the macro expander places the residual code
back into the program body.

\begin{Figure}\begin{Centerfigure}\begin{FigureInside}\begin{SInsetFlow}\begin{bigtabular}{@{\bigtableleftpad}l@{}l@{}l@{}l@{}l@{}l@{}l@{}}
\hbox{ } &
\hbox{ } &
\hbox{ } &
\begin{RktBlk}\begin{tabular}[c]{@{}l@{}}
\hbox{\mbox{\hphantom{\Scribtexttt{x}}}\Scribtexttt{1}} \\
\hbox{\mbox{\hphantom{\Scribtexttt{x}}}\Scribtexttt{2}} \\
\hbox{\mbox{\hphantom{\Scribtexttt{x}}}\Scribtexttt{3}} \\
\hbox{\mbox{\hphantom{\Scribtexttt{x}}}\Scribtexttt{4}} \\
\hbox{\mbox{\hphantom{\Scribtexttt{x}}}\Scribtexttt{5}} \\
\hbox{\mbox{\hphantom{\Scribtexttt{x}}}\Scribtexttt{6}} \\
\hbox{\mbox{\hphantom{\Scribtexttt{x}}}\Scribtexttt{7}} \\
\hbox{\mbox{\hphantom{\Scribtexttt{x}}}\Scribtexttt{8}}\end{tabular}\end{RktBlk} &
\hbox{ } &
\hbox{ } &
\begin{RktBlk}\begin{tabular}[c]{@{}l@{}}
\hbox{\RktPn{(}\RktSym{define{-}syntax}\mbox{\hphantom{\Scribtexttt{x}}}\RktSym{\#\%editor}} \\
\hbox{\mbox{\hphantom{\Scribtexttt{xx}}}\RktPn{(}\RktSym{syntax{-}parser}} \\
\hbox{\mbox{\hphantom{\Scribtexttt{xxxx}}}\RktPn{[}\RktPn{(}\RktSym{{\char`\_}}\mbox{\hphantom{\Scribtexttt{x}}}\RktPn{(}\RktSym{module}\mbox{\hphantom{\Scribtexttt{x}}}\RktSym{name}\RktPn{)}\mbox{\hphantom{\Scribtexttt{x}}}\RktSym{body}\RktPn{)}} \\
\hbox{\mbox{\hphantom{\Scribtexttt{xxxxx}}}\RktPn{(}\RktSym{define/syntax{-}parse}\mbox{\hphantom{\Scribtexttt{x}}}\RktSym{elaborator}} \\
\hbox{\mbox{\hphantom{\Scribtexttt{xxxxxxx}}}\RktPn{(}\RktSym{forge{-}identifier}\mbox{\hphantom{\Scribtexttt{x}}}\RktRdr{\#{\textquotesingle}}\RktSym{module}\mbox{\hphantom{\Scribtexttt{x}}}\RktRdr{\#{\textquotesingle}}\RktSym{name}\RktPn{)}\RktPn{)}} \\
\hbox{\mbox{\hphantom{\Scribtexttt{xxxxx}}}\RktPn{(}\RktSym{define/syntax{-}parse}\mbox{\hphantom{\Scribtexttt{x}}}\RktSym{state}} \\
\hbox{\mbox{\hphantom{\Scribtexttt{xxxxxxx}}}\RktPn{(}\RktSym{deserialize{-}state}\mbox{\hphantom{\Scribtexttt{x}}}\RktRdr{\#{\textquotesingle}}\RktSym{body}\RktPn{)}\RktPn{)}} \\
\hbox{\mbox{\hphantom{\Scribtexttt{xxxxx}}}\RktRdr{\#{\textquotesingle}}\RktPn{(}\RktSym{elaborator}\mbox{\hphantom{\Scribtexttt{x}}}\RktSym{state}\RktPn{)}\RktPn{]}\RktPn{)}\RktPn{)}}\end{tabular}\end{RktBlk}\end{bigtabular}\end{SInsetFlow}\end{FigureInside}\end{Centerfigure}

\Centertext{\Legend{\FigureTarget{\label{t:x28counter_x28x22figurex22_x22figx3aeditorx2dmacrox22x29x29}\textsf{Fig.}~\textsf{10}. }{t:x28counter_x28x22figurex22_x22figx3aeditorx2dmacrox22x29x29}\textsf{Implementation of the \RktSym{\#\%editor} macro}}}\end{Figure}

\Ssubsubsectionstarx{Cooperating With an IDE}{Cooperating With an IDE}\label{t:x28part_x22ide2x22x29}

Finally, the prototype exploits DrRacket{'}s plug{-}in API for the event handling
\Autobibref{~[\hyperref[t:x28autobib_x22Robert_Bruce_Findler_and_PLTDrRacketx3a_Programming_EnvironmentPLT_Design_Incx2ex2c_PLTx2dTRx2d2010x2d22010httpsx3ax2fx2fracketx2dlangx2eorgx2ftr2x2fx22x29]{\AutobibLink{Findler and PLT}} \hyperref[t:x28autobib_x22Robert_Bruce_Findler_and_PLTDrRacketx3a_Programming_EnvironmentPLT_Design_Incx2ex2c_PLTx2dTRx2d2010x2d22010httpsx3ax2fx2fracketx2dlangx2eorgx2ftr2x2fx22x29]{\AutobibLink{\Thyperref{2010}{autobiblab:20}}}]} and its Cairo{-}based drawing API for editor rendering.  A specially designed
plug{-}in connects interactive{-}syntax extensions to the IDE. It inserts menu
entries that programmers can use to instantiate interactive{-}syntax extensions
and insert them into specific points into code (figure~\hyperref[t:x28counter_x28x22figurex22_x22figx3ainsertx22x29x29]{\FigureRef{11}{t:x28counter_x28x22figurex22_x22figx3ainsertx22x29x29}}). The plug{-}in also assists with
saving and retrieving modules that contain editors. When a developer saves a
file, the plugin serializes all instances of interactive syntax extensions
into \RktMeta{}\RktPn{\#editor()()}\RktMeta{} blocks; conversely, when a
developer opens such a module, it uses the language{'}s parser to scan the file
for \RktMeta{}\RktPn{\#editor()()}\RktMeta{} blocks and informs the IDE
about them.

\begin{Figure}\begin{Centerfigure}\begin{FigureInside}\raisebox{-0.1082568807339328bp}{\makebox[240.0bp][l]{\includegraphics[trim=2.4000000000000004 2.4000000000000004 2.4000000000000004 2.4000000000000004]{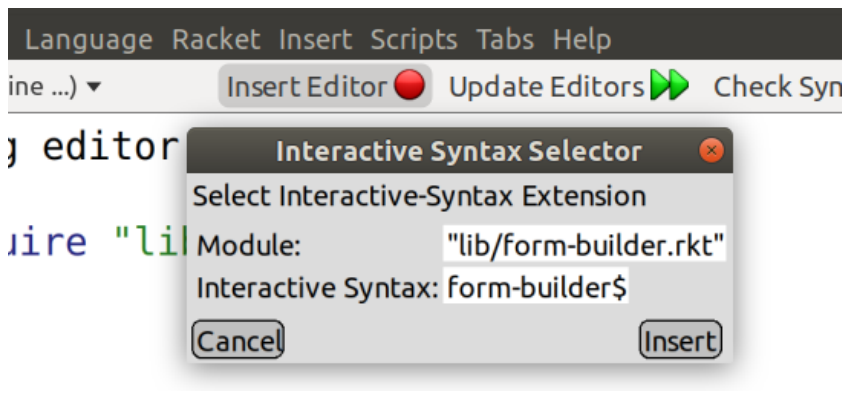}}}
\identity{\vspace{-0.3cm}}\end{FigureInside}\end{Centerfigure}

\Centertext{\Legend{\FigureTarget{\label{t:x28counter_x28x22figurex22_x22figx3ainsertx22x29x29}\textsf{Fig.}~\textsf{11}. }{t:x28counter_x28x22figurex22_x22figx3ainsertx22x29x29}\textsf{Inserting an editor inside of DrRacket}}}\end{Figure}

Technically, the prototype relies on Racket{'}s GUI toolbox
and sandboxes mechanism\Autobibref{~[\hyperref[t:x28autobib_x22Matthew_Flattx2c_Robert_Bruce_Findlerx2c_and_John_ClementsGUIx3a_Racket_Graphics_ToolkitPLT_Design_Incx2ex2c_PLTx2dTRx2d2010x2d32010httpsx3ax2fx2fracketx2dlangx2eorgx2ftr3x2fx22x29]{\AutobibLink{Flatt et al\Sendabbrev{.}}} \hyperref[t:x28autobib_x22Matthew_Flattx2c_Robert_Bruce_Findlerx2c_and_John_ClementsGUIx3a_Racket_Graphics_ToolkitPLT_Design_Incx2ex2c_PLTx2dTRx2d2010x2d32010httpsx3ax2fx2fracketx2dlangx2eorgx2ftr3x2fx22x29]{\AutobibLink{\Thyperref{2010}{autobiblab:23}}}]}. Specifically, the
Racket evaluator provides controlled channels for sandboxed
namespaces to connect to the rest of the Racket runtime
system. The Racket GUI toolbox already supports graphics
within textual programs via the snip API. DrRacket supplies
a drawing context to snips and passes user events to them.
However, snips are extra{-}linguistic. They are thus
IDE{-}specific elements and not elements of the language the
programmer edits. They were DrRacket{'}s first attempt at
mixing graphical and textual programming, but made any file
that used them unreadable outside of DrRacket. The prototype
bridges the gap between these two GUI elements so that
editors remain language elements yet connect to the IDE
smoothly.

\identity{\begin{wrapfigure}{r}{1.5in}\vspace{-0.4cm}}\raisebox{-0.7999999999999972bp}{\makebox[72.8bp][l]{\includegraphics[trim=2.4000000000000004 2.4000000000000004 2.4000000000000004 2.4000000000000004]{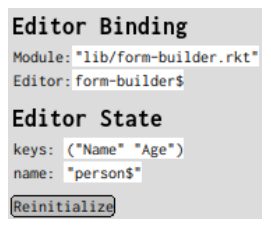}}}\identity{\vspace{-0.3cm} \end{wrapfigure}}

The prototype accommodates failures with a simple fallback editor
GUI. If a developer were to inject a typo into an
\RktMeta{}\RktPn{\#editor()()}\RktMeta{}
via Emacs or were to move the file that contains an
interactive{-}syntax extension, the prototype does not
crash. Instead it hands control to a form editor, which
displays a small default GUI whose fields show the editor{'}s
binding and state information shown on the right. Here the
form builder is found in \RktVal{"lib/form{-}builder{\hbox{\texttt{.}}}rkt"},
which provides \RktSym{form{-}builder\$} as an identifier. The
editor has two state fields: \RktSym{name} and
\RktSym{keys}. Clicking on the
\BeginAccSupp{method=plain,ActualText={reinitialize},space}\raisebox{-0.3999999999999986bp}{\makebox[30.400000000000006bp][l]{\includegraphics[trim=2.4000000000000004 2.4000000000000004 2.4000000000000004 2.4000000000000004]{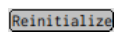}}}\EndAccSupp{}
button tells the runtime system to make another attempt at re{-}initializing the
editor with these values.

\sectionNewpage

\Ssection{Related Work}{Related Work}\label{t:x28part_x22Relatedx5fWorkx22x29}

Three years ago at ICFP in Oxford, \Autobibref{\hyperref[t:x28autobib_x22Leif_Andersenx2c_Stephen_Changx2c_and_Matthias_FelleisenSuper_8_Languages_for_Making_Movies_x28Functional_Pearlx29Proceedings_of_the_ACM_on_Programming_Languages_1x28International_Conference_on_Functional_Programmingx29x2c_ppx2e_30x2d1x2dx2d30x2d292017httpsx3ax2fx2fdoix2eorgx2f10x2e1145x2f3110274x22x29]{\AutobibLink{Andersen et al\Sendabbrev{.}}}~[\hyperref[t:x28autobib_x22Leif_Andersenx2c_Stephen_Changx2c_and_Matthias_FelleisenSuper_8_Languages_for_Making_Movies_x28Functional_Pearlx29Proceedings_of_the_ACM_on_Programming_Languages_1x28International_Conference_on_Functional_Programmingx29x2c_ppx2e_30x2d1x2dx2d30x2d292017httpsx3ax2fx2fdoix2eorgx2f10x2e1145x2f3110274x22x29]{\AutobibLink{\Thyperref{2017}{autobiblab:2}}}]} presented the Video language.
Video mixes text and editable video clips so
that users can easily script productions, both textually and interactively.
While this work on Video is a bespoke production, it raises the question of
how to generalize this idea of graphical elements embedded in scripts and
programs. Summarily speaking, this work empowers developers to

\noindent \begin{itemize}\atItemizeStart

\item create their own interactive{-}syntax extensions

\item abstract with a plain syntax extension over interactive{-}syntax extensions

\item abstract with an interactive{-}syntax extension over interactive{-}syntax extensions\end{itemize}

\noindent In other words, interactive{-}syntax extensions are a proper
part of the programming language in contrast to IDE{-}specific
plugins, such as Video{'}s.

In addition to drawing inspiration from the work of Andersen
et al., this project also draws on ideas from research on edit
time code, programming systems, and non{-}standard forms of
editing.

\Ssubsection{Edit Time}{Edit Time}\label{t:x28part_x22Editx5fTimex22x29}

Two rather distinct pieces of work combine edit{-}time computation with a
form of programming. The first is due to Erdweg in the context of the
Spoofax language workbench project and is truly about general{-}purpose
programming languages. The second is Microsoft{'}s mixing of textual and
graphical {``}programs{''} in the productivity suite.

Like Racket, Spoofax\Autobibref{~[\hyperref[t:x28autobib_x22Lennart_Cx2e_Lx2e_Kats_and_Eelco_VisserThe_Spoofax_Language_WorkbenchIn_Procx2e_Objectx2dOriented_Programmingx2c_Systemsx2c_Languages_x26_Applicationsx2c_ppx2e_444x2dx2d4632010httpsx3ax2fx2fdoix2eorgx2f10x2e1145x2f1932682x2e1869497x22x29]{\AutobibLink{Kats and Visser}} \hyperref[t:x28autobib_x22Lennart_Cx2e_Lx2e_Kats_and_Eelco_VisserThe_Spoofax_Language_WorkbenchIn_Procx2e_Objectx2dOriented_Programmingx2c_Systemsx2c_Languages_x26_Applicationsx2c_ppx2e_444x2dx2d4632010httpsx3ax2fx2fdoix2eorgx2f10x2e1145x2f1932682x2e1869497x22x29]{\AutobibLink{\Thyperref{2010}{autobiblab:33}}}]} is a framework
for developing programming languages. \Autobibref{\hyperref[t:x28autobib_x22Sebastian_Erdwegx2c_Lennart_Cx2e_Lx2e_Katsx2c_Tillmann_Rendelx2c_Christian_Kxe4stnerx2c_Klaus_Ostermannx2c_and_Eelco_VisserGrowing_a_Language_Environment_with_Editor_LibrariesIn_Procx2e_Generative_Programming_and_Component_Engineeringx2c_ppx2e_167x2dx2d1762011httpsx3ax2fx2fdoix2eorgx2f10x2e1145x2f2189751x2e2047891x22x29]{\AutobibLink{Erdweg et al\Sendabbrev{.}}}~[\hyperref[t:x28autobib_x22Sebastian_Erdwegx2c_Lennart_Cx2e_Lx2e_Katsx2c_Tillmann_Rendelx2c_Christian_Kxe4stnerx2c_Klaus_Ostermannx2c_and_Eelco_VisserGrowing_a_Language_Environment_with_Editor_LibrariesIn_Procx2e_Generative_Programming_and_Component_Engineeringx2c_ppx2e_167x2dx2d1762011httpsx3ax2fx2fdoix2eorgx2f10x2e1145x2f2189751x2e2047891x22x29]{\AutobibLink{\Thyperref{2011}{autobiblab:17}}}]} recognizes that when
developers grow programming languages, they would also like to grow the IDE
support. For example, a new language feature may require a new static analysis or
refactoring transformations, and these tools should cooperate with the language{'}s
IDE. They therefore propose a framework for creating edit{-}time libraries. In
essence such libraries would connect the language implementation with the IDE and
specifically the IDE tool suite. Like Video, these libraries are
IDE plugins and thus extra{-}linguistic.

Microsoft Office plugins, called VSTO Add{-}ins\Autobibref{~[\hyperref[t:x28autobib_x22MicrosoftOffice_and_SharePoint_Development_in_Visual_StudioRetrieved_Januaryx2c_20192019httpsx3ax2fx2fdocsx2emicrosoftx2ecomx2fenx2dusx2fvisualstudiox2fvstox2fofficex2dandx2dsharepointx2ddevelopmentx2dinx2dvisualx2dstudiox3fviewx3dvsx2d2017x22x29]{\AutobibLink{Microsoft}} \hyperref[t:x28autobib_x22MicrosoftOffice_and_SharePoint_Development_in_Visual_StudioRetrieved_Januaryx2c_20192019httpsx3ax2fx2fdocsx2emicrosoftx2ecomx2fenx2dusx2fvisualstudiox2fvstox2fofficex2dandx2dsharepointx2ddevelopmentx2dinx2dvisualx2dstudiox3fviewx3dvsx2d2017x22x29]{\AutobibLink{\Thyperref{2019}{autobiblab:37}}}]}, allow authors to
create new types of documents and embed them into other documents. A developer
might make a music type{-}setting editor, which another might use to put music
notation into a PowerPoint presentation. Even though this tool set lives in the
.NET framework, it is an extra{-}linguistic idea and does not allow developers to
build programming abstractions.

\Ssubsection{Graphical and Live Languages}{Graphical and Live Languages}\label{t:x28part_x22Graphicalx5fandx5fLivex5fLanguagesx22x29}

Several programming \emph{systems} have
enabled a mixture of some graphical and textual programming for
decades. The four most prominent examples are Boxer, Hypercard, Scratch, and
Smalltalk.

Boxer\Autobibref{~[\hyperref[t:x28autobib_x22Andrea_Ax2e_diSessa_and_Harold_AbelsonBoxerx3a_A_Reconstructible_Computational_MediumCommunications_of_the_ACM_29x289x29x2c_ppx2e_859x2dx2d8681986httpsx3ax2fx2fdoix2eorgx2f10x2e1145x2f6592x2e6595x22x29]{\AutobibLink{diSessa and Abelson}} \hyperref[t:x28autobib_x22Andrea_Ax2e_diSessa_and_Harold_AbelsonBoxerx3a_A_Reconstructible_Computational_MediumCommunications_of_the_ACM_29x289x29x2c_ppx2e_859x2dx2d8681986httpsx3ax2fx2fdoix2eorgx2f10x2e1145x2f6592x2e6595x22x29]{\AutobibLink{\Thyperref{1986}{autobiblab:12}}}]} allows developers to embed GUI elements
within other GUI elements ({``}boxing{''}), to name such GUI
elements, and to refer to these names in program code. That
is, {``}programs{''} consist of graphical renderings of GUI
objects and program text (inside the boxes). For example, a
Boxer programmer could create a box that contains an image
of a Tsuro tile, name it, and refer to this name in a unit
test in a surrounding box. Boxer does \emph{not} satisfy any
of the other desiderata in \ChapRef{\SectionNumberLink{t:x28part_x22secx3adesignx22x29}{2}}{The Design Space}. In
particular, it has poor support for creating new
abstractions with regard to the GUI elements.

Scratch\Autobibref{~[\hyperref[t:x28autobib_x22Mitchel_Resnickx2c_John_Maloneyx2c_Andrxe9s_Monroyx2dHernxe1ndezx2c_Natalie_Ruskx2c_Evelyn_Eastmondx2c_Karen_Brennanx2c_Amon_Millnerx2c_Eric_Rosenbaumx2c_Jay_Silverx2c_Brian_Silvermanx2c_and_Yasmin_KafaiScratchx3a_Programming_for_AllCommunications_of_the_ACM_52x2811x29x2c_ppx2e_60x2dx2d672009httpsx3ax2fx2fdoix2eorgx2f10x2e1145x2f1592761x2e1592779x22x29]{\AutobibLink{Resnick et al\Sendabbrev{.}}} \hyperref[t:x28autobib_x22Mitchel_Resnickx2c_John_Maloneyx2c_Andrxe9s_Monroyx2dHernxe1ndezx2c_Natalie_Ruskx2c_Evelyn_Eastmondx2c_Karen_Brennanx2c_Amon_Millnerx2c_Eric_Rosenbaumx2c_Jay_Silverx2c_Brian_Silvermanx2c_and_Yasmin_KafaiScratchx3a_Programming_for_AllCommunications_of_the_ACM_52x2811x29x2c_ppx2e_60x2dx2d672009httpsx3ax2fx2fdoix2eorgx2f10x2e1145x2f1592761x2e1592779x22x29]{\AutobibLink{\Thyperref{2009}{autobiblab:45}}}]}, also an MIT product, is a
fully graphical language system, with wide applications in
education. In Scratch, users write their programs by
snapping graphical blocks together. These blocks resemble
puzzle pieces and snapping them together creates
syntactically valid programs. Scratch offers limited, but growing\Autobibref{~[\hyperref[t:x28autobib_x22Brian_Harvey_and_Jens_Mxf6nigBringing_x5cx22No_Ceilingx5cx22_to_Scratchx3a_Can_One_Language_Serve_Kids_and_Computer_Scientistsx3fIn_Procx2e_Constructionismx2c_ppx2e_1x2dx2d102010x22x29]{\AutobibLink{Harvey and M\"{o}nig}} \hyperref[t:x28autobib_x22Brian_Harvey_and_Jens_Mxf6nigBringing_x5cx22No_Ceilingx5cx22_to_Scratchx3a_Can_One_Language_Serve_Kids_and_Computer_Scientistsx3fIn_Procx2e_Constructionismx2c_ppx2e_1x2dx2d102010x22x29]{\AutobibLink{\Thyperref{2010}{autobiblab:30}}}]},
capabilities for a programmer to make new block types.

Hypercard\Autobibref{~[\hyperref[t:x28autobib_x22Danny_GoodmanThe_Complete_Hypercard_HandbookBantam_Computer_Books1988x22x29]{\AutobibLink{Goodman}} \hyperref[t:x28autobib_x22Danny_GoodmanThe_Complete_Hypercard_HandbookBantam_Computer_Books1988x22x29]{\AutobibLink{\Thyperref{1988}{autobiblab:29}}}]} gives users a graphical interface to make
interactive documents. Authors have used hypercard to create everything
from user interfaces to adventure games. While hypercard has been used
in a wide variety of domains, it is not a general{-}purpose language.

Smalltalk\Autobibref{~[\hyperref[t:x28autobib_x22Alexandre_Bergelx2c_Damien_Cassoux2c_Stxe9phane_Ducassex2c_and_Jannik_LavalDeep_into_PharoSquare_Bracket_Associates2013x22x29]{\AutobibLink{Bergel et al\Sendabbrev{.}}} \hyperref[t:x28autobib_x22Alexandre_Bergelx2c_Damien_Cassoux2c_Stxe9phane_Ducassex2c_and_Jannik_LavalDeep_into_PharoSquare_Bracket_Associates2013x22x29]{\AutobibLink{\Thyperref{2013}{autobiblab:5}}}; \hyperref[t:x28autobib_x22Adele_Goldberg_and_David_RobsonSmalltalkx2d80x3a_The_Language_and_Its_ImplementationAddisonx2dWesley_Longman_Publishing_Co1983x22x29]{\AutobibLink{Goldberg and Robson}} \hyperref[t:x28autobib_x22Adele_Goldberg_and_David_RobsonSmalltalkx2d80x3a_The_Language_and_Its_ImplementationAddisonx2dWesley_Longman_Publishing_Co1983x22x29]{\AutobibLink{\Thyperref{1983}{autobiblab:28}}}; \hyperref[t:x28autobib_x22Daniel_Ingallsx2c_Krzysztof_Palaczx2c_Stephen_Uhlerx2c_Antero_Taivalsaarix2c_and_Tommi_MikkonenThe_Lively_Kernel_A_Selfx2dsupporting_System_on_a_Web_PageIn_Procx2e_Selfx2dSustaining_Systemsx2c_ppx2e_31x2dx2d502008httpsx3ax2fx2fdoix2eorgx2f10x2e1007x2f978x2d3x2d540x2d89275x2d5x5f2x22x29]{\AutobibLink{Ingalls et al\Sendabbrev{.}}} \hyperref[t:x28autobib_x22Daniel_Ingallsx2c_Krzysztof_Palaczx2c_Stephen_Uhlerx2c_Antero_Taivalsaarix2c_and_Tommi_MikkonenThe_Lively_Kernel_A_Selfx2dsupporting_System_on_a_Web_PageIn_Procx2e_Selfx2dSustaining_Systemsx2c_ppx2e_31x2dx2d502008httpsx3ax2fx2fdoix2eorgx2f10x2e1007x2f978x2d3x2d540x2d89275x2d5x5f2x22x29]{\AutobibLink{\Thyperref{2008}{autobiblab:32}}}; \hyperref[t:x28autobib_x22Clemens_Nx2e_Klokmosex2c_James_Rx2e_Eaganx2c_Siemen_Baaderx2c_Wendy_Mackayx2c_and_Michel_Beaudouinx2dLafonWebstratesx3a_Shareable_Dynamic_MediaIn_Procx2e_ACM_Symposium_on_User_Interface_Software_and_Technologyx2c_ppx2e_280x2dx2d2902015httpsx3ax2fx2fdoix2eorgx2f10x2e1145x2f2807442x2e2807446x22x29]{\AutobibLink{Klokmose et al\Sendabbrev{.}}} \hyperref[t:x28autobib_x22Clemens_Nx2e_Klokmosex2c_James_Rx2e_Eaganx2c_Siemen_Baaderx2c_Wendy_Mackayx2c_and_Michel_Beaudouinx2dLafonWebstratesx3a_Shareable_Dynamic_MediaIn_Procx2e_ACM_Symposium_on_User_Interface_Software_and_Technologyx2c_ppx2e_280x2dx2d2902015httpsx3ax2fx2fdoix2eorgx2f10x2e1145x2f2807442x2e2807446x22x29]{\AutobibLink{\Thyperref{2015}{autobiblab:34}}}; \hyperref[t:x28autobib_x22Roman_Rxe4dlex2c_Midas_Nouwensx2c_Kristian_Antonsenx2c_James_Rx2e_Eaganx2c_and_Clemens_Nx2e_KlokmoseCodestratesx3a_Literate_Computing_with_WebstratesIn_Procx2e_ACM_Symposium_on_User_Interface_Software_and_Technologyx2c_ppx2e_715x2dx2d7252017httpsx3ax2fx2fdoix2eorgx2f10x2e1145x2f3126594x2e3126642x22x29]{\AutobibLink{R\"{a}dle et al\Sendabbrev{.}}} \hyperref[t:x28autobib_x22Roman_Rxe4dlex2c_Midas_Nouwensx2c_Kristian_Antonsenx2c_James_Rx2e_Eaganx2c_and_Clemens_Nx2e_KlokmoseCodestratesx3a_Literate_Computing_with_WebstratesIn_Procx2e_ACM_Symposium_on_User_Interface_Software_and_Technologyx2c_ppx2e_715x2dx2d7252017httpsx3ax2fx2fdoix2eorgx2f10x2e1145x2f3126594x2e3126642x22x29]{\AutobibLink{\Thyperref{2017}{autobiblab:46}}}]} supports direct manipulation of GUI objects, often
called live programming. Rather than
separating code from objects, Smalltalk programs exist in a shared
environment, the Morphic\Autobibref{~[\hyperref[t:x28autobib_x22John_Maloneyx2c_Kimberly_Mx2e_Rosex2c_and_Walt_Disney_ImagineeringAn_Introduction_to_Morphicx3a_The_Squeak_User_Interface_FrameworkIn_Squeakx3a_Open_Personal_Computing_and_Multimediax2c_ppx2e_39x2dx2d77_Pearson2001x22x29]{\AutobibLink{Maloney et al\Sendabbrev{.}}} \hyperref[t:x28autobib_x22John_Maloneyx2c_Kimberly_Mx2e_Rosex2c_and_Walt_Disney_ImagineeringAn_Introduction_to_Morphicx3a_The_Squeak_User_Interface_FrameworkIn_Squeakx3a_Open_Personal_Computing_and_Multimediax2c_ppx2e_39x2dx2d77_Pearson2001x22x29]{\AutobibLink{\Thyperref{2001}{autobiblab:36}}}]} user interface.  Programmers
can visualize GUI objects, inspect and modify their code component, and
re{-}connect them to the program.  No Smalltalk systems truly accommodate
general{-}purpose graphical{-}oriented programming as a primary mode, however.

GRAIL\Autobibref{~[\hyperref[t:x28autobib_x22Tx2e_Ox2e_Ellisx2c_Jx2e_Fx2e_Heafnerx2c_and_Wx2e_Lx2e_SibleyThe_GRAIL_Language_and_OperationsRAND_Corporationx2c_RMx2d6001x2dARPA1969httpsx3ax2fx2fdoix2eorgx2f10x2e7249x2fRM6001x22x29]{\AutobibLink{Ellis et al\Sendabbrev{.}}} \hyperref[t:x28autobib_x22Tx2e_Ox2e_Ellisx2c_Jx2e_Fx2e_Heafnerx2c_and_Wx2e_Lx2e_SibleyThe_GRAIL_Language_and_OperationsRAND_Corporationx2c_RMx2d6001x2dARPA1969httpsx3ax2fx2fdoix2eorgx2f10x2e7249x2fRM6001x22x29]{\AutobibLink{\Thyperref{1969}{autobiblab:16}\AutobibLink{a},\AutobibLink{b}}}]} is possibly one of
the oldest examples of graphical syntax. It allows users
to create and program with graphical flow diagrams. Despite
the apparent limitations of this domain, GRAIL was powerful
enough to be implemented using itself.

Notebooks\Autobibref{~[\hyperref[t:x28autobib_x22Jeremy_AshkenasObservablex3a_The_User_ManualRetrieved_Februaryx2c_20202019httpsx3ax2fx2fobservablehqx2ecomx2fx40observablehqx2fuserx2dmanualx22x29]{\AutobibLink{Ashkenas}} \hyperref[t:x28autobib_x22Jeremy_AshkenasObservablex3a_The_User_ManualRetrieved_Februaryx2c_20202019httpsx3ax2fx2fobservablehqx2ecomx2fx40observablehqx2fuserx2dmanualx22x29]{\AutobibLink{\Thyperref{2019}{autobiblab:3}}}; \hyperref[t:x28autobib_x22Lx2e_Bernardinx2c_Px2e_Chinx2c_Px2e_DeMarcox2c_Kx2e_Ox2e_Geddesx2c_Dx2e_Ex2e_Gx2e_Harex2c_Kx2e_Mx2e_Healx2c_Gx2e_Labahnx2c_Jx2e_Px2e_Mayx2c_Jx2e_McCarronx2c_Mx2e_Bx2e_Monaganx2c_Dx2e_Ohashix2c_and_Sx2e_Mx2e_VorkoetterMaple_Programming_GuideMaplesoft2012x22x29]{\AutobibLink{Bernardin et al\Sendabbrev{.}}} \hyperref[t:x28autobib_x22Lx2e_Bernardinx2c_Px2e_Chinx2c_Px2e_DeMarcox2c_Kx2e_Ox2e_Geddesx2c_Dx2e_Ex2e_Gx2e_Harex2c_Kx2e_Mx2e_Healx2c_Gx2e_Labahnx2c_Jx2e_Px2e_Mayx2c_Jx2e_McCarronx2c_Mx2e_Bx2e_Monaganx2c_Dx2e_Ohashix2c_and_Sx2e_Mx2e_VorkoetterMaple_Programming_GuideMaplesoft2012x22x29]{\AutobibLink{\Thyperref{2012}{autobiblab:6}}}; \hyperref[t:x28autobib_x22Fernando_Perez_and_Brian_Ex2e_GrangerIPythonx3a_A_System_for_Interactive_Scientific_ComputingComputing_in_Science_and_Engineering_9x283x29x2c_ppx2e_21x2dx2d292007httpsx3ax2fx2fdoix2eorgx2f10x2e1109x2fMCSEx2e2007x2e53x22x29]{\AutobibLink{Perez and Granger}} \hyperref[t:x28autobib_x22Fernando_Perez_and_Brian_Ex2e_GrangerIPythonx3a_A_System_for_Interactive_Scientific_ComputingComputing_in_Science_and_Engineering_9x283x29x2c_ppx2e_21x2dx2d292007httpsx3ax2fx2fdoix2eorgx2f10x2e1109x2fMCSEx2e2007x2e53x22x29]{\AutobibLink{\Thyperref{2007}{autobiblab:43}}}; \hyperref[t:x28autobib_x22Stephen_WolframThe_Mathematica_BookFourth_editionx2e_Cambridge_University_Press1988x22x29]{\AutobibLink{Wolfram}} \hyperref[t:x28autobib_x22Stephen_WolframThe_Mathematica_BookFourth_editionx2e_Cambridge_University_Press1988x22x29]{\AutobibLink{\Thyperref{1988}{autobiblab:51}}}]}
and Webstrates\Autobibref{~[\hyperref[t:x28autobib_x22Clemens_Nx2e_Klokmosex2c_James_Rx2e_Eaganx2c_Siemen_Baaderx2c_Wendy_Mackayx2c_and_Michel_Beaudouinx2dLafonWebstratesx3a_Shareable_Dynamic_MediaIn_Procx2e_ACM_Symposium_on_User_Interface_Software_and_Technologyx2c_ppx2e_280x2dx2d2902015httpsx3ax2fx2fdoix2eorgx2f10x2e1145x2f2807442x2e2807446x22x29]{\AutobibLink{Klokmose et al\Sendabbrev{.}}} \hyperref[t:x28autobib_x22Clemens_Nx2e_Klokmosex2c_James_Rx2e_Eaganx2c_Siemen_Baaderx2c_Wendy_Mackayx2c_and_Michel_Beaudouinx2dLafonWebstratesx3a_Shareable_Dynamic_MediaIn_Procx2e_ACM_Symposium_on_User_Interface_Software_and_Technologyx2c_ppx2e_280x2dx2d2902015httpsx3ax2fx2fdoix2eorgx2f10x2e1145x2f2807442x2e2807446x22x29]{\AutobibLink{\Thyperref{2015}{autobiblab:34}}}; \hyperref[t:x28autobib_x22Roman_Rxe4dlex2c_Midas_Nouwensx2c_Kristian_Antonsenx2c_James_Rx2e_Eaganx2c_and_Clemens_Nx2e_KlokmoseCodestratesx3a_Literate_Computing_with_WebstratesIn_Procx2e_ACM_Symposium_on_User_Interface_Software_and_Technologyx2c_ppx2e_715x2dx2d7252017httpsx3ax2fx2fdoix2eorgx2f10x2e1145x2f3126594x2e3126642x22x29]{\AutobibLink{R\"{a}dle et al\Sendabbrev{.}}} \hyperref[t:x28autobib_x22Roman_Rxe4dlex2c_Midas_Nouwensx2c_Kristian_Antonsenx2c_James_Rx2e_Eaganx2c_and_Clemens_Nx2e_KlokmoseCodestratesx3a_Literate_Computing_with_WebstratesIn_Procx2e_ACM_Symposium_on_User_Interface_Software_and_Technologyx2c_ppx2e_715x2dx2d7252017httpsx3ax2fx2fdoix2eorgx2f10x2e1145x2f3126594x2e3126642x22x29]{\AutobibLink{\Thyperref{2017}{autobiblab:46}}}]} are
essentially a modern reincarnation of this model, except
that they use a read{-}eval{-}print loop approach to object
manipulation rather than the GUI{-}based one, made so
attractive by the Morphic framework. These systems do not
permit domain{-}specific syntax extensions.

\Ssubsection{Projectional and Bidirectional Editing}{Projectional and Bidirectional Editing}\label{t:x28part_x22Projectionalx5fandx5fBidirectionalx5fEditingx22x29}

Bidirectional editors attempt to present two editable views
for a program that developers can manipulate in lockstep.
Sketch{-}n{-}Sketch\Autobibref{~[\hyperref[t:x28autobib_x22Ravi_Chughx2c_Brian_Hempelx2c_Mitchell_Spradlinx2c_and_Jacob_AlbersProgrammatic_and_Direct_Manipulationx2c_Together_at_LastIn_Procx2e_Programming_Languages_Design_and_Implementationx2c_ppx2e_341x2dx2d3542016httpsx3ax2fx2fdoix2eorgx2f10x2e1145x2f2980983x2e2908103x22x29]{\AutobibLink{Chugh et al\Sendabbrev{.}}} \hyperref[t:x28autobib_x22Ravi_Chughx2c_Brian_Hempelx2c_Mitchell_Spradlinx2c_and_Jacob_AlbersProgrammatic_and_Direct_Manipulationx2c_Together_at_LastIn_Procx2e_Programming_Languages_Design_and_Implementationx2c_ppx2e_341x2dx2d3542016httpsx3ax2fx2fdoix2eorgx2f10x2e1145x2f2980983x2e2908103x22x29]{\AutobibLink{\Thyperref{2016}{autobiblab:8}}}; \hyperref[t:x28autobib_x22Brian_Hempelx2c_Justin_Lubinx2c_Grace_Lux2c_and_Ravi_ChughDeucex3a_A_Lightweight_User_Interface_for_Structured_EditingIn_Procx2e_International_Conference_on_Software_Engineeringx2c_ppx2e_654x2dx2d6642018httpsx3ax2fx2fdoix2eorgx2f10x2e1145x2f3180155x2e3180165x22x29]{\AutobibLink{Hempel et al\Sendabbrev{.}}} \hyperref[t:x28autobib_x22Brian_Hempelx2c_Justin_Lubinx2c_Grace_Lux2c_and_Ravi_ChughDeucex3a_A_Lightweight_User_Interface_for_Structured_EditingIn_Procx2e_International_Conference_on_Software_Engineeringx2c_ppx2e_654x2dx2d6642018httpsx3ax2fx2fdoix2eorgx2f10x2e1145x2f3180155x2e3180165x22x29]{\AutobibLink{\Thyperref{2018}{autobiblab:31}}}]}, for
example, allows programmers to create SVG{-}like pictures both
programmatically with text, and by directly manipulating the
picture. Another example is Dreamweaver\Autobibref{~[\hyperref[t:x28autobib_x22AdobeAdobe_Dreamweaver_CC_HelpRetrieved_Mayx2c_20202019httpsx3ax2fx2fhelpxx2eadobex2ecomx2fpdfx2fdreamweaverx5freferencex2epdfx22x29]{\AutobibLink{Adobe}} \hyperref[t:x28autobib_x22AdobeAdobe_Dreamweaver_CC_HelpRetrieved_Mayx2c_20202019httpsx3ax2fx2fhelpxx2eadobex2ecomx2fpdfx2fdreamweaverx5freferencex2epdfx22x29]{\AutobibLink{\Thyperref{2019}{autobiblab:1}}}]},
which allows authors to create web pages directly, and drop
down to HTML when needed. Changes made in one view propagate
back to the other, keeping them in sync. We conjecture that an interactive{-}syntax
mechanism like ours could be used to implement such a
bidirectional editing system. Likewise, a bidirectional
editing capability would improve the process of creating
interactive{-}syntax extensions.

Wizards and code completion tools, such as Graphite\Autobibref{~[\hyperref[t:x28autobib_x22Cyrus_Omarx2c_YoungSeok_Yoonx2c_Thomas_Dx2e_LaTozax2c_and_Brad_Ax2e_MyersActive_Code_CompletionIn_Procx2e_International_Conference_on_Software_Engineeringx2c_ppx2e_859x2dx2d8692012x22x29]{\AutobibLink{Omar et al\Sendabbrev{.}}} \hyperref[t:x28autobib_x22Cyrus_Omarx2c_YoungSeok_Yoonx2c_Thomas_Dx2e_LaTozax2c_and_Brad_Ax2e_MyersActive_Code_CompletionIn_Procx2e_International_Conference_on_Software_Engineeringx2c_ppx2e_859x2dx2d8692012x22x29]{\AutobibLink{\Thyperref{2012}{autobiblab:40}}}]},
preform this task in one direction. A small graphical UI can generate
textual code for a programmer. However, once finished, the programmer
cannot return to the graphical UI from text.

Projectional editing aims to give programmers the ability to edit programs
visually.\NoteBox{\NoteContent{Intentional
Software\Autobibref{~[\hyperref[t:x28autobib_x22Charles_Simonyix2c_Magnus_Christersonx2c_and_Shane_CliffordIntentional_SoftwareACM_SIGPLAN_Notices_41x2810x29x2c_ppx2e_451x2dx2d4642006httpsx3ax2fx2fdoix2eorgx2f10x2e1145x2f1167515x2e1167511x22x29]{\AutobibLink{Simonyi et al\Sendabbrev{.}}} \hyperref[t:x28autobib_x22Charles_Simonyix2c_Magnus_Christersonx2c_and_Shane_CliffordIntentional_SoftwareACM_SIGPLAN_Notices_41x2810x29x2c_ppx2e_451x2dx2d4642006httpsx3ax2fx2fdoix2eorgx2f10x2e1145x2f1167515x2e1167511x22x29]{\AutobibLink{\Thyperref{2006}{autobiblab:47}}}]} has similar goals, but there is almost no concrete
information in the literature about this project.}}
Indeed, in this world, there are no programs per se, only
graphically presented abstract syntax trees (AST), which a developer can
edit and manipulate. The system can then render the ASTs as conventional
program text. The most well{-}known system is
MPS\Autobibref{~[\hyperref[t:x28autobib_x22Vaclav_Pechx2c_Alex_Shatalinx2c_and_Markus_VoelterJetBrains_MPS_as_a_Tool_for_Extending_JavaIn_Procx2e_Principles_and_Practice_of_Programming_in_Javax2c_ppx2e_165x2dx2d1682013httpsx3ax2fx2fdoix2eorgx2f10x2e1145x2f2500828x2e2500846x22x29]{\AutobibLink{Pech et al\Sendabbrev{.}}} \hyperref[t:x28autobib_x22Vaclav_Pechx2c_Alex_Shatalinx2c_and_Markus_VoelterJetBrains_MPS_as_a_Tool_for_Extending_JavaIn_Procx2e_Principles_and_Practice_of_Programming_in_Javax2c_ppx2e_165x2dx2d1682013httpsx3ax2fx2fdoix2eorgx2f10x2e1145x2f2500828x2e2500846x22x29]{\AutobibLink{\Thyperref{2013}{autobiblab:42}}}; \hyperref[t:x28autobib_x22Markus_Voelter_and_Sascha_LissonSupporting_Diverse_Notations_in_MPSx2019_Projectional_EditorIn_Procx2e_International_Workshop_on_The_Globalization_of_Modeling_Languages2014x22x29]{\AutobibLink{Voelter and Lisson}} \hyperref[t:x28autobib_x22Markus_Voelter_and_Sascha_LissonSupporting_Diverse_Notations_in_MPSx2019_Projectional_EditorIn_Procx2e_International_Workshop_on_The_Globalization_of_Modeling_Languages2014x22x29]{\AutobibLink{\Thyperref{2014}{autobiblab:48}}}]}. It has been used to create large
non{-}textual programming systems\Autobibref{~[\hyperref[t:x28autobib_x22Markus_Voelterx2c_Daniel_Ratiux2c_Bernhard_Schaetzx2c_and_Bernd_Kolbmbeddrx3a_an_Extensible_Cx2dbased_Programming_Language_and_IDE_for_Embedded_SystemsIn_Procx2e_Conference_on_Systemsx2c_Programmingx2c_and_Applicationsx3a_Software_for_Humanityx2c_ppx2e_121x2dx2d1402012httpsx3ax2fx2fdoix2eorgx2f10x2e1145x2f2384716x2e2384767x22x29]{\AutobibLink{Voelter et al\Sendabbrev{.}}} \hyperref[t:x28autobib_x22Markus_Voelterx2c_Daniel_Ratiux2c_Bernhard_Schaetzx2c_and_Bernd_Kolbmbeddrx3a_an_Extensible_Cx2dbased_Programming_Language_and_IDE_for_Embedded_SystemsIn_Procx2e_Conference_on_Systemsx2c_Programmingx2c_and_Applicationsx3a_Software_for_Humanityx2c_ppx2e_121x2dx2d1402012httpsx3ax2fx2fdoix2eorgx2f10x2e1145x2f2384716x2e2384767x22x29]{\AutobibLink{\Thyperref{2012}{autobiblab:49}}}]}. Unlike
interactive{-}syntax extensions, projectional editors must be modified in
their host editors and always demand separated edit{-}time and run{-}time
modules. Such a separation means all editors must be attached to a program
project, they cannot be constructed locally within a file. It therefore is rather
difficult to abstract over them.

Barista\Autobibref{~[\hyperref[t:x28autobib_x22Amy_Ko_and_Brad_Ax2e_MyersBaristax3a_An_Implementation_Framework_for_Enabling_New_Toolsx2c_Interaction_Techniques_and_Views_in_Code_EditorsIn_Procx2e_Conference_on_Human_Factors_in_Computing_Systemsx2c_ppx2e_387x2dx2d3962006httpsx3ax2fx2fdoix2eorgx2f10x2e1145x2f1124772x2e1124831x22x29]{\AutobibLink{Ko and Myers}} \hyperref[t:x28autobib_x22Amy_Ko_and_Brad_Ax2e_MyersBaristax3a_An_Implementation_Framework_for_Enabling_New_Toolsx2c_Interaction_Techniques_and_Views_in_Code_EditorsIn_Procx2e_Conference_on_Human_Factors_in_Computing_Systemsx2c_ppx2e_387x2dx2d3962006httpsx3ax2fx2fdoix2eorgx2f10x2e1145x2f1124772x2e1124831x22x29]{\AutobibLink{\Thyperref{2006}{autobiblab:35}}}]} is a framework that lets programmers mix
textual and visual programs. The graphical extensions,
however, are tied to the Barista framework, rather than the
programs themselves. Like MPS, Barista saves the ASTs for a
program, rather than the raw text.

The Hazel project and Livelits\Autobibref{~[\hyperref[t:x28autobib_x22Cyrus_Omarx2c_Nick_Collinsx2c_David_Moonx2c_Ian_Voyseyx2c_and_Ravi_ChughLivelitsx3a_Filling_Typed_Holes_with_Live_GUIs_x28Extended_Abstractx29In_Procx2e_Workshop_on_Typex2ddriven_Development2019x22x29]{\AutobibLink{Omar et al\Sendabbrev{.}}} \hyperref[t:x28autobib_x22Cyrus_Omarx2c_Nick_Collinsx2c_David_Moonx2c_Ian_Voyseyx2c_and_Ravi_ChughLivelitsx3a_Filling_Typed_Holes_with_Live_GUIs_x28Extended_Abstractx29In_Procx2e_Workshop_on_Typex2ddriven_Development2019x22x29]{\AutobibLink{\Thyperref{2019}{autobiblab:39}}}]} are also closely
related to interactive{-}syntax extensions. Like editors, the Livelits
proposal aims to let programmers embed graphical syntax into their code.
In contrast to interactive{-}syntax extensions, which use phases to support
editor instantiation and manipulation, the proposed Livelits will employ
typed{-}hole editing. Finally, while the Livelits proposal is just a two{-}page
blueprint, we conjecture that these constructs will not be deployed in the
same range of linguistic contexts as interactive{-}syntax extensions (see
\ChapRef{\SectionNumberLink{t:x28part_x22secx3aexamplesx22x29}{4}}{Evaluating a Plethora of Examples}).

\sectionNewpage

\Ssection{From Limitations to Future Work}{From Limitations to Future Work}\label{t:x28part_x22secx3alimitationsx22x29}

The prototype falls short of the plan laid out in \ChapRef{\SectionNumberLink{t:x28part_x22secx3adesignx22x29}{2}}{The Design Space}, though
the shortcomings are of a non{-}essential technical nature, not principled ones:

\noindent \begin{enumerate}\atItemizeStart

\item The prototype partially re{-}uses Racket{'}s GUI library in the context of
interactive{-}syntax definitions. At the moment, the prototype relies on a
Racket{-}coded GUI tailored to the interactive{-}syntax system,
meaning developers cannot use all GUIs for both syntax extensions and the
application itself. The shortcoming is due to two intertwined technical reasons:
the setup of Racket{'}s GUI library and DrRacket{'}s use of an editor canvas, which
cannot embed controls and other window areas.

\item The prototype does \emph{not} validate the usability of the construction
across different visual IDEs. It does accommodate the use of DrRacket and plain
text editors such as Emacs or Vim. While the textual rendering of editors is
syntactically constrained, a developer who prefers a text editor can still work
on code and even read embedded interactive{-}syntax.
Due to the design choice of relying on a Racket{-}based GUI for editors, though,
interactive syntax will work in any IDE that can run Racket code and grant
access to a drawing context such as a canvas.  For details on how to engineer a
general solution, see \SecRef{\SectionNumberLink{t:x28part_x22secx3apartsx22x29}{5.1}}{Ingredients for an Interactive{-}Syntax System}.

\item A minor shortcoming concerns editors that contain text fields into which
developers enter code. In the current prototype, these text fields are just
widgets that permit plain text editing. With some amount of labor,
an interactive{-}syntax extension could use a miniature version of DrRacket so
that developers would not just edit plain text, but code, in these places.

\item Finally, the use of Racket{'}s sub{-}modules to implement an edit{-}time phase
falls short of the language{'}s standard meta{-}programming ideals. While they
\emph{mostly} work correctly for mapping editors to code, the solution exhibits
hygiene problems in some corner cases. Furthermore, while some meta{-}programming
extensions in conventional languages, e.g., Rust, do implement hygienic expansion,
others completely fail in this regard, e.g., Scala, which may cause additional
problems in adapting this idea to different language contexts.\end{enumerate}

\noindent All of these limitations naturally point to future investigations.

Besides these technical investigations, the idea also demands a user{-}facing
evaluation in addition to the expressiveness evaluation presented in
\ChapRef{\SectionNumberLink{t:x28part_x22secx3aexamplesx22x29}{4}}{Evaluating a Plethora of Examples}. Here are some questions for such a study:

\noindent \begin{itemize}\atItemizeStart

\item How quickly do developers identify situations where the use of interactive
syntax might benefit their successors?

\item How much more difficult is it to articulate code as interactive syntax than text?

\item Is it easier to comprehend code formulated with interactive syntax instead of text?\end{itemize}

\noindent Answers may simultaneously confirm the conjecture behind the design of
interactive syntax and point to technical problems in existing systems.
The authors will attempt to answer these question after hardening the prototype into an easily usable system.

\sectionNewpage

\Ssection{Conclusion}{Conclusion}\label{t:x28part_x22Conclusionx22x29}

Linear text is the most widely embraced means for writing down programs.
But, we also know that in many contexts a picture is worth a thousand
words. Developers know this, which is why ASCII diagrams accompany many
programs in comments and why type{-}set documentation comes with elaborate
diagrams and graphics. Developers and their support staff create these
comments and documents because they accept the idea that code is a message
to some future programmer who will have to understand and modify it.

If we wish to combine the productivity of text{-}oriented programming with
the power of pictures, we must extend our textual programming languages
with graphical syntax. A fixed set of graphical syntaxes or static images
do not suffice, however. We must equip developers with the expressive power
to create interactive graphics for the problems that they are working on
and integrate these graphical pieces of program directly into the
code. Concisely put, turning comments into executable code is the only way to
keep comments in sync with code.

When a developer invests energy into interactive GUI code, this effort
must pay off. Hence a developer should be able to exploit elements of the
user{-}interface code in interactive{-}syntax extensions. Conversely, any
investment into GUI elements for an interactive{-}syntax extension must carry
over to the actual user{-}interface code for a software system.

Finally, good developers build reusable abstractions. In
this spirit, an interactive{-}syntax extension mechanism must
come with the power to abstract over interactive{-}syntax
extensions with an interactive{-}syntax extension. If this is
available to developers, they may soon offer complete libraries of
interactive{-}syntax building blocks.

Our paper presents the design, implementation, and evaluation of the first
interactive{-}syntax extensions mechanism that mostly satisfies all of these
criteria. While the implementation is a prototype, it is robust enough to
demonstrate the broad applicability of the idea with examples from
algorithms, compilers, file systems, networking, as well as some narrow
domains such as circuit simulation and game program development. In terms
of linguistics, the prototype can already accommodate interactive syntax
for visual data objects, complex patterns, sophisticated templates, and
meta forms. We consider it a promising step
towards a true synthesis of text and {``}moving{''} pictures.

\sectionNewpage

\Ssectionstarx{Acknowledgments}{Acknowledgments}\label{t:x28part_x22Acknowledgmentsx22x29}

This research was partially supported by NSF grants 1823244
and 20050550. We also thank Benjamin Chung, Benjamin
Greenman, Elizabeth Grimm, Jason Hemann, Shriram
Krishnamurthi, and Ming{-}Ho Yee for useful discussions and
feedback on early drafts of this paper.

\identity{\appendix}

\sectionNewpage

\Ssection{Submodules in Racket, a Brief Review}{Submodules in Racket, a Brief Review}\label{t:x28part_x22secx3asubmodulesx22x29}

Racket{'}s notion of submodule\Autobibref{~[\hyperref[t:x28autobib_x22Matthew_FlattSubmodules_in_Racketx2c_You_Want_it_Whenx2c_Againx3fIn_Procx2e_Generative_Programmingx3a_Concepts_x26_Experiencesx2c_ppx2e_13x2dx2d222013httpsx3ax2fx2fdoix2eorgx2f10x2e1145x2f2517208x2e2517211x22x29]{\AutobibLink{Flatt}} \hyperref[t:x28autobib_x22Matthew_FlattSubmodules_in_Racketx2c_You_Want_it_Whenx2c_Againx3fIn_Procx2e_Generative_Programmingx3a_Concepts_x26_Experiencesx2c_ppx2e_13x2dx2d222013httpsx3ax2fx2fdoix2eorgx2f10x2e1145x2f2517208x2e2517211x22x29]{\AutobibLink{\Thyperref{2013}{autobiblab:22}}}]} is the key to the
implementation of an edit phase.
In Racket, submodules exist to organize a single file{-}size module into separate
entities. A submodule does not get evaluated unless it is explicitly
required, which is possible locally as well as from another file.

The
introduction of submodules was motivated by two major desires: to designate some part
of the code as \RktModLink{\RktSym{main}} and to include exemplary unit tests right next to
a function definition. Because submodules are separated from the surrounding module,
adding such tests has no  impact on the size or running time of the main module
itself.

Consider this example:
\identity{~\\\begin{minipage}[c]{0.9\textwidth}}

\begin{SCodeFlow}\begin{RktBlk}\begin{SingleColumn}\Smaller{\Smaller{\mbox{\hphantom{\Scribtexttt{x}}}\Scribtexttt{1}\mbox{\hphantom{\Scribtexttt{x}}}}}\RktCmt{;;}\mbox{\hphantom{\Scribtexttt{x}}}\RktCmt{inc}\mbox{\hphantom{\Scribtexttt{x}}}\RktCmt{{\hbox{\texttt{:}}}}\mbox{\hphantom{\Scribtexttt{x}}}\RktCmt{[Box}\mbox{\hphantom{\Scribtexttt{x}}}\RktCmt{Integer]}\mbox{\hphantom{\Scribtexttt{x}}}\RktCmt{{-}{\Stttextmore}}\mbox{\hphantom{\Scribtexttt{x}}}\RktCmt{Void}\RktMeta{}

\Smaller{\Smaller{\mbox{\hphantom{\Scribtexttt{x}}}\Scribtexttt{2}\mbox{\hphantom{\Scribtexttt{x}}}}}\RktMeta{}\RktCmt{;;}\mbox{\hphantom{\Scribtexttt{x}}}\RktCmt{increment}\mbox{\hphantom{\Scribtexttt{x}}}\RktCmt{the}\mbox{\hphantom{\Scribtexttt{x}}}\RktCmt{content}\mbox{\hphantom{\Scribtexttt{x}}}\RktCmt{of}\mbox{\hphantom{\Scribtexttt{x}}}\RktCmt{the}\mbox{\hphantom{\Scribtexttt{x}}}\RktCmt{given}\mbox{\hphantom{\Scribtexttt{x}}}\RktCmt{box}\mbox{\hphantom{\Scribtexttt{x}}}\RktCmt{by}\mbox{\hphantom{\Scribtexttt{x}}}\RktCmt{1}\RktMeta{}

\Smaller{\Smaller{\mbox{\hphantom{\Scribtexttt{x}}}\Scribtexttt{3}\mbox{\hphantom{\Scribtexttt{x}}}}}\RktMeta{~}

\Smaller{\Smaller{\mbox{\hphantom{\Scribtexttt{x}}}\Scribtexttt{4}\mbox{\hphantom{\Scribtexttt{x}}}}}\RktMeta{}\RktPn{(}\RktSym{module+}\RktMeta{}\mbox{\hphantom{\Scribtexttt{x}}}\RktMeta{}\RktSym{test}\RktMeta{}

\Smaller{\Smaller{\mbox{\hphantom{\Scribtexttt{x}}}\Scribtexttt{5}\mbox{\hphantom{\Scribtexttt{x}}}}}\RktMeta{}\mbox{\hphantom{\Scribtexttt{xx}}}\RktMeta{}\RktPn{(}\RktSym{let}\RktMeta{}\mbox{\hphantom{\Scribtexttt{x}}}\RktMeta{}\RktPn{(}\RktPn{[}\RktSym{x}\RktMeta{}\mbox{\hphantom{\Scribtexttt{x}}}\RktMeta{}\RktPn{(}\RktSym{box}\RktMeta{}\mbox{\hphantom{\Scribtexttt{x}}}\RktMeta{}\RktVal{0}\RktPn{)}\RktPn{]}\RktPn{)}\RktMeta{}

\Smaller{\Smaller{\mbox{\hphantom{\Scribtexttt{x}}}\Scribtexttt{6}\mbox{\hphantom{\Scribtexttt{x}}}}}\RktMeta{}\mbox{\hphantom{\Scribtexttt{xxxx}}}\RktMeta{}\RktPn{(}\RktSym{check{-}equal{\hbox{\texttt{?}}}}\RktMeta{}\mbox{\hphantom{\Scribtexttt{x}}}\RktMeta{}\RktPn{(}\RktSym{unbox}\RktMeta{}\mbox{\hphantom{\Scribtexttt{x}}}\RktMeta{}\RktSym{x}\RktPn{)}\RktMeta{}\mbox{\hphantom{\Scribtexttt{x}}}\RktMeta{}\RktVal{0}\RktPn{)}\RktMeta{}

\Smaller{\Smaller{\mbox{\hphantom{\Scribtexttt{x}}}\Scribtexttt{7}\mbox{\hphantom{\Scribtexttt{x}}}}}\RktMeta{}\mbox{\hphantom{\Scribtexttt{xxxx}}}\RktMeta{}\RktPn{(}\RktSym{inc}\RktMeta{}\mbox{\hphantom{\Scribtexttt{x}}}\RktMeta{}\RktSym{x}\RktPn{)}\RktMeta{}

\Smaller{\Smaller{\mbox{\hphantom{\Scribtexttt{x}}}\Scribtexttt{8}\mbox{\hphantom{\Scribtexttt{x}}}}}\RktMeta{}\mbox{\hphantom{\Scribtexttt{xxxx}}}\RktMeta{}\RktPn{(}\RktSym{check{-}equal{\hbox{\texttt{?}}}}\RktMeta{}\mbox{\hphantom{\Scribtexttt{x}}}\RktMeta{}\RktPn{(}\RktSym{unbox}\RktMeta{}\mbox{\hphantom{\Scribtexttt{x}}}\RktMeta{}\RktSym{x}\RktPn{)}\RktMeta{}\mbox{\hphantom{\Scribtexttt{x}}}\RktMeta{}\RktVal{1}\RktPn{)}\RktPn{)}\RktPn{)}\RktMeta{}

\Smaller{\Smaller{\mbox{\hphantom{\Scribtexttt{x}}}\Scribtexttt{9}\mbox{\hphantom{\Scribtexttt{x}}}}}\RktMeta{~}

\Smaller{\Smaller{\Scribtexttt{10}\mbox{\hphantom{\Scribtexttt{x}}}}}\RktMeta{}\RktPn{(}\RktSym{define}\RktMeta{}\mbox{\hphantom{\Scribtexttt{x}}}\RktMeta{}\RktPn{(}\RktSym{inc}\RktMeta{}\mbox{\hphantom{\Scribtexttt{x}}}\RktMeta{}\RktSym{counter}\RktPn{)}\RktMeta{}

\Smaller{\Smaller{\Scribtexttt{11}\mbox{\hphantom{\Scribtexttt{x}}}}}\RktMeta{}\mbox{\hphantom{\Scribtexttt{xx}}}\RktMeta{}\RktPn{(}\RktSym{set{-}box{\hbox{\texttt{!}}}}\RktMeta{}\mbox{\hphantom{\Scribtexttt{x}}}\RktMeta{}\RktSym{counter}\RktMeta{}\mbox{\hphantom{\Scribtexttt{x}}}\RktMeta{}\RktPn{(}\RktSym{add1}\RktMeta{}\mbox{\hphantom{\Scribtexttt{x}}}\RktMeta{}\RktPn{(}\RktSym{unbox}\RktMeta{}\mbox{\hphantom{\Scribtexttt{x}}}\RktMeta{}\RktSym{counter}\RktPn{)}\RktPn{)}\RktPn{)}\RktPn{)}\RktMeta{}\end{SingleColumn}\end{RktBlk}\end{SCodeFlow}

\noindent \identity{\end{minipage}\\[.2cm]}
The unit{-}test module creates a fresh counter box, initialized to \RktVal{0},
increments it, and checks the value again. The programmer can evaluate these
unit tests explicitly from another module or with a command{-}line tool:

\noindent \begin{SCodeFlow}\begin{SVerbatim}\begin{SingleColumn}\Scribtexttt{[linux] \$ raco test inc{\hbox{\texttt{.}}}rkt}

\Scribtexttt{raco test{\hbox{\texttt{:}}} (submod "inc{\hbox{\texttt{.}}}rkt" test)}

\Scribtexttt{2 tests passed}\end{SingleColumn}\end{SVerbatim}\end{SCodeFlow}

\noindent When \RktSym{inc} is imported from another module, the unit tests are neither
loaded nor run.

\sectionNewpage

\Ssectionstarx{References}{References}\label{t:x28part_x22docx2dbibliographyx22x29}

\begin{AutoBibliography}\begin{bigtabular}{@{\bigtableleftpad}l@{}l@{}}
\hbox{\label{autobiblab:1}} &
\hbox{\label{t:x28autobib_x22AdobeAdobe_Dreamweaver_CC_HelpRetrieved_Mayx2c_20202019httpsx3ax2fx2fhelpxx2eadobex2ecomx2fpdfx2fdreamweaverx5freferencex2epdfx22x29}\Autobibentry{Adobe. Adobe Dreamweaver CC Help. Retrieved May, 2020, 2019. \href{https://helpx.adobe.com/pdf/dreamweaver_reference.pdf}{\Snolinkurl{https://helpx.adobe.com/pdf/dreamweaver_reference.pdf}}}} \\
\hbox{\label{autobiblab:2}} &
\hbox{\label{t:x28autobib_x22Leif_Andersenx2c_Stephen_Changx2c_and_Matthias_FelleisenSuper_8_Languages_for_Making_Movies_x28Functional_Pearlx29Proceedings_of_the_ACM_on_Programming_Languages_1x28International_Conference_on_Functional_Programmingx29x2c_ppx2e_30x2d1x2dx2d30x2d292017httpsx3ax2fx2fdoix2eorgx2f10x2e1145x2f3110274x22x29}\Autobibentry{Leif Andersen, Stephen Chang, and Matthias Felleisen. Super 8 Languages for Making Movies (Functional Pearl). \textit{Proceedings of the ACM on Programming Languages} 1(International Conference on Functional Programming), pp. 30{-}1{--}30{-}29, 2017. \href{https://doi.org/10.1145/3110274}{\Snolinkurl{https://doi.org/10.1145/3110274}}}} \\
\hbox{\label{autobiblab:3}} &
\hbox{\label{t:x28autobib_x22Jeremy_AshkenasObservablex3a_The_User_ManualRetrieved_Februaryx2c_20202019httpsx3ax2fx2fobservablehqx2ecomx2fx40observablehqx2fuserx2dmanualx22x29}\Autobibentry{Jeremy Ashkenas. Observable: The User Manual. Retrieved February, 2020, 2019. \href{https://observablehq.com/@observablehq/user-manual}{\Snolinkurl{https://observablehq.com/@observablehq/user-manual}}}} \\
\hbox{\label{autobiblab:4}} &
\hbox{\label{t:x28autobib_x22Rudolf_BayerSymmetric_Binary_Bx2dTreesx3a_Data_Structure_and_Maintenance_AlgorithmsActa_Informatica_1x284x29x2c_ppx2e_290x2dx2d3061972httpsx3ax2fx2fdoix2eorgx2f10x2e1007x2fBF00289509x22x29}\Autobibentry{Rudolf Bayer. Symmetric Binary B{-}Trees: Data Structure and Maintenance Algorithms. \textit{Acta Informatica} 1(4), pp. 290{--}306, 1972. \href{https://doi.org/10.1007/BF00289509}{\Snolinkurl{https://doi.org/10.1007/BF00289509}}}} \\
\hbox{\label{autobiblab:5}} &
\hbox{\label{t:x28autobib_x22Alexandre_Bergelx2c_Damien_Cassoux2c_Stxe9phane_Ducassex2c_and_Jannik_LavalDeep_into_PharoSquare_Bracket_Associates2013x22x29}\Autobibentry{Alexandre Bergel, Damien Cassou, St\'{e}phane Ducasse, and Jannik Laval. Deep into Pharo. Square Bracket Associates, 2013.}} \\
\hbox{\label{autobiblab:6}} &
\hbox{\label{t:x28autobib_x22Lx2e_Bernardinx2c_Px2e_Chinx2c_Px2e_DeMarcox2c_Kx2e_Ox2e_Geddesx2c_Dx2e_Ex2e_Gx2e_Harex2c_Kx2e_Mx2e_Healx2c_Gx2e_Labahnx2c_Jx2e_Px2e_Mayx2c_Jx2e_McCarronx2c_Mx2e_Bx2e_Monaganx2c_Dx2e_Ohashix2c_and_Sx2e_Mx2e_VorkoetterMaple_Programming_GuideMaplesoft2012x22x29}\Autobibentry{L. Bernardin, P. Chin, P. DeMarco, K. O. Geddes, D. E. G. Hare, K. M. Heal, G. Labahn, J. P. May, J. McCarron, M. B. Monagan, D. Ohashi, and S. M. Vorkoetter. Maple Programming Guide. Maplesoft, 2012.}} \\
\hbox{\label{autobiblab:7}} &
\hbox{\label{t:x28autobib_x22Marat_Boshernitsan_and_Michael_Sx2e_DownesVisual_Programming_Languagesx3a_a_SurveyEECS_Departmentx2c_University_of_Californiax2c_Berkeleyx2c_UCBx2fCSDx2d04x2d13682004httpx3ax2fx2fwww2x2eeecsx2eberkeleyx2eedux2fPubsx2fTechRptsx2f2004x2f6201x2ehtmlx22x29}\Autobibentry{Marat Boshernitsan and Michael S. Downes. Visual Programming Languages: a Survey. EECS Department, University of California, Berkeley, UCB/CSD{-}04{-}1368, 2004. \href{http://www2.eecs.berkeley.edu/Pubs/TechRpts/2004/6201.html}{\Snolinkurl{http://www2.eecs.berkeley.edu/Pubs/TechRpts/2004/6201.html}}}} \\
\hbox{\label{autobiblab:8}} &
\hbox{\label{t:x28autobib_x22Ravi_Chughx2c_Brian_Hempelx2c_Mitchell_Spradlinx2c_and_Jacob_AlbersProgrammatic_and_Direct_Manipulationx2c_Together_at_LastIn_Procx2e_Programming_Languages_Design_and_Implementationx2c_ppx2e_341x2dx2d3542016httpsx3ax2fx2fdoix2eorgx2f10x2e1145x2f2980983x2e2908103x22x29}\Autobibentry{Ravi Chugh, Brian Hempel, Mitchell Spradlin, and Jacob Albers. Programmatic and Direct Manipulation, Together at Last. In \textit{Proc. Programming Languages Design and Implementation}, pp. 341{--}354, 2016. \href{https://doi.org/10.1145/2980983.2908103}{\Snolinkurl{https://doi.org/10.1145/2980983.2908103}}}} \\
\hbox{\label{autobiblab:9}} &
\hbox{\label{t:x28autobib_x22Hubert_Comonx2c_Max_Dauchetx2c_Remi_Gilleronx2c_Florent_Jacquemardx2c_Denis_Lugiezx2c_Christof_Lxf6dingx2c_Sophie_Tisonx2c_and_Marc_TommasiTree_Automata_Techniques_and_Applications2007httpx3ax2fx2ftatax2egforgex2einriax2efrx2fx22x29}\Autobibentry{Hubert Comon, Max Dauchet, Remi Gilleron, Florent Jacquemard, Denis Lugiez, Christof L\"{o}ding, Sophie Tison, and Marc Tommasi. Tree Automata Techniques and Applications. 2007. \href{http://tata.gforge.inria.fr/}{\Snolinkurl{http://tata.gforge.inria.fr/}}}} \\
\hbox{\label{autobiblab:10}} &
\hbox{\label{t:x28autobib_x22Gregory_Cooper_and_Shriram_KrishnamurthiEmbedding_Dynamic_Dataflow_in_a_Callx2dbyx2dValue_LanguageIn_Procx2e_European_Symposium_on_Programmingx2c_ppx2e_294x2dx2d3082004httpsx3ax2fx2fdoix2eorgx2f10x2e1007x2f11693024x5f20x22x29}\Autobibentry{Gregory Cooper and Shriram Krishnamurthi. Embedding Dynamic Dataflow in a Call{-}by{-}Value Language. In \textit{Proc. European Symposium on Programming}, pp. 294{--}308, 2004. \href{https://doi.org/10.1007/11693024_20}{\Snolinkurl{https://doi.org/10.1007/11693024_20}}}} \\
\hbox{\label{autobiblab:11}} &
\hbox{\label{t:x28autobib_x22Thomas_Hx2e_Cormenx2c_Charles_Ex2e_Leisersonx2c_Ronald_Lx2e_Rivestx2c_and_Clifford_SteinIntroduction_to_Algorithmsx2c_Third_EditionMIT_Press2009x22x29}\Autobibentry{Thomas H. Cormen, Charles E. Leiserson, Ronald L. Rivest, and Clifford Stein. \textit{Introduction to Algorithms, Third Edition}. MIT Press, 2009.}} \\
\hbox{\label{autobiblab:12}} &
\hbox{\label{t:x28autobib_x22Andrea_Ax2e_diSessa_and_Harold_AbelsonBoxerx3a_A_Reconstructible_Computational_MediumCommunications_of_the_ACM_29x289x29x2c_ppx2e_859x2dx2d8681986httpsx3ax2fx2fdoix2eorgx2f10x2e1145x2f6592x2e6595x22x29}\Autobibentry{Andrea A. diSessa and Harold Abelson. Boxer: A Reconstructible Computational Medium. \textit{Communications of the ACM} 29(9), pp. 859{--}868, 1986. \href{https://doi.org/10.1145/6592.6595}{\Snolinkurl{https://doi.org/10.1145/6592.6595}}}} \\
\hbox{\label{autobiblab:13}} &
\hbox{\label{t:x28autobib_x22Rx2e_Kent_DybvigThe_Development_of_Chez_SchemeIn_Procx2e_International_Conference_on_Functional_Programmingx2c_ppx2e_1x2dx2d122006httpsx3ax2fx2fdoix2eorgx2f10x2e1145x2f1160074x2e1159805x22x29}\Autobibentry{R. Kent Dybvig. The Development of Chez Scheme. In \textit{Proc. International Conference on Functional Programming}, pp. 1{--}12, 2006. \href{https://doi.org/10.1145/1160074.1159805}{\Snolinkurl{https://doi.org/10.1145/1160074.1159805}}}} \\
\hbox{\label{autobiblab:14}} &
\hbox{\label{t:x28autobib_x22Andrew_Dx2e_Eisenberg_and_Gregor_KiczalesExpressive_Programs_Through_Presentation_ExtensionIn_Procx2e_International_Conference_on_Aspectx2dOriented_Software_Developmentx2c_ppx2e_73x2dx2d842007httpsx3ax2fx2fdoix2eorgx2f10x2e1145x2f1218563x2e1218573x22x29}\Autobibentry{Andrew D. Eisenberg and Gregor Kiczales. Expressive Programs Through Presentation Extension. In \textit{Proc. International Conference on Aspect{-}Oriented Software Development}, pp. 73{--}84, 2007. \href{https://doi.org/10.1145/1218563.1218573}{\Snolinkurl{https://doi.org/10.1145/1218563.1218573}}}} \\
\hbox{\label{autobiblab:16}} &
\hbox{\label{t:x28autobib_x22Tx2e_Ox2e_Ellisx2c_Jx2e_Fx2e_Heafnerx2c_and_Wx2e_Lx2e_SibleyThe_GRAIL_Language_and_OperationsRAND_Corporationx2c_RMx2d6001x2dARPA1969httpsx3ax2fx2fdoix2eorgx2f10x2e7249x2fRM6001x22x29}\Autobibentry{T. O. Ellis, J. F. Heafner, and W. L. Sibley. The GRAIL Language and Operations. RAND Corporation, RM{-}6001{-}ARPA, 1969a. \href{https://doi.org/10.7249/RM6001}{\Snolinkurl{https://doi.org/10.7249/RM6001}}}} \\
\hbox{\label{autobiblab:g264496}} &
\hbox{\label{t:x28autobib_x22Tx2e_Ox2e_Ellisx2c_Jx2e_Fx2e_Heafnerx2c_and_Wx2e_Lx2e_SibleyThe_Grail_Projectx3a_An_Experiment_in_Manx2dMachine_CommunicationsRAND_Corporationx2c_RMx2d5999x2dARPA1969httpsx3ax2fx2fwwwx2erandx2eorgx2fpubsx2fresearchx5fmemorandax2fRM5999x2ehtmlx22x29}\Autobibentry{T. O. Ellis, J. F. Heafner, and W. L. Sibley. The Grail Project: An Experiment in Man{-}Machine Communications. RAND Corporation, RM{-}5999{-}ARPA, 1969b. \href{https://www.rand.org/pubs/research_memoranda/RM5999.html}{\Snolinkurl{https://www.rand.org/pubs/research_memoranda/RM5999.html}}}} \\
\hbox{\label{autobiblab:17}} &
\hbox{\label{t:x28autobib_x22Sebastian_Erdwegx2c_Lennart_Cx2e_Lx2e_Katsx2c_Tillmann_Rendelx2c_Christian_Kxe4stnerx2c_Klaus_Ostermannx2c_and_Eelco_VisserGrowing_a_Language_Environment_with_Editor_LibrariesIn_Procx2e_Generative_Programming_and_Component_Engineeringx2c_ppx2e_167x2dx2d1762011httpsx3ax2fx2fdoix2eorgx2f10x2e1145x2f2189751x2e2047891x22x29}\Autobibentry{Sebastian Erdweg, Lennart C. L. Kats, Tillmann Rendel, Christian K\"{a}stner, Klaus Ostermann, and Eelco Visser. Growing a Language Environment with Editor Libraries. In \textit{Proc. Generative Programming and Component Engineering}, pp. 167{--}176, 2011. \href{https://doi.org/10.1145/2189751.2047891}{\Snolinkurl{https://doi.org/10.1145/2189751.2047891}}}} \\
\hbox{\label{autobiblab:18}} &
\hbox{\label{t:x28autobib_x22Matthias_Felleisenx2c_Robert_Bruce_Findlerx2c_Matthew_Flattx2c_Shriram_Krishnamurthix2c_Eli_Barzilayx2c_Jay_McCarthyx2c_and_Sam_Tobinx2dHochstadtA_Programmable_Programming_LanguageCommunications_of_the_ACM_61x283x29x2c_ppx2e_62x2dx2d712018httpsx3ax2fx2fdoix2eorgx2f10x2e1145x2f3127323x22x29}\Autobibentry{Matthias Felleisen, Robert Bruce Findler, Matthew Flatt, Shriram Krishnamurthi, Eli Barzilay, Jay McCarthy, and Sam Tobin{-}Hochstadt. A Programmable Programming Language. \textit{Communications of the ACM} 61(3), pp. 62{--}71, 2018. \href{https://doi.org/10.1145/3127323}{\Snolinkurl{https://doi.org/10.1145/3127323}}}} \\
\hbox{\label{autobiblab:19}} &
\hbox{\label{t:x28autobib_x22Robert_Bruce_Findlerx2c_John_Clementsx2c_Cormac_Flanaganx2c_Matthew_Flattx2c_Shriram_Krishnamurthix2c_Paul_Stecklerx2c_and_Matthias_FelleisenDrSchemex3a_A_Programming_Environment_for_SchemeJournal_of_Functional_Programming_12x282x29x2c_ppx2e_159x2dx2d1822002httpsx3ax2fx2fdoix2eorgx2f10x2e1017x2fS0956796801004208x22x29}\Autobibentry{Robert Bruce Findler, John Clements, Cormac Flanagan, Matthew Flatt, Shriram Krishnamurthi, Paul Steckler, and Matthias Felleisen. DrScheme: A Programming Environment for Scheme. \textit{Journal of Functional Programming} 12(2), pp. 159{--}182, 2002. \href{https://doi.org/10.1017/S0956796801004208}{\Snolinkurl{https://doi.org/10.1017/S0956796801004208}}}} \\
\hbox{\label{autobiblab:20}} &
\hbox{\label{t:x28autobib_x22Robert_Bruce_Findler_and_PLTDrRacketx3a_Programming_EnvironmentPLT_Design_Incx2ex2c_PLTx2dTRx2d2010x2d22010httpsx3ax2fx2fracketx2dlangx2eorgx2ftr2x2fx22x29}\Autobibentry{Robert Bruce Findler and PLT. DrRacket: Programming Environment. PLT Design Inc., PLT{-}TR{-}2010{-}2, 2010. \href{https://racket-lang.org/tr2/}{\Snolinkurl{https://racket-lang.org/tr2/}}}} \\
\hbox{\label{autobiblab:21}} &
\hbox{\label{t:x28autobib_x22Matthew_FlattComposable_and_Compilable_Macrosx2c_You_Want_It_Whenx3fIn_Procx2e_International_Conference_on_Functional_Programmingx2c_ppx2e_72x2dx2d832002x22x29}\Autobibentry{Matthew Flatt. Composable and Compilable Macros, You Want It When? In \textit{Proc. International Conference on Functional Programming}, pp. 72{--}83, 2002.}} \\
\hbox{\label{autobiblab:22}} &
\hbox{\label{t:x28autobib_x22Matthew_FlattSubmodules_in_Racketx2c_You_Want_it_Whenx2c_Againx3fIn_Procx2e_Generative_Programmingx3a_Concepts_x26_Experiencesx2c_ppx2e_13x2dx2d222013httpsx3ax2fx2fdoix2eorgx2f10x2e1145x2f2517208x2e2517211x22x29}\Autobibentry{Matthew Flatt. Submodules in Racket, You Want it When, Again? In \textit{Proc. Generative Programming: Concepts \& Experiences}, pp. 13{--}22, 2013. \href{https://doi.org/10.1145/2517208.2517211}{\Snolinkurl{https://doi.org/10.1145/2517208.2517211}}}} \\
\hbox{\label{autobiblab:23}} &
\hbox{\label{t:x28autobib_x22Matthew_Flattx2c_Robert_Bruce_Findlerx2c_and_John_ClementsGUIx3a_Racket_Graphics_ToolkitPLT_Design_Incx2ex2c_PLTx2dTRx2d2010x2d32010httpsx3ax2fx2fracketx2dlangx2eorgx2ftr3x2fx22x29}\Autobibentry{Matthew Flatt, Robert Bruce Findler, and John Clements. GUI: Racket Graphics Toolkit. PLT Design Inc., PLT{-}TR{-}2010{-}3, 2010. \href{https://racket-lang.org/tr3/}{\Snolinkurl{https://racket-lang.org/tr3/}}}} \\
\hbox{\label{autobiblab:24}} &
\hbox{\label{t:x28autobib_x22Matthew_Flattx2c_Robert_Bruce_Findlerx2c_and_Matthias_FelleisenScheme_with_Classesx2c_Mixinsx2c_and_TraitsIn_Procx2e_Asian_Symposium_Programming_Languages_and_Systemsx2c_ppx2e_270x2dx2d2892006x22x29}\Autobibentry{Matthew Flatt, Robert Bruce Findler, and Matthias Felleisen. Scheme with Classes, Mixins, and Traits. In \textit{Proc. Asian Symposium Programming Languages and Systems}, pp. 270{--}289, 2006.}} \\
\hbox{\label{autobiblab:25}} &
\hbox{\label{t:x28autobib_x22Matthew_Flatt_and_PLTReferencex3a_RacketPLT_Design_Incx2ex2c_PLTx2dTRx2d2010x2d12010httpsx3ax2fx2fracketx2dlangx2eorgx2ftr1x2fx22x29}\Autobibentry{Matthew Flatt and PLT. Reference: Racket. PLT Design Inc., PLT{-}TR{-}2010{-}1, 2010. \href{https://racket-lang.org/tr1/}{\Snolinkurl{https://racket-lang.org/tr1/}}}} \\
\hbox{\label{autobiblab:26}} &
\hbox{\label{t:x28autobib_x22Robert_Fourerx2c_David_Mx2e_Gayx2c_and_Brian_Wx2e_KernighanAMPLx3a_A_Modeling_Language_for_Mathematical_Programming2nd_editionx2e_Cengage_Learning2002httpsx3ax2fx2famplx2ecomx2fresourcesx2fthex2damplx2dbookx2fx22x29}\Autobibentry{Robert Fourer, David M. Gay, and Brian W. Kernighan. AMPL: A Modeling Language for Mathematical Programming. 2nd edition. Cengage Learning, 2002. \href{https://ampl.com/resources/the-ampl-book/}{\Snolinkurl{https://ampl.com/resources/the-ampl-book/}}}} \\
\hbox{\label{autobiblab:27}} &
\hbox{\label{t:x28autobib_x22Gx2e_Wx2e_Frenchx2c_Jx2e_Rx2e_Kennawayx2c_and_Ax2e_Mx2e_DayPrograms_as_Visualx2c_Interactive_DocumentsJournal_of_Softwarex3a_Practice_and_Experience_44x288x29x2c_ppx2e_911x2dx2d9302014httpsx3ax2fx2fdoix2eorgx2f10x2e1002x2fspex2e2182x22x29}\Autobibentry{G. W. French, J. R. Kennaway, and A. M. Day. Programs as Visual, Interactive Documents. \textit{Journal of Software: Practice and Experience} 44(8), pp. 911{--}930, 2014. \href{https://doi.org/10.1002/spe.2182}{\Snolinkurl{https://doi.org/10.1002/spe.2182}}}} \\
\hbox{\label{autobiblab:28}} &
\hbox{\label{t:x28autobib_x22Adele_Goldberg_and_David_RobsonSmalltalkx2d80x3a_The_Language_and_Its_ImplementationAddisonx2dWesley_Longman_Publishing_Co1983x22x29}\Autobibentry{Adele Goldberg and David Robson. Smalltalk{-}80: The Language and Its Implementation. Addison{-}Wesley Longman Publishing Co, 1983.}} \\
\hbox{\label{autobiblab:29}} &
\hbox{\label{t:x28autobib_x22Danny_GoodmanThe_Complete_Hypercard_HandbookBantam_Computer_Books1988x22x29}\Autobibentry{Danny Goodman. The Complete Hypercard Handbook. Bantam Computer Books, 1988.}} \\
\hbox{\label{autobiblab:30}} &
\hbox{\label{t:x28autobib_x22Brian_Harvey_and_Jens_Mxf6nigBringing_x5cx22No_Ceilingx5cx22_to_Scratchx3a_Can_One_Language_Serve_Kids_and_Computer_Scientistsx3fIn_Procx2e_Constructionismx2c_ppx2e_1x2dx2d102010x22x29}\Autobibentry{Brian Harvey and Jens M\"{o}nig. Bringing "No Ceiling" to Scratch: Can One Language Serve Kids and Computer Scientists? In \textit{Proc. Constructionism}, pp. 1{--}10, 2010.}} \\
\hbox{\label{autobiblab:31}} &
\hbox{\label{t:x28autobib_x22Brian_Hempelx2c_Justin_Lubinx2c_Grace_Lux2c_and_Ravi_ChughDeucex3a_A_Lightweight_User_Interface_for_Structured_EditingIn_Procx2e_International_Conference_on_Software_Engineeringx2c_ppx2e_654x2dx2d6642018httpsx3ax2fx2fdoix2eorgx2f10x2e1145x2f3180155x2e3180165x22x29}\Autobibentry{Brian Hempel, Justin Lubin, Grace Lu, and Ravi Chugh. Deuce: A Lightweight User Interface for Structured Editing. In \textit{Proc. International Conference on Software Engineering}, pp. 654{--}664, 2018. \href{https://doi.org/10.1145/3180155.3180165}{\Snolinkurl{https://doi.org/10.1145/3180155.3180165}}}} \\
\hbox{\label{autobiblab:32}} &
\hbox{\label{t:x28autobib_x22Daniel_Ingallsx2c_Krzysztof_Palaczx2c_Stephen_Uhlerx2c_Antero_Taivalsaarix2c_and_Tommi_MikkonenThe_Lively_Kernel_A_Selfx2dsupporting_System_on_a_Web_PageIn_Procx2e_Selfx2dSustaining_Systemsx2c_ppx2e_31x2dx2d502008httpsx3ax2fx2fdoix2eorgx2f10x2e1007x2f978x2d3x2d540x2d89275x2d5x5f2x22x29}\Autobibentry{Daniel Ingalls, Krzysztof Palacz, Stephen Uhler, Antero Taivalsaari, and Tommi Mikkonen. The Lively Kernel A Self{-}supporting System on a Web Page. In \textit{Proc. Self{-}Sustaining Systems}, pp. 31{--}50, 2008. \href{https://doi.org/10.1007/978-3-540-89275-5_2}{\Snolinkurl{https://doi.org/10.1007/978-3-540-89275-5_2}}}} \\
\hbox{\label{autobiblab:33}} &
\hbox{\label{t:x28autobib_x22Lennart_Cx2e_Lx2e_Kats_and_Eelco_VisserThe_Spoofax_Language_WorkbenchIn_Procx2e_Objectx2dOriented_Programmingx2c_Systemsx2c_Languages_x26_Applicationsx2c_ppx2e_444x2dx2d4632010httpsx3ax2fx2fdoix2eorgx2f10x2e1145x2f1932682x2e1869497x22x29}\Autobibentry{Lennart C. L. Kats and Eelco Visser. The Spoofax Language Workbench. In \textit{Proc. Object{-}Oriented Programming, Systems, Languages \& Applications}, pp. 444{--}463, 2010. \href{https://doi.org/10.1145/1932682.1869497}{\Snolinkurl{https://doi.org/10.1145/1932682.1869497}}}} \\
\hbox{\label{autobiblab:34}} &
\hbox{\label{t:x28autobib_x22Clemens_Nx2e_Klokmosex2c_James_Rx2e_Eaganx2c_Siemen_Baaderx2c_Wendy_Mackayx2c_and_Michel_Beaudouinx2dLafonWebstratesx3a_Shareable_Dynamic_MediaIn_Procx2e_ACM_Symposium_on_User_Interface_Software_and_Technologyx2c_ppx2e_280x2dx2d2902015httpsx3ax2fx2fdoix2eorgx2f10x2e1145x2f2807442x2e2807446x22x29}\Autobibentry{Clemens N. Klokmose, James R. Eagan, Siemen Baader, Wendy Mackay, and Michel Beaudouin{-}Lafon. Webstrates: Shareable Dynamic Media. In \textit{Proc. ACM Symposium on User Interface Software and Technology}, pp. 280{--}290, 2015. \href{https://doi.org/10.1145/2807442.2807446}{\Snolinkurl{https://doi.org/10.1145/2807442.2807446}}}} \\
\hbox{\label{autobiblab:35}} &
\hbox{\label{t:x28autobib_x22Amy_Ko_and_Brad_Ax2e_MyersBaristax3a_An_Implementation_Framework_for_Enabling_New_Toolsx2c_Interaction_Techniques_and_Views_in_Code_EditorsIn_Procx2e_Conference_on_Human_Factors_in_Computing_Systemsx2c_ppx2e_387x2dx2d3962006httpsx3ax2fx2fdoix2eorgx2f10x2e1145x2f1124772x2e1124831x22x29}\Autobibentry{Amy Ko and Brad A. Myers. Barista: An Implementation Framework for Enabling New Tools, Interaction Techniques and Views in Code Editors. In \textit{Proc. Conference on Human Factors in Computing Systems}, pp. 387{--}396, 2006. \href{https://doi.org/10.1145/1124772.1124831}{\Snolinkurl{https://doi.org/10.1145/1124772.1124831}}}} \\
\hbox{\label{autobiblab:36}} &
\hbox{\label{t:x28autobib_x22John_Maloneyx2c_Kimberly_Mx2e_Rosex2c_and_Walt_Disney_ImagineeringAn_Introduction_to_Morphicx3a_The_Squeak_User_Interface_FrameworkIn_Squeakx3a_Open_Personal_Computing_and_Multimediax2c_ppx2e_39x2dx2d77_Pearson2001x22x29}\Autobibentry{John Maloney, Kimberly M. Rose, and Walt Disney Imagineering. An Introduction to Morphic: The Squeak User Interface Framework. In \textit{Squeak: Open Personal Computing and Multimedia}, pp. 39{--}77 Pearson, 2001.}} \\
\hbox{\label{autobiblab:37}} &
\hbox{\label{t:x28autobib_x22MicrosoftOffice_and_SharePoint_Development_in_Visual_StudioRetrieved_Januaryx2c_20192019httpsx3ax2fx2fdocsx2emicrosoftx2ecomx2fenx2dusx2fvisualstudiox2fvstox2fofficex2dandx2dsharepointx2ddevelopmentx2dinx2dvisualx2dstudiox3fviewx3dvsx2d2017x22x29}\Autobibentry{Microsoft. Office and SharePoint Development in Visual Studio. Retrieved January, 2019, 2019. \href{https://docs.microsoft.com/en-us/visualstudio/vsto/office-and-sharepoint-development-in-visual-studio?view=vs-2017}{\Snolinkurl{https://docs.microsoft.com/en-us/visualstudio/vsto/office-and-sharepoint-development-in-visual-studio?view=vs-2017}}}} \\
\hbox{\label{autobiblab:38}} &
\hbox{\label{t:x28autobib_x22Chris_OkasakiRedx2dblack_Trees_in_a_Functional_SettingJournal_of_Functional_Programming_9x284x29x2c_ppx2e_471x2dx2d4771999httpsx3ax2fx2fdoix2eorgx2f10x2e1017x2fS0956796899003494x22x29}\Autobibentry{Chris Okasaki. Red{-}black Trees in a Functional Setting. \textit{Journal of Functional Programming} 9(4), pp. 471{--}477, 1999. \href{https://doi.org/10.1017/S0956796899003494}{\Snolinkurl{https://doi.org/10.1017/S0956796899003494}}}} \\
\hbox{\label{autobiblab:39}} &
\hbox{\label{t:x28autobib_x22Cyrus_Omarx2c_Nick_Collinsx2c_David_Moonx2c_Ian_Voyseyx2c_and_Ravi_ChughLivelitsx3a_Filling_Typed_Holes_with_Live_GUIs_x28Extended_Abstractx29In_Procx2e_Workshop_on_Typex2ddriven_Development2019x22x29}\Autobibentry{Cyrus Omar, Nick Collins, David Moon, Ian Voysey, and Ravi Chugh. Livelits: Filling Typed Holes with Live GUIs (Extended Abstract). In \textit{Proc. Workshop on Type{-}driven Development}, 2019.}} \\
\hbox{\label{autobiblab:40}} &
\hbox{\label{t:x28autobib_x22Cyrus_Omarx2c_YoungSeok_Yoonx2c_Thomas_Dx2e_LaTozax2c_and_Brad_Ax2e_MyersActive_Code_CompletionIn_Procx2e_International_Conference_on_Software_Engineeringx2c_ppx2e_859x2dx2d8692012x22x29}\Autobibentry{Cyrus Omar, YoungSeok Yoon, Thomas D. LaToza, and Brad A. Myers. Active Code Completion. In \textit{Proc. International Conference on Software Engineering}, pp. 859{--}869, 2012.}} \\
\hbox{\label{autobiblab:41}} &
\hbox{\label{t:x28autobib_x22Mark_OvermarsTeaching_Computer_Science_Through_Game_DesignComputer_37x284x29x2c_ppx2e_81x2dx2d832004httpsx3ax2fx2fdoix2eorgx2f10x2e1109x2fMCx2e2004x2e1297314x22x29}\Autobibentry{Mark Overmars. Teaching Computer Science Through Game Design. \textit{Computer} 37(4), pp. 81{--}83, 2004. \href{https://doi.org/10.1109/MC.2004.1297314}{\Snolinkurl{https://doi.org/10.1109/MC.2004.1297314}}}} \\
\hbox{\label{autobiblab:42}} &
\hbox{\label{t:x28autobib_x22Vaclav_Pechx2c_Alex_Shatalinx2c_and_Markus_VoelterJetBrains_MPS_as_a_Tool_for_Extending_JavaIn_Procx2e_Principles_and_Practice_of_Programming_in_Javax2c_ppx2e_165x2dx2d1682013httpsx3ax2fx2fdoix2eorgx2f10x2e1145x2f2500828x2e2500846x22x29}\Autobibentry{Vaclav Pech, Alex Shatalin, and Markus Voelter. JetBrains MPS as a Tool for Extending Java. In \textit{Proc. Principles and Practice of Programming in Java}, pp. 165{--}168, 2013. \href{https://doi.org/10.1145/2500828.2500846}{\Snolinkurl{https://doi.org/10.1145/2500828.2500846}}}} \\
\hbox{\label{autobiblab:43}} &
\hbox{\label{t:x28autobib_x22Fernando_Perez_and_Brian_Ex2e_GrangerIPythonx3a_A_System_for_Interactive_Scientific_ComputingComputing_in_Science_and_Engineering_9x283x29x2c_ppx2e_21x2dx2d292007httpsx3ax2fx2fdoix2eorgx2f10x2e1109x2fMCSEx2e2007x2e53x22x29}\Autobibentry{Fernando Perez and Brian E. Granger. IPython: A System for Interactive Scientific Computing. \textit{Computing in Science and Engineering} 9(3), pp. 21{--}29, 2007. \href{https://doi.org/10.1109/MCSE.2007.53}{\Snolinkurl{https://doi.org/10.1109/MCSE.2007.53}}}} \\
\hbox{\label{autobiblab:44}} &
\hbox{\label{t:x28autobib_x22Jon_PostelTransmission_Control_ProtocolInternet_Engineering_Task_Forcex2c_RFC_7931981httpsx3ax2fx2ftoolsx2eietfx2eorgx2fhtmlx2frfc793x22x29}\Autobibentry{Jon Postel. Transmission Control Protocol. Internet Engineering Task Force, RFC 793, 1981. \href{https://tools.ietf.org/html/rfc793}{\Snolinkurl{https://tools.ietf.org/html/rfc793}}}} \\
\hbox{\label{autobiblab:45}} &
\hbox{\label{t:x28autobib_x22Mitchel_Resnickx2c_John_Maloneyx2c_Andrxe9s_Monroyx2dHernxe1ndezx2c_Natalie_Ruskx2c_Evelyn_Eastmondx2c_Karen_Brennanx2c_Amon_Millnerx2c_Eric_Rosenbaumx2c_Jay_Silverx2c_Brian_Silvermanx2c_and_Yasmin_KafaiScratchx3a_Programming_for_AllCommunications_of_the_ACM_52x2811x29x2c_ppx2e_60x2dx2d672009httpsx3ax2fx2fdoix2eorgx2f10x2e1145x2f1592761x2e1592779x22x29}\Autobibentry{Mitchel Resnick, John Maloney, Andr\'{e}s Monroy{-}Hern\'{a}ndez, Natalie Rusk, Evelyn Eastmond, Karen Brennan, Amon Millner, Eric Rosenbaum, Jay Silver, Brian Silverman, and Yasmin Kafai. Scratch: Programming for All. \textit{Communications of the ACM} 52(11), pp. 60{--}67, 2009. \href{https://doi.org/10.1145/1592761.1592779}{\Snolinkurl{https://doi.org/10.1145/1592761.1592779}}}} \\
\hbox{\label{autobiblab:46}} &
\hbox{\label{t:x28autobib_x22Roman_Rxe4dlex2c_Midas_Nouwensx2c_Kristian_Antonsenx2c_James_Rx2e_Eaganx2c_and_Clemens_Nx2e_KlokmoseCodestratesx3a_Literate_Computing_with_WebstratesIn_Procx2e_ACM_Symposium_on_User_Interface_Software_and_Technologyx2c_ppx2e_715x2dx2d7252017httpsx3ax2fx2fdoix2eorgx2f10x2e1145x2f3126594x2e3126642x22x29}\Autobibentry{Roman R\"{a}dle, Midas Nouwens, Kristian Antonsen, James R. Eagan, and Clemens N. Klokmose. Codestrates: Literate Computing with Webstrates. In \textit{Proc. ACM Symposium on User Interface Software and Technology}, pp. 715{--}725, 2017. \href{https://doi.org/10.1145/3126594.3126642}{\Snolinkurl{https://doi.org/10.1145/3126594.3126642}}}} \\
\hbox{\label{autobiblab:47}} &
\hbox{\label{t:x28autobib_x22Charles_Simonyix2c_Magnus_Christersonx2c_and_Shane_CliffordIntentional_SoftwareACM_SIGPLAN_Notices_41x2810x29x2c_ppx2e_451x2dx2d4642006httpsx3ax2fx2fdoix2eorgx2f10x2e1145x2f1167515x2e1167511x22x29}\Autobibentry{Charles Simonyi, Magnus Christerson, and Shane Clifford. Intentional Software. \textit{ACM SIGPLAN Notices} 41(10), pp. 451{--}464, 2006. \href{https://doi.org/10.1145/1167515.1167511}{\Snolinkurl{https://doi.org/10.1145/1167515.1167511}}}} \\
\hbox{\label{autobiblab:48}} &
\hbox{\label{t:x28autobib_x22Markus_Voelter_and_Sascha_LissonSupporting_Diverse_Notations_in_MPSx2019_Projectional_EditorIn_Procx2e_International_Workshop_on_The_Globalization_of_Modeling_Languages2014x22x29}\Autobibentry{Markus Voelter and Sascha Lisson. Supporting Diverse Notations in MPS{'} Projectional Editor. In \textit{Proc. International Workshop on The Globalization of Modeling Languages}, 2014.}} \\
\hbox{\label{autobiblab:49}} &
\hbox{\label{t:x28autobib_x22Markus_Voelterx2c_Daniel_Ratiux2c_Bernhard_Schaetzx2c_and_Bernd_Kolbmbeddrx3a_an_Extensible_Cx2dbased_Programming_Language_and_IDE_for_Embedded_SystemsIn_Procx2e_Conference_on_Systemsx2c_Programmingx2c_and_Applicationsx3a_Software_for_Humanityx2c_ppx2e_121x2dx2d1402012httpsx3ax2fx2fdoix2eorgx2f10x2e1145x2f2384716x2e2384767x22x29}\Autobibentry{Markus Voelter, Daniel Ratiu, Bernhard Schaetz, and Bernd Kolb. mbeddr: an Extensible C{-}based Programming Language and IDE for Embedded Systems. In \textit{Proc. Conference on Systems, Programming, and Applications: Software for Humanity}, pp. 121{--}140, 2012. \href{https://doi.org/10.1145/2384716.2384767}{\Snolinkurl{https://doi.org/10.1145/2384716.2384767}}}} \\
\hbox{\label{autobiblab:50}} &
\hbox{\label{t:x28autobib_x22Holger_Vogtx2c_Marcel_Hendrixx2c_and_Paolo_NenziNgspice_Users_ManualNGSPICEx2c_302019httpx3ax2fx2fngspicex2esourceforgex2enetx2fdocsx2fngspicex2d30x2dmanualx2epdfx22x29}\Autobibentry{Holger Vogt, Marcel Hendrix, and Paolo Nenzi. Ngspice Users Manual. NGSPICE, 30, 2019. \href{http://ngspice.sourceforge.net/docs/ngspice-30-manual.pdf}{\Snolinkurl{http://ngspice.sourceforge.net/docs/ngspice-30-manual.pdf}}}} \\
\hbox{\label{autobiblab:51}} &
\hbox{\label{t:x28autobib_x22Stephen_WolframThe_Mathematica_BookFourth_editionx2e_Cambridge_University_Press1988x22x29}\Autobibentry{Stephen Wolfram. The Mathematica Book. Fourth edition. Cambridge University Press, 1988.}}\end{bigtabular}\end{AutoBibliography}

\postDoc
\end{document}